\documentclass[12pt]{article}
\usepackage{fullpage}
\usepackage{times}

\usepackage[
pageanchor=true,
plainpages=false,
pdfpagelabels, %
bookmarks,
bookmarksnumbered,
pdfborder=0 0 0,
pdfpagemode=UseOutlines]{hyperref}

\usepackage{doi}

\usepackage{amsthm}
\def\qedhere{}

\theoremstyle{plain}
\newtheorem{theorem}{Theorem}
\newtheorem{lemma}[theorem]{Lemma}

\newtheorem{corollary}[theorem]{Corollary}

\theoremstyle{definition}
\newtheorem{definition}{Definition}
\theoremstyle{remark}
\newtheorem{remark}{Remark}
\newtheorem{example}{Example}

\newif\iflongversion
\longversiontrue

\newcommand*{\contextapp}[2]{#1\text{\footnotesize$($}#2\text{\footnotesize$)$}}%

\usepackage{amsmath}
\RequirePackage{stmaryrd} %

\usepackage{tikz}
\usetikzlibrary{matrix}

\tikzstyle{transition}=[-stealth]
\tikzstyle{similar state}=[double]
\tikzstyle{transition exists}=[dashed]
\tikzstyle{equivalence}=[<->,double]

\iflongversion
\def\qedhere{}
\usepackage[keepproof]{endproof2}
\else
\usepackage[noproof]{endproof2}
\fi

\usepackage[bbsets,Dfprime]{math}
\usepackage[modernsign,substopindex,shortmquant,mquantifiertype,mconnectiveformal,bracketinterpret,postfixinterpret,bracketmodalinterpret,setfixinterpret,modifopindex,seqarrow,seqoptional,sidenotecalculus,abbrseqcontext,shortterms,nosigmaterms,novarterms]{logic}
\usepackage[pretest,nocommandblocks]{progreg}
\usepackage[bracketinterpret,postfixinterpret,bracketmodalinterpret]{dL}

\usepackage{prettyref}
\newcommand{\rref}[2][]{\prettyref{#2}}
\newrefformat{sec}{Section\,\ref{#1}}
\newrefformat{def}{Def.\,\ref{#1}}
\newrefformat{thm}{Theorem\,\ref{#1}}
\newrefformat{prop}{Proposition\,\ref{#1}}
\newrefformat{lem}{Lemma\,\ref{#1}}
\newrefformat{cor}{Corollary\,\ref{#1}}
\newrefformat{ex}{Example\,\ref{#1}}
\newrefformat{tab}{Table\,\ref{#1}}
\newrefformat{fig}{Fig.\,\ref{#1}}
\newrefformat{rem}{Remark\,\ref{#1}}

\newcommand{\bebecomes}{\mathrel{::=}}
\newcommand{\alternative}{~|~}
\providecommand{\dfn}[2][]{\emph{#2}}

\newcommand*{\genDE}[1]{\theta}%
\newcommand{\ivr}{\psi}

\let\alignOrg\align
\def\align{\vspace{-0.5\baselineskip}\alignOrg}

\newcommand{\der}[1]{(#1)'}%

\newcommand{\admissible}{\text{admissible}\xspace}

\newcommand{\solf}{\bar{x}}%

\renewcommand{\linterpretationsname}{\mathcal{S}}

\newcommand{\I}{\vdLint[const=I,state=\nu]}
\newcommand{\It}{\vdLint[const=I,state=\omega]}
\newcommand{\If}{\DALint[const=I,flow=\varphi]}
\newcommand*{\Iff}[1][\zeta]{\vdLint[const=I,state=\varphi(#1)]}%

\newcommand{\Ia}{\iadjointSubst{\sigma}{\I}}%
\newcommand{\Ita}{\iadjointSubst{\sigma}{\It}}%
\newcommand{\Iminner}{\imodif[const]{\I}{\,\usarg}{d}}%

\newcommand{\Ialt}{\vdLint[const=J,state=\tilde{\nu}]}%
\newcommand{\Italt}{\vdLint[const=J,state=\tilde{\omega}]}%

  \makeatletter
  \renewcommand{\iadjointSubst}[2]{%
    \useinterpretation{#2}%
    \edef\tmpadjointconst{{#1}^*_{\Interpretation@state}{\Interpretation@const}}%
    \iconcat[const=\tmpadjointconst]{#2}
  }
  \makeatother

\newcommand{\applyusubst}[2]{#1(#2)}%

\newcommand{\usubst}[3][]{\subst[#1]{#2(\usarg)}{#3(\usarg)}}%
\newcommand{\preusubst}[2][]{#1}%

\newcommand{\usubstlist}[1]{\{#1\}}
\newcommand{\usubstmod}[2]{#1\mapsto#2}%

\newcommand{\usarg}{\usebox{\USarg}}%
\newcommand{\uscarg}{\usebox{\UScarg}}%
\newcommand{\usall}{\bar{x}}
\newcommand*{\allvars}{\mathcal{V}}

\newcommand*{\intsigns}[1]{\Sigma(#1)}
\newcommand*{\vars}[1]{\mathop{\text{V}}(#1)}
\newcommand*{\freevars}[1]{\mathop{\text{FV}}(#1)}
\newcommand*{\boundvars}[1]{\mathop{\text{BV}}(#1)}
\newcommand*{\mustboundvars}[1]{\mathop{\text{MBV}}(#1)}
\newcommand*{\replacees}[1]{#1}

\newcommand{\predicational}{quantifier\xspace}%
\newcommand{\Predicational}{Quantifier\xspace}%

\usepackage{ifpdf}
\ifpdf
\fi

\usepackage{fancyhdr}
\title{A Uniform Substitution Calculus\\for Differential Dynamic Logic}
\author{Andr\'e Platzer\thanks{
  Computer Science Department, Carnegie Mellon University, Pittsburgh, USA
  aplatzer@cs.cmu.edu
} 
}

\begin{document}
\maketitle
\allowdisplaybreaks
\thispagestyle{empty}

\begin{abstract}
This paper introduces a new proof calculus for \emph{differential dynamic logic} (\dL) that is entirely based on \emph{uniform substitution}, a proof rule that substitutes a formula for a predicate symbol everywhere.
Uniform substitutions make it possible to rely on \emph{axioms} rather than axiom schemata, substantially simplifying implementations.
Instead of subtle schema variables and soundness-critical side conditions on the occurrence patterns of variables, the resulting calculus adopts only a finite number of ordinary \dL formulas as axioms.
The static semantics of differential dynamic logic is captured exclusively in uniform substitutions and bound variable renamings as opposed to being spread in delicate ways across the prover implementation.
In addition to sound uniform substitutions, this paper introduces \emph{differential forms} for differential dynamic logic that make it possible to internalize differential invariants, differential substitutions, and derivations as first-class axioms in \dL.
\\[\medskipamount]
\end{abstract}

\section{Introduction}

\emph{Differential dynamic logic} (\dL) \cite{DBLP:journals/jar/Platzer08,DBLP:conf/lics/Platzer12b} is a logic for proving correctness properties of hybrid systems.
It has a sound and complete proof calculus relative to differential equations \cite{DBLP:journals/jar/Platzer08,DBLP:conf/lics/Platzer12b} and a sound and complete proof calculus relative to discrete systems \cite{DBLP:conf/lics/Platzer12b}.
Both sequent calculi \cite{DBLP:journals/jar/Platzer08} and Hilbert-type axiomatizations \cite{DBLP:conf/lics/Platzer12b} have been presented for \dL but only the former has been implemented.
The implementation of \dL's sequent calculus in \KeYmaera\iflongversion \cite{DBLP:conf/cade/PlatzerQ08}\fi makes it straightforward for users to prove properties of hybrid systems, because it provides rules performing natural decompositions for each operator.
The downside is that the implementation of the rule schemata and their side conditions on occurrence constraints and relations of reading and writing of variables as well as rule applications in context is nontrivial and inflexible in \KeYmaera.

The goal of this paper is to identify how to make it straightforward to implement the axioms and proof rules of differential dynamic logic by writing down a finite list of \emph{axioms} (concrete formulas, not axiom schemata that represent an infinite list of axioms subject to sophisticated soundness-critical schema variable matching implementations).
They require multiple axioms to be combined with one another to obtain the effect that a user would want for proving a hybrid system conjecture.
This paper argues that this is still a net win for hybrid systems, because a substantially simpler prover core is easier to implement correctly, and the need to combine multiple axioms to obtain user-level proof steps can be achieved equally well by appropriate tactics, which are not soundness-critical.

To achieve this goal, this paper follows observations for differential game logic \cite{DBLP:journals/corr/Platzer14:dGL} that highlight the significance and elegance of \emph{uniform substitutions}, a classical proof rule for first-order logic \cite[\S35,40]{Church_1956}.
Uniform substitutions uniformly instantiate predicate and function symbols with formulas and terms, respectively, as functions of their  arguments.
In the presence of the nontrivial binding structure that nondeterminism and differential equations of hybrid programs induce for the dynamic modalities of differential dynamic logic, flexible but sound uniform substitutions become more complex for \dL, but can still be read off elegantly from its static semantics.
In fact, \dL's static semantics is solely captured\footnote{
This approach is dual to other successful ways of solving the intricacies and subtleties of substitutions \cite{DBLP:journals/jsl/Church40,DBLP:journals/jsl/Henkin53} by imposing occurrence side conditions on axiom schemata and proof rules, which is what uniform substitutions can get rid of.
}
in the implementation of uniform substitution (and bound variable renaming), thereby leading to a completely modular proof calculus.

This paper introduces a static and dynamic semantics for \emph{differential-form} \dL, proves coincidence lemmas and uniform substitution lemmas, culminating in a soundness proof for uniform substitutions (\rref{sec:usubst}).
It exploits the new \emph{differential forms} that this paper adds to \dL for internalizing differential invariants \cite{DBLP:journals/logcom/Platzer10}, differential cuts \cite{DBLP:journals/logcom/Platzer10,DBLP:journals/lmcs/Platzer12},  differential ghosts \cite{DBLP:journals/lmcs/Platzer12}, differential substitutions, total differentials and Lie-derivations \cite{DBLP:journals/logcom/Platzer10,DBLP:journals/lmcs/Platzer12} as first-class citizens in \dL, culminating in entirely modular axioms for differential equations and a superbly modular soundness proof (\rref{sec:dL-axioms}).
This approach is to be contrasted with earlier approaches for differential invariants that were based on complex built-in rules \cite{DBLP:journals/logcom/Platzer10,DBLP:journals/lmcs/Platzer12}.
The relationship to related work from previous presentations of differential dynamic logic \cite{DBLP:journals/jar/Platzer08,DBLP:conf/lics/Platzer12b} continues to apply except that \dL now internalizes differential equation reasoning axiomatically via differential forms.

\newsavebox{\Rval}%
\sbox{\Rval}{$\scriptstyle\mathbb{R}$}
\irlabel{qear|\usebox{\Rval}}

\newsavebox{\USarg}%
\sbox{\USarg}{$\boldsymbol{\cdot}$}

\newsavebox{\UScarg}%
\sbox{\UScarg}{$\boldsymbol{\_}$}

\newsavebox{\Lightningval}%
\sbox{\Lightningval}{$\scriptstyle\textcolor{red}{\lightning}$}
\irlabel{clash|clash\usebox{\Lightningval}} %
\irlabel{unsound|\usebox{\Lightningval}}  %

\section{Differential-Form Differential Dynamic Logic}
\subsection{Syntax}
Formulas and hybrid programs (\HPs) of \dL are defined by simultaneous induction based on the following definition of terms.
Similar simultaneous inductions are used throughout the proofs for \dL.
The set of all \emph{variables} is $\allvars$.
For any $V\subseteq\allvars$ is \(\D{V}\mdefeq\{\D{x} : x\in V\}\) the set of \emph{differential symbols} $\D{x}$ for the variables in $V$.
Function symbols are written $f,g,h$, predicate symbols $p,q,r$, and variables $x,y,z\in\allvars$ with differential symbols $\D{x},\D{y},\D{z}\in\D{\allvars}$.
Program constants are $a,b,c$.

\begin{definition}[Terms]
\emph{Terms} are defined by this grammar
(with $\theta,\eta,\theta_1,\dots,\theta_k$ as terms, $x\in\allvars$ as variable, $\D{x}\in\D{\allvars}$ differential symbol, and $f$ function symbol):
\[
  \theta,\eta ~\bebecomes~
  x
  \alternative \D{x}
  \alternative
  f(\theta_1,\dots,\theta_k)
  \alternative
  \theta+\eta
  \alternative
  \theta\cdot\eta
  \alternative \der{\theta}
\]
\end{definition}
Number literals such as 0,1 are allowed as function symbols without arguments that are always interpreted as the numbers they denote.
Beyond differential symbols $\D{x}$, \emph{differential-form \dL} allows \emph{differentials} \(\der{\theta}\) of terms $\theta$ as terms for the purpose of axiomatically internalizing reasoning about differential equations.
\begin{definition}[Hybrid program]
\emph{Hybrid programs} (\HPs) are defined by the following grammar (with $\alpha,\beta$ as \HPs, program constant $a$, variable $x$, term $\theta$ possibly containing $x$, and formula $\ivr$ of first-order logic of real arithmetic):
\[
  \alpha,\beta ~\bebecomes~
  a\alternative
  \pupdate{\pumod{x}{\theta}}
  \alternative \Dupdate{\Dumod{\D{x}}{\theta}}
  \alternative
  \ptest{\ivr}
  \alternative
  \pevolvein{\D{x}=\genDE{x}}{\ivr}
  \alternative
  \alpha\cup\beta
  \alternative
  \alpha;\beta
  \alternative
  \prepeat{\alpha}
\]
\end{definition}
\emph{Assignments} \m{\pupdate{\pumod{x}{\theta}}} of $\theta$ to variable $x$, \emph{tests} \m{\ptest{\ivr}} of the formula $\ivr$ in the current state, \emph{differential equations} \(\pevolvein{\D{x}=\genDE{x}}{\ivr}\) restricted to the evolution domain constraint $\ivr$, \emph{nondeterministic choices} \(\pchoice{\alpha}{\beta}\), \emph{sequential compositions} \(\alpha;\beta\), and \emph{nondeterministic repetition} \(\prepeat{\alpha}\) are as usual in \dL \cite{DBLP:journals/jar/Platzer08,DBLP:conf/lics/Platzer12b}.
The effect of the \emph{differential assignment} \m{\Dupdate{\Dumod{\D{x}}{\theta}}} to differential symbol $\D{x}$ is similar to the effect of the assignment \m{\pupdate{\pumod{x}{\theta}}} to variable $x$, except that it changes the value of the differential symbol $\D{x}$ around instead of the value of $x$.
It is not to be confused with the differential equation \(\pevolve{\D{x}=\genDE{x}}\), which will follow said differential equation continuously for an arbitrary amount of time.
The differential assignment \m{\Dupdate{\Dumod{\D{x}}{\theta}}}, instead, only assigns the value of $\theta$ to the differential symbol $\D{x}$ discretely once at an instant of time.
Program constants $a$ are uninterpreted, i.e.\ their behavior depends on the interpretation in the same way that the values of function symbols  $f$ and predicate symbols $p$ depends on their interpretation.

\begin{definition}[\dL formula]
The \emph{formulas of (differential-form) differential dynamic logic} ({\dL}) are defined by the grammar
(with \dL formulas $\phi,\psi$, terms $\theta,\eta,\theta_1,\dots,\theta_k$, predicate symbol $p$, \predicational symbol $C$, variable $x$, \HP $\alpha$):
  \[
  \phi,\psi ~\bebecomes~
  \theta\geq\eta \alternative
  p(\theta_1,\dots,\theta_k) \alternative
  \contextapp{C}{\phi} \alternative
  \lnot \phi \alternative
  \phi \land \psi \alternative
  \lforall{x}{\phi} \alternative 
  \lexists{x}{\phi} \alternative
  \dbox{\alpha}{\phi}
  \alternative \ddiamond{\alpha}{\phi}
  \]
\end{definition}
Operators $>,\leq,<,\lor,\limply,\lbisubjunct$ are definable, e.g., \(\phi\limply\psi\) as \(\lnot(\phi\land\lnot\psi)\).
Likewise \(\dbox{\alpha}{\phi}\) is equivalent to \(\lnot\ddiamond{\alpha}{\lnot\phi}\) and \(\lforall{x}{\phi}\) equivalent to \(\lnot\lexists{x}{\lnot\phi}\).
The modal formula \(\dbox{\alpha}{\phi}\) expresses that $\phi$ holds after all runs of $\alpha$, while the dual \(\ddiamond{\alpha}{\phi}\) expresses that there is a run of $\alpha$ after which $\phi$ holds.
\emph{\Predicational symbols} $C$ (with formula $\phi$ as argument), i.e.\ higher-order predicate symbols that bind all variables of $\phi$, are unnecessary but internalize contextual congruence reasoning efficiently.

\subsection{Dynamic Semantics}

A state is a mapping from variables $\allvars$ 
and differential symbols $\D{\allvars}$ to $\reals$.
The set of states is denoted \(\linterpretations{\Sigma}{V}\).
Let
\m{\iget[state]{\imodif[state]{\I}{x}{r}}} denote the state that agrees with state~$\iget[state]{\I}$ except for the value of variable~\m{x}, which is changed to~\m{r \in \reals}, and accordingly for \m{\iget[state]{\imodif[state]{\I}{x'}{r}}}. %
The interpretation of a function symbol $f$ with arity $n$ (i.e.\ with $n$ arguments) is a smooth function \(\iget[const]{\I}(f):\reals^n\to\reals\) of $n$ arguments.

\begin{definition}[Semantics of terms] \label{def:dL-valuationTerm}
For each interpretation $\iget[const]{\I}$, the \emph{semantics of a term} $\theta$ in a state $\iget[state]{\I}\in\linterpretations{\Sigma}{V}$ is its value in $\reals$.
It is defined inductively as follows
\begin{compactenum}
\item \m{\ivaluation{\I}{x} = \iget[state]{\I}(x)} for variable $x\in\allvars$
\item \m{\ivaluation{\I}{\D{x}} = \iget[state]{\I}(\D{x})} for differential symbol $\D{x}\in\D{\allvars}$
\item \(\ivaluation{\I}{f(\theta_1,\dots,\theta_k)} = \iget[const]{\I}(f)\big(\ivaluation{\I}{\theta_1},\dots,\ivaluation{\I}{\theta_k}\big)\) for function symbol $f$
\item \m{\ivaluation{\I}{\theta+\eta} = \ivaluation{\I}{\theta} + \ivaluation{\I}{\eta}}
\item \m{\ivaluation{\I}{\theta\cdot\eta} = \ivaluation{\I}{\theta} \cdot \ivaluation{\I}{\eta}}
\item
\m{
\ivaluation{\I}{\der{\theta}}
\displaystyle
=
\newcommand{\Idot}{\vdLint[const=I,state=]}
\sum_x \iget[state]{\I}(\D{x}) \Dp[x]{\ivaluation{\Idot}{\theta}}(\iget[state]{\I})
=
\def\Im{\imodif[state]{\I}{x}{X}}%
\sum_x \iget[state]{\I}(\D{x})
 \itimes \Dp[X]{\ivaluation{\Im}{\theta}}
}
\end{compactenum}
\end{definition}

\noindent
Time-derivatives are undefined in an isolated state $\iget[state]{\I}$.
The clou is that differentials can still be given a local semantics:
\m{\ivaluation{\I}{\der{\theta}}} is the sum of all (analytic) spatial partial derivatives of the value of $\theta$ by all variables $x$ (or rather their values $X$) multiplied by the corresponding tangent described by the value $\iget[state]{\I}(\D{x})$ of differential symbol $\D{x}$.
That sum over all variables $x\in\allvars$ has finite support, because $\theta$ only mentions finitely many variables $x$ and the partial derivative by variables $x$ that do not occur in $\theta$ is 0.
The spatial derivatives exist since $\ivaluation{\I}{\theta}$ is a composition of smooth functions, so smooth.
Thus, the semantics of \m{\ivaluation{\I}{\der{\theta}}} is the \emph{differential}\footnote{%
A slight abuse of notation rewrites the differential as \(\ivalues{\I}{\der{\theta}} = d\ivalues{\I}{\theta} = \sum_{i=1}^n \Dp[x^i]{\ivalues{\I}{\theta}} dx^i\)
when $x^1,\dots,x^n$ are the variables in $\theta$ and their differentials \(dx^i\) form the basis of the cotangent space, which, when evaluated at a point $\iget[state]{\I}$ whose values \(\iget[state]{\I}(\D{x})\) determine the tangent vector alias vector field, coincides with \rref{def:dL-valuationTerm}.
}
of (the value of) $\theta$, hence a differential one-form giving a real value for each tangent vector (i.e.\ vector field) described by the values \(\iget[state]{\I}(\D{x})\).
The values \m{\iget[state]{\I}(\D{x})} of the differential symbols $\D{x}$ describe an arbitrary tangent vector or vector field.
Along the flow of (the vector field of a) differential equation, though, the value of the differential \m{\der{\theta}} coincides with the analytic time-derivative of $\theta$ (\rref{lem:differentialLemma}).
The interpretation of predicate symbol $p$ with arity $n$ is an $n$-ary relation \(\iget[const]{\I}(p)\subseteq\reals^n\).
The interpretation of \predicational symbol $C$ is a functional \(\iget[const]{\I}(C)\) mapping subsets \(M\subseteq\linterpretations{\Sigma}{V}\) to subsets \(\iget[const]{\I}(C)(M)\subseteq\linterpretations{\Sigma}{V}\).

\begin{definition}[\dL semantics] \label{def:dL-valuation}
The \emph{semantics of a \dL formula} $\phi$, for each interpretation $\iget[const]{\I}$ with a corresponding set of states $\linterpretations{\Sigma}{V}$, is the subset \m{\imodel{\I}{\phi}\subseteq\linterpretations{\Sigma}{V}} of states in which $\phi$ is true.
It is defined inductively as follows
\begin{compactenum}
\item \(\imodel{\I}{\theta\geq\eta} = \{\iget[state]{\I} \in \linterpretations{\Sigma}{V} \with \ivaluation{\I}{\theta}\geq\ivaluation{\I}{\eta}\}\)
\item \(\imodel{\I}{p(\theta_1,\dots,\theta_k)} = \{\iget[state]{\I} \in \linterpretations{\Sigma}{V} \with (\ivaluation{\I}{\theta_1},\dots,\ivaluation{\I}{\theta_k})\in\iget[const]{\I}(p)\}\)
\item \(\imodel{\I}{\contextapp{C}{\phi}} = \iget[const]{\I}(C)\big(\imodel{\I}{\phi}\big)\) for \predicational symbol $C$
\item \(\imodel{\I}{\lnot\phi} = \scomplement{(\imodel{\I}{\phi})}
= \linterpretations{\Sigma}{V}\setminus\imodel{\I}{\phi}\) 
\item \(\imodel{\I}{\phi\land\psi} = \imodel{\I}{\phi} \cap \imodel{\I}{\psi}\)
\item
\(\def\Im{\imodif[state]{\I}{x}{r}}%
\imodel{\I}{\lexists{x}{\phi}} =  \{\iget[state]{\I} \in \linterpretations{\Sigma}{V} \with \iget[state]{\Im} \in \imodel{\I}{\phi} ~\text{for some}~r\in\reals\}\)

\item \(\imodel{\I}{\ddiamond{\alpha}{\phi}} = \iaccess[\alpha]{\I}\compose\imodel{\I}{\phi}\)
\(=\{\iget[state]{\I} \with \imodels{\It}{\phi} ~\text{for some}~\iget[state]{\It}~\text{such that}~\iaccessible[\alpha]{\I}{\It}\}\)

\item \(\imodel{\I}{\dbox{\alpha}{\phi}} = \imodel{\I}{\lnot\ddiamond{\alpha}{\lnot\phi}}\)
\(=\{\iget[state]{\I} \with \imodels{\It}{\phi} ~\text{for all}~\iget[state]{\It}~\text{such that}~\iaccessible[\alpha]{\I}{\It}\}\)

\end{compactenum}
A \dL formula $\phi$ is \emph{valid in $\iget[const]{\I}$}, written \m{\iget[const]{\I}\models{\phi}}, iff \m{\imodel{\I}{\phi}=\linterpretations{\Sigma}{V}}, i.e.\ \m{\imodels{\I}{\phi}} for all $\iget[state]{\I}$.
Formula $\phi$ is \emph{valid}, written \m{\entails\phi}, iff \m{\iget[const]{\I}\models{\phi}} for all interpretations $\iget[const]{\I}$.
\end{definition}

\noindent
The interpretation of a program constant $a$ is a state-transition relation \(\iget[const]{\I}(a)\subseteq\linterpretations{\Sigma}{V}\times\linterpretations{\Sigma}{V}\),
where \(\related{\iget[const]{\I}(a)}{\iget[state]{\I}}{\iget[state]{\It}}\) iff $a$ can run from initial state $\iget[state]{\I}$ to final state $\iget[state]{\It}$.

\begin{definition}[Transition semantics of \HPs] \label{def:HP-transition}
For each interpretation $\iget[const]{\I}$, each \HP $\alpha$ is interpreted semantically as a binary transition relation \m{\iaccess[\alpha]{\I}\subseteq\linterpretations{\Sigma}{V}\times\linterpretations{\Sigma}{V}} on states, defined inductively by
\begin{compactenum}
\item \m{\iaccess[a]{\I} = \iget[const]{\I}(a)} for program constants $a$
\item
\(\def\Im{\imodif[state]{\I}{x}{r}}%
\iaccess[\pupdate{\pumod{x}{\theta}}]{\I} = \{(\iget[state]{\I},\iget[state]{\Im}) \with r=\ivaluation{\I}{\theta}\}
= \{(\iget[state]{\I},\iget[state]{\It}) \with 
 \iget[state]{\It}=\iget[state]{\I}~\text{except}~\ivaluation{\It}{x}=\ivaluation{\I}{\theta}\}
\)

\item
\(\def\Im{\imodif[state]{\I}{x'}{r}}%
\iaccess[\Dupdate{\Dumod{\D{x}}{\theta}}]{\I} = \{(\iget[state]{\I},\iget[state]{\Im}) \with r=\ivaluation{\I}{\theta}\}
= \{(\iget[state]{\I},\iget[state]{\It}) \with 
 \iget[state]{\It}=\iget[state]{\I}~\text{except}~\ivaluation{\It}{\D{x}}=\ivaluation{\I}{\theta}\}
\)

\item \m{\iaccess[\ptest{\ivr}]{\I} = \{(\iget[state]{\I},\iget[state]{\I}) \with \imodels{\I}{\ivr}\}}
\item
  \m{\iaccess[\pevolvein{\D{x}=\genDE{x}}{\ivr}]{\I} = \{(\iget[state]{\I},\iget[state]{\It}) \with
  \imodels{\If}{\D{x}=\genDE{x}\land\ivr}},
  i.e.
  \(\imodels{\Iff[\zeta]}{\D{x}=\genDE{x}\land\ivr}\)
  for all \(0\leq \zeta\leq r\),
  for some function \m{\iget[flow]{\If}:[0,r]\to\linterpretations{\Sigma}{V}} of some duration $r$ for which all
  \(\iget[state]{\Iff[\zeta]}(\D{x}) = \D[t]{\iget[state]{\Iff[t]}(x)}(\zeta)\) exist
  and \(\iget[state]{\I}=\iget[state]{\Iff[0]}\) on $\scomplement{\{\D{x}\}}$ and \(\iget[state]{\It}=\iget[state]{\Iff[r]}\)$\}$;
   i.e., $\iget[flow]{\If}$ solves the differential equation
   and satisfies $\ivr$ at all times.
   In case $r=0$, the only condition is that \(\iget[state]{\I}=\iget[state]{\It}\) on $\scomplement{\{\D{x}\}}$ and \(\iget[state]{\It}(\D{x})=\ivaluation{\It}{\genDE{x}}\) and \(\imodels{\It}{\ivr}\).

\item \m{\iaccess[\pchoice{\alpha}{\beta}]{\I} = \iaccess[\alpha]{\I} \cup \iaccess[\beta]{\I}}

\item
\newcommand{\Iz}{\iconcat[state=\mu]{\I}}
\m{\iaccess[\alpha;\beta]{\I} = \iaccess[\alpha]{\I} \compose\iaccess[\beta]{\I}}
\(= \{(\iget[state]{\I},\iget[state]{\It}) : (\iget[state]{\I},\iget[state]{\Iz}) \in \iaccess[\alpha]{\I},  (\iget[state]{\Iz},\iget[state]{\It}) \in \iaccess[\beta]{\I}\}\)

\item \m{\iaccess[\prepeat{\alpha}]{\I} = \displaystyle
\closureTransitive{\big(\iaccess[\alpha]{\I}\big)}
=
\cupfold_{n\in\naturals}\iaccess[{\prepeat[n]{\alpha}}]{\I}} 
with \m{\prepeat[n+1]{\alpha} \mequiv \prepeat[n]{\alpha};\alpha} and \m{\prepeat[0]{\alpha}\mequiv\,\ptest{\ltrue}}
\end{compactenum}
where $\closureTransitive{\rho}$ denotes the reflexive transitive closure of relation $\rho$.
\end{definition}
The initial values \(\iget[state]{\I}(\D{x})\) of differential symbols $\D{x}$ do \emph{not} influence the behavior of\\ \(\iaccessible[\pevolvein{\D{x}=\genDE{x}}{\ivr}]{\I}{\It}\), because they may not be compatible with the time-derivatives for the differential equation, e.g. in \m{\Dupdate{\Dumod{\D{x}}{1}};\pevolve{\D{x}=2}}, with a $\D{x}$ mismatch.
\iflongversion
The final values \(\iget[state]{\It}(\D{x})\) will coincide with the derivatives, though.
\fi

Functions and predicates are interpreted by $\iget[const]{\I}$ and are only influenced indirectly by $\iget[state]{\I}$ through the values of their arguments.
So \(p(e)\limply\dbox{\pupdate{\pumod{x}{x+1}}}{p(e)}\) is valid if $x$ is not in $e$ since the change in $x$ does not change whether $p(e)$ is true (\rref{lem:coincidence-term}).
By contrast \(p(x)\limply\dbox{\pupdate{\pumod{x}{x+1}}}{p(x)}\) is invalid, since it is false when \(\iget[const]{\I}(p)=\{d \with d\leq5\}\) and \(\iget[state]{\I}(x)=4.5\).
If the semantics of $p$ were to depend on the state $\iget[state]{\I}$, then there would be no discernible relationship between the truth-values of $p$ in different states, so not even \(p\limply\dbox{\pupdate{\pumod{x}{x+1}}}{p}\) would be valid.

\subsection{Static Semantics}

The static semantics of \dL and \HPs defines some aspects of their behavior that can be read off directly from their syntactic structure without running their programs or evaluating their dynamical effects.
The most important aspects of the static semantics concern free or bound occurrences of variables\iflongversion (which are closely related to the notions of scope and definitions/uses in compilers)\fi.
Bound variables $x$ are those that are bound by $\lforall{x}{}$or $\lexists{x}{}$, but also those that are bound by modalities such as \(\dbox{\pupdate{\pumod{x}{5y}}}{}\)
or \(\ddiamond{\pevolve{\D{x}=1}}{}\)
or \(\dbox{\pchoice{\pumod{x}{1}}{\pevolve{\D{x}=1}}}{}\)
or \(\dbox{\pchoice{\pumod{x}{1}}{\ptest{\ltrue}}}{}\).

The notions of free and bound variables are defined by simultaneous induction in the subsequent definitions: free variables for terms ($\freevars{\theta}$), formulas ($\freevars{\phi}$), and \HPs ($\freevars{\alpha}$), as well as bound variables for formulas ($\boundvars{\phi}$) and for \HPs ($\boundvars{\alpha}$).
For \HPs, there will be a need to distinguish must-bound variables ($\mustboundvars{\alpha}$) that are bound/written to on all executions of $\alpha$ from (may-)bound variables ($\boundvars{\alpha}$) which are bound on some (not necessarily all) execution paths of $\alpha$, such as in \(\dbox{\pchoice{\pumod{x}{1}}{(\pupdate{\pumod{x}{0}};\pupdate{\pumod{y}{x+1}})}}{}\), which has bound variables \(\{x,y\}\) but must-bound variables only $\{x\}$, because $y$ is not written to in the first choice.

\begin{definition}[Bound variable] \label{def:boundvars}
  The set $\boundvars{\phi}\subseteq\allvars\cup\D{\allvars}$ of \emph{bound variables} of \dL formula $\phi$ is defined inductively as
  \begin{align*}
  \boundvars{\theta\geq\eta} = \boundvars{p(\theta_1,\dots,\theta_k)} &= \emptyset\\
  \boundvars{\contextapp{C}{\phi}} &= \allvars\cup\D{\allvars}\\ %
  \boundvars{\lnot\phi} &= \boundvars{\phi}\\
  \boundvars{\phi\land\psi} &= \boundvars{\phi}\cup\boundvars{\psi}\\
  \boundvars{\lforall{x}{\phi}} = \boundvars{\lexists{x}{\phi}} &= \{x\}\cup\boundvars{\phi}\\
  \boundvars{\dbox{\alpha}{\phi}} = \boundvars{\ddiamond{\alpha}{\phi}} &= \boundvars{\alpha}\cup\boundvars{\phi}
  \end{align*}
\end{definition}

\begin{definition}[Free variable] \label{def:freevars}
  The set $\freevars{\theta}\subseteq\allvars\cup\D{\allvars}$ of \emph{free variables} of term $\theta$, i.e.\ those that occur in $\theta$, is defined inductively as
  \begin{align*}
  \freevars{x} &= \{x\}\\
  \freevars{\D{x}} &= \{\D{x}\}\\
  \freevars{f(\theta_1,\dots,\theta_k)} &= \freevars{\theta_1}\cup\dots\cup\freevars{\theta_k}\\
  \freevars{\theta+\eta} = \freevars{\theta\cdot\eta} &= \freevars{\theta}\cup\freevars{\eta}
  \\
  \freevars{\der{\theta}} &= \freevars{\theta} \cup \D{\freevars{\theta}}
  \end{align*}
  The set $\freevars{\phi}$ of \emph{free variables} of \dL formula $\phi$, i.e.\ all those that occur in $\phi$ outside the scope of quantifiers or modalities binding it, is defined inductively as
  \begin{align*}
  \freevars{\theta\geq\eta} &= \freevars{\theta}\cup\freevars{\eta}\\
  \freevars{p(\theta_1,\dots,\theta_k)} &= \freevars{\theta_1}\cup\dots\cup\freevars{\theta_k}\\
  \freevars{\contextapp{C}{\phi}} &= \allvars\cup\D{\allvars}\\ %
  \freevars{\lnot\phi} &= \freevars{\phi}\\
  \freevars{\phi\land\psi} &= \freevars{\phi}\cup\freevars{\psi}\\
  \freevars{\lforall{x}{\phi}} = \freevars{\lexists{x}{\phi}} &= \freevars{\phi}\setminus\{x\}\\
  \freevars{\dbox{\alpha}{\phi}} = \freevars{\ddiamond{\alpha}{\phi}} &= \freevars{\alpha}\cup(\freevars{\phi}\setminus\mustboundvars{\alpha})
  \end{align*}
\end{definition}
Soundness requires that
\(\freevars{\dbox{\alpha}{\phi}}\) is not defined as \(\freevars{\alpha}\cup(\freevars{\phi}\setminus\boundvars{\alpha})\),
otherwise \(\dbox{\pchoice{\pupdate{\pumod{x}{1}}}{\pupdate{\pumod{y}{2}}}}{x\geq1}\) would have no free variables,
but its truth-value depends on the initial value of $x$,
demanding \(\freevars{\dbox{\pchoice{\pupdate{\pumod{x}{1}}}{\pupdate{\pumod{y}{2}}}}{x\geq1}}=\{x\}\).
\iflongversion
  The simpler definition 
  \(\freevars{\dbox{\alpha}{\phi}} = \freevars{\alpha}\cup\freevars{\phi}\)
  would be correct, but the results would be less precise then.
  Likewise for $\ddiamond{\alpha}{\phi}$.
\fi
\iflongversion
Soundness requires \(\freevars{\der{\theta}}\) not to be defined as \(\D{\freevars{\theta}}\),
because
the value of \(\der{x y}\) depends on $\{x,\D{x},y,\D{y}\}$, since \(\der{x y}\) equals \(\D{x} y + x \D{y}\) (\rref{lem:derivationLemma}).
\fi

The static semantics defines which variables are free so may be read ($\freevars{\alpha}$), which are bound ($\boundvars{\alpha}$) so may be written to somewhere in $\alpha$, and which are must-bound ($\mustboundvars{\alpha}$) so must be written to on all execution paths of $\alpha$.

\begin{definition}[Bound variable] \label{def:boundvars-HP}
  The set $\boundvars{\alpha}\subseteq\allvars\cup\D{\allvars}$ of \emph{bound variables} of \HP $\alpha$, i.e.\ all those that may potentially be written to, is defined inductively:
  \begin{align*}
  \boundvars{a} &= \allvars\cup\D{\allvars} &&\text{for program constant $a$}\\
  \boundvars{\pupdate{\pumod{x}{\theta}}} &= \{x\}
  \\
  \boundvars{\Dupdate{\Dumod{\D{x}}{\theta}}} &= \{\D{x}\}
  \\
  \boundvars{\ptest{\ivr}} &= \emptyset
  \\
  \boundvars{\pevolvein{\D{x}=\genDE{x}}{\ivr}} &= \{x,\D{x}\}
  \\
  \boundvars{\alpha\cup\beta} = \boundvars{\alpha;\beta} &= \boundvars{\alpha}\cup\boundvars{\beta}
  \\
  \boundvars{\prepeat{\alpha}} &= \boundvars{\alpha}
  \end{align*}
\end{definition}

\begin{definition}[Must-bound variable] \label{def:mustboundvar}
  The set $\mustboundvars{\alpha}\subseteq\boundvars{\alpha}\subseteq\allvars\cup\D{\allvars}$ of \emph{must-bound variables} of \HP $\alpha$, i.e.\ all those that must be written to on all paths of $\alpha$, is defined inductively as
  \begin{align*}
  \mustboundvars{a} &= \emptyset &&\text{for program constant $a$}\\
    \mustboundvars{\alpha} &= \boundvars{\alpha} &&\text{for other atomic \HPs $\alpha$}
    \\
  \mustboundvars{\alpha\cup\beta} &= \mustboundvars{\alpha}\cap\mustboundvars{\beta}
  \\
  \mustboundvars{\alpha;\beta} &= \mustboundvars{\alpha}\cup\mustboundvars{\beta}
  \\
  \mustboundvars{\prepeat{\alpha}} &= \emptyset
  \end{align*}
\end{definition}
\iflongversion
Obviously, \(\mustboundvars{\alpha}\subseteq\boundvars{\alpha}\).
If $\alpha$ is only built by sequential compositions from atomic programs without program constants, then \(\mustboundvars{\alpha}=\boundvars{\alpha}\).
\fi

\begin{definition}[Free variable] \label{def:freevars-HP}
  The set $\freevars{\alpha}\subseteq\allvars\cup\D{\allvars}$ of \emph{free variables} of \HP $\alpha$, i.e.\ all those that may potentially be read, is defined inductively as
  \begin{align*}
  \freevars{a} &= \allvars\cup\D{\allvars} &&\hspace{-10pt}\text{for program constant $a$}\\
  \freevars{\pupdate{\pumod{x}{\theta}}}
  = \freevars{\Dupdate{\Dumod{\D{x}}{\theta}}} &= \freevars{\theta}
  \\
  \freevars{\ptest{\ivr}} &= \freevars{\ivr}
  \\
  \freevars{\pevolvein{\D{x}=\genDE{x}}{\ivr}} &= \{x\}\cup\freevars{\genDE{x}}\cup\freevars{\ivr}
  \\
  \freevars{\pchoice{\alpha}{\beta}} &= \freevars{\alpha}\cup\freevars{\beta}
  \\
  \freevars{\alpha;\beta} &= \freevars{\alpha}\cup(\freevars{\beta}\setminus\mustboundvars{\alpha})
  \\
  \freevars{\prepeat{\alpha}} &= \freevars{\alpha}
  \end{align*}
\iflongversion
The \emph{variables} of \HP $\alpha$, whether free or bound, are \(\vars{\alpha}=\freevars{\alpha}\cup\boundvars{\alpha}\).
\fi
\end{definition}
\iflongversion
  The simpler definition 
  \(\freevars{\alpha\cup\beta} = \freevars{\alpha}\cup\freevars{\beta}\)
  would be correct, but the results would be less precise then.
\fi
Unlike $x$, the left-hand side $\D{x}$ of differential equations is not added to the free variables of \(\freevars{\pevolvein{\D{x}=\genDE{x}}{\ivr}}\), because its behavior does not depend on the initial value of differential symbols $\D{x}$, only the initial value of $x$.
  Free and bound variables are the set of all variables $\allvars$ and differential symbols $\D{\allvars}$ for program constants $a$, because their effect depends on the interpretation $\iget[const]{\I}$, so may read and write all \(\freevars{a}=\boundvars{a}=\allvars\cup\D{\allvars}\) but not on all paths \(\mustboundvars{a}=\emptyset\).
  Subsequent results about free and bound variables are, thus, vacuously true when program constants occur.
Corresponding observations hold for \predicational symbols.

The static semantics defines which variables are readable or writable.
There may not be any run of $\alpha$ in which a variable is read or written to.
If $x\not\in\freevars{\alpha}$, though, then $\alpha$  cannot read the value of $x$.
If $x\not\in\boundvars{\alpha}$, it cannot write to $x$.
\iflongversion
\rref{def:freevars-HP} is parsimonious.
For example, $x$ is not a free variable of the following program
\[
(\pchoice{\pupdate{\pumod{x}{1}}}{\pupdate{\pumod{x}{2}}}); \pupdate{\pumod{z}{x+y}}
\]
because $x$ is never actually read, since $x$ must have been defined on \emph{every} execution path of the first part before being read by the second part.
No execution of the above program, thus, depends on the initial value of $x$, which is why it is not a free variable.
This would have been different for the simpler definition
\[
\freevars{\alpha;\beta} = \freevars{\alpha}\cup\freevars{\beta}
\]
There is a limit to the precision with which any static analysis can determine which variables are really read or written \cite{DBLP:journals/ams/Rice53}.
The static semantics in \rref{def:freevars-HP} will, e.g., call $x$ a free variable of the following program even though no execution could read it, because they fail test $\ptest{\lfalse}$ when running the branch reading $x$:
\[
\pupdate{\pumod{z}{0}}; \prepeat{(\ptest{\lfalse};\pupdate{\pumod{z}{z+x}})}
\]
\fi

The \dfn{signature}, i.e.\ set of function, predicate, \predicational symbols, and program constants in $\phi$ is denoted by \(\intsigns{\phi}\) (accordingly for terms and programs).
It is defined like $\freevars{\phi}$ except that all occurrences are free.
Variables in $\allvars\cup\D{\allvars}$ are interpreted by state $\iget[state]{\I}$.
The symbols in $\intsigns{\phi}$ are interpreted by interpretation $\iget[const]{\I}$.
\subsection{Correctness of Static Semantics}
The following result reflects that \HPs have bounded effect: for a variable $x$ to be modified during a run of $\alpha$, $x$ needs the be a bound variable in \HP $\alpha$, i.e.\ \(x\in\boundvars{\alpha}\), so that $\alpha$ can write to $x$.
The converse is not true, because $\alpha$ may bind a variable $x$, e.g. by having an assignment to $x$, that never actually changes the value of $x$, such as \(\pupdate{\pumod{x}{x}}\) or because the assignment can never be executed.
\iflongversion
The following program, for example, binds $x$ but will never change the value of $x$ because there is no way of satisfying the test $\ptest{\lfalse}$:
\(
\pchoice{(\ptest{\lfalse}; \pupdate{\pumod{x}{42}})}{\pupdate{\pumod{z}{x+1}}}
\).
\fi
\begin{lemma}[Bound effect lemma] \label{lem:bound}
  If \(\iaccessible[\alpha]{\I}{\It}\), then \(\iget[state]{\I}=\iget[state]{\It}\) on $\scomplement{\boundvars{\alpha}}$.
\end{lemma}
\begin{proofatend}
The proof is by a straightforward structural induction on $\alpha$.
\begin{compactitem}
\item Since \(\boundvars{a} = \allvars\cup\D{\allvars}\), the statement is vacuously true for program constant $a$, because \(\scomplement{\boundvars{a}}=\emptyset\).

\item \m{\iaccessible[\pupdate{\pumod{x}{\theta}}]{\I}{\It} = \{(\iget[state]{\I},\iget[state]{\It}) \with \iget[state]{\It}=\iget[state]{\I}~\text{except that}~\ivaluation{\It}{x}=\ivaluation{\I}{\theta}\}} 
implies that \(\iget[state]{\I}=\iget[state]{\It}\) except for \(\{x\}=\boundvars{\pupdate{\pumod{x}{\theta}}}\).

\item \m{\iaccessible[\Dupdate{\Dumod{\D{x}}{\theta}}]{\I}{\It} = \{(\iget[state]{\I},\iget[state]{\It}) \with \iget[state]{\It}=\iget[state]{\I}~\text{except that}~\ivaluation{\It}{\D{x}}=\ivaluation{\I}{\theta}\}} 
implies that \(\iget[state]{\I}=\iget[state]{\It}\) except for \(\{\D{x}\}=\boundvars{\Dupdate{\Dumod{\D{x}}{\theta}}}\).

\item \m{\iaccessible[\ptest{\ivr}]{\I}{\I} = \{(\iget[state]{\I},\iget[state]{\I}) \with \imodels{\I}{\ivr} ~\text{i.e.}~ \iget[state]{\I}\in\imodel{\I}{\ivr}\}}
fits to \(\boundvars{\ptest{\ivr}}=\emptyset\)

\item
  \m{\iaccessible[\pevolvein{\D{x}=\genDE{x}}{\ivr}]{\I}{\It}} 
  implies that \(\iget[state]{\I}=\iget[state]{\It}\) except for the variables with differential equations, which are \(\{x,\D{x}\}=\boundvars{\pevolvein{\D{x}=\genDE{x}}{\ivr}}\)
  observing that \(\iget[state]{\I}(\D{x})\) and \(\iget[state]{\It}(\D{x})\) may differ by \rref{def:HP-transition}.

\item \m{\iaccessible[\pchoice{\alpha}{\beta}]{\I}{\It} = \iaccess[\alpha]{\I} \cup \iaccess[\beta]{\I}}
implies \(\iaccessible[\alpha]{\I}{\It}\) or \(\iaccessible[\beta]{\I}{\It}\),
which, by induction hypothesis,
implies \(\iget[state]{\I}=\iget[state]{\It}\) on $\scomplement{\boundvars{\alpha}}$
or
\(\iget[state]{\I}=\iget[state]{\It}\) on $\scomplement{\boundvars{\beta}}$, respectively.
Either case implies \(\iget[state]{\I}=\iget[state]{\It}\) on $\scomplement{(\boundvars{\alpha}\cup\boundvars{\beta})}=\scomplement{\boundvars{\pchoice{\alpha}{\beta}}}$.

\item
\newcommand{\Iz}{\dLint[state=\mu]}%
\m{\iaccessible[\alpha;\beta]{\I}{\It} = \iaccess[\alpha]{\I} \compose\iaccess[\beta]{\I}},
i.e.\ there is a $\iget[state]{\Iz}$ such that \(\iaccessible[\alpha]{\I}{\Iz}\) and \(\iaccessible[\beta]{\Iz}{\It}\).
So, by induction hypothesis, \(\iget[state]{\I}=\iget[state]{\Iz}\) on $\scomplement{\boundvars{\alpha}}$ and \(\iget[state]{\Iz}=\iget[state]{\It}\) on $\scomplement{\boundvars{\beta}}$.
By transitivity, \(\iget[state]{\I}=\iget[state]{\It}\) on $\scomplement{(\boundvars{\alpha}\cup\boundvars{\beta})}=\scomplement{\boundvars{\alpha;\beta}}$.

\item
\renewcommand{\Iz}[1][]{\dLint[state=\nu_{#1}]}%
\m{\iaccessible[\prepeat{\alpha}]{\I}{\It} = \displaystyle\cupfold_{n\in\naturals}\iaccess[{\prepeat[n]{\alpha}}]{\I}}, then there is an $n\in\naturals$ and a sequence \(\iget[state]{\Iz[0]}=\iget[state]{\I},\iget[state]{\Iz[1]},\dots,\iget[state]{\Iz[n]}=\iget[state]{\It}\) such that \(\iaccessible[\alpha]{\Iz[i]}{\Iz[i+1]}\) for all $i<n$.
By $n$ uses of the induction hypothesis, \(\iget[state]{\Iz[i]}=\iget[state]{\Iz[i+1]}\) on $\scomplement{\boundvars{\alpha}}$ for all $i<n$.
Thus, \(\iget[state]{\I}=\iget[state]{\Iz[0]}=\iget[state]{\Iz[n]}=\iget[state]{\It}\) on $\scomplement{\boundvars{\alpha}}=\scomplement{\boundvars{\prepeat{\alpha}}}$.
\qedhere
\end{compactitem}
\end{proofatend}
Similarly, only $\boundvars{\phi}$ change their value during the evaluation of formulas.

The value of a term only depends on the values of its free variables.
When evaluating a term $\theta$ in two states $\iget[state]{\I}$, $\iget[state]{\Ialt}$ that differ widely but agree on the free variables $\freevars{\theta}$ of $\theta$, the values of $\theta$ in both states coincide.
Accordingly for different interpretations \(\iget[const]{\I},\iget[const]{\Ialt}\) that agree on the symbols $\intsigns{\theta}$ that occur in $\theta$.

\begin{lemma}[Coincidence lemma] \label{lem:coincidence-term}
  If \(\iget[state]{\I}=\iget[state]{\Ialt}\) on $\freevars{\theta}$
  and \(\iget[const]{\I}=\iget[const]{\Ialt}\) on $\intsigns{\theta}$, then
  \m{\ivaluation{\I}{\theta}=\ivaluation{\Ialt}{\theta}}.
\end{lemma}
\begin{proofatend}
The proof is by structural induction on $\theta$.
\begin{compactitem}
\item \m{\ivaluation{\I}{x} = \iget[state]{\I}(x)}
= \(\iget[state]{\Ialt}(x) = \ivaluation{\Ialt}{x}\) for variable $x$ since \(\iget[state]{\I}=\iget[state]{\Ialt}\) on \(\freevars{x}=\{x\}\).

\item \m{\ivaluation{\I}{\D{x}} = \iget[state]{\I}(\D{x})}
= \(\iget[state]{\Ialt}(\D{x}) = \ivaluation{\Ialt}{\D{x}}\) for differential symbol $\D{x}$ since \(\iget[state]{\I}=\iget[state]{\Ialt}\) on \(\freevars{\D{x}}=\{\D{x}\}\).

\item \m{\ivaluation{\I}{f(\theta_1,\dots,\theta_k)} = \iget[const]{\I}(f)(\ivaluation{\I}{\theta_1},\dots,\ivaluation{\I}{\theta_k})}
\(\stackrel{\text{IH}}{=} \iget[const]{\Ialt}(f)({\ivaluation{\Ialt}{\theta_1},\dots,\ivaluation{\Ialt}{\theta_k}})
= \ivaluation{\Ialt}{f(\theta_1,\dots,\theta_k)}\)
by induction hypothesis, because \(\freevars{\theta_i}\subseteq\freevars{f(\theta_1,\dots,\theta_k)}\)
and $\iget[const]{\I}$ and $\iget[const]{\Ialt}$ were assumed to agree on the function symbol $f$ that occurs in the term.

\item \m{\ivaluation{\I}{\theta+\eta} = \ivaluation{\I}{\theta} + \ivaluation{\I}{\eta}}
\(\stackrel{\text{IH}}{=} \ivaluation{\Ialt}{\theta} + \ivaluation{\Ialt}{\eta} = \ivaluation{\Ialt}{\theta+\eta}\)
by induction hypothesis, because \(\freevars{\theta}\subseteq\freevars{\theta+\eta}\) and \(\freevars{\eta}\subseteq\freevars{\theta+\eta}\).

\item \m{\ivaluation{\I}{\theta\cdot\eta} = \ivaluation{\I}{\theta} \cdot \ivaluation{\I}{\eta}}
\(\stackrel{\text{IH}}{=} \ivaluation{\Ialt}{\theta} \cdot \ivaluation{\Ialt}{\eta} = \ivaluation{\Ialt}{\theta\cdot\eta}\)
by induction hypothesis, because \(\freevars{\theta}\subseteq\freevars{\theta\cdot\eta}\) and \(\freevars{\eta}\subseteq\freevars{\theta\cdot\eta}\).

\item
{\def\Im{\imodif[state]{\I}{x}{X}}%
\def\Ialtm{\vdLint[const=J,state=\modif{\tilde{\nu}}{x}{X}]}%
\m{\ivaluation{\I}{\der{\theta}}
\displaystyle
= \sum_x \iget[state]{\I}(\D{x}) \itimes \Dp[X]{\ivaluation{\Im}{\theta}}
= \sum_x \iget[state]{\Ialt}(\D{x}) \itimes \Dp[X]{\ivaluation{\Im}{\theta}}
\stackrel{\text{IH}}{=} \sum_x \iget[state]{\Ialt}(\D{x}) \itimes \Dp[X]{\ivaluation{\Ialtm}{\theta}}
}
since \(\iget[state]{\I}=\iget[state]{\Ialt}\) on $\freevars{\der{\theta}}$,
which includes all differential symbols $\D{x}$ for all \(x\in\freevars{\theta}\) (the others have partial derivative 0 so do not contribute to the sum),
and by induction hypothesis on the simpler term $\theta$, because \(\freevars{\theta}\subseteq\freevars{\der{\theta}}\).
Note that partial derivatives are functional, i.e. the partial derivatives by $X$ of \(\ivaluation{\Im}{\theta}\) and \(\ivaluation{\Ialtm}{\theta}\) agree since,
by induction hypothesis, \(\ivaluation{\Im}{\theta}=\ivaluation{\Ialtm}{\theta}\) for all $X$ since \(\iget[state]{\Im}=\iget[state]{\Ialtm}\) on \(\{x\}\cup\freevars{\theta}\) since $x$ is interpreted to be $X$ in both states and \(\iget[state]{\I}=\iget[state]{\Ialt}\) on $\freevars{\theta}$ already.
}
\qedhere
\end{compactitem}
\end{proofatend}

By a more subtle argument, the values of \dL formulas also only depend on the values of their free variables.
When evaluating \dL formula $\phi$ in two states $\iget[state]{\I}$, $\iget[state]{\Ialt}$ that differ but agree on the free variables $\freevars{\phi}$ of $\phi$, the (truth) values of $\phi$ in both states coincide.
\rref{lem:coincidence} and~\ref{lem:coincidence-HP} are proved by simultaneous induction.
\begin{lemma}[Coincidence lemma] \label{lem:coincidence}
  If \(\iget[state]{\I}=\iget[state]{\Ialt}\) on $\freevars{\phi}$
  and \(\iget[const]{\I}=\iget[const]{\Ialt}\) on $\intsigns{\phi}$, then
  \m{\imodels{\I}{\phi}} iff \m{\imodels{\Ialt}{\phi}}.
\end{lemma}
\begin{proofatend}
The proof is by structural induction on $\phi$.
\begin{enumerate}
\item \(\imodels{\I}{p(\theta_1,\dots,\theta_k)}\) iff \((\ivaluation{\I}{\theta_1},\dots,\ivaluation{\I}{\theta_k})\in\iget[const]{\I}(p)\)
iff \((\ivaluation{\Ialt}{\theta_1},\dots,\ivaluation{\Ialt}{\theta_k})\in\iget[const]{\Ialt}(p)\) iff \(\imodels{\Ialt}{p(\theta_1,\dots,\theta_k)}\)
by \rref{lem:coincidence-term} since $\freevars{\theta_i}\subseteq\freevars{p(\theta_1,\dots,\theta_k)}$
and $\iget[const]{\I}$ and $\iget[const]{\Ialt}$ were assumed to agree on the function symbol $p$ that occurs in the formula.

\item \(\imodels{\I}{\theta\geq\eta}\) iff \(\ivaluation{\I}{\theta}\geq\ivaluation{\I}{\eta}\)
iff \(\ivaluation{\Ialt}{\theta}\geq\ivaluation{\Ialt}{\eta}\) iff \(\imodels{\Ialt}{\theta\geq\eta}\)
by \rref{lem:coincidence-term} since $\freevars{\theta}\cup\freevars{\eta}\subseteq\freevars{\theta\geq\eta}$
and the interpretation of $\geq$ is fixed.

\item \(\imodels{\I}{\contextapp{C}{\phi}} = \iget[const]{\I}(C)\big(\imodel{\I}{\phi}\big)\)
iff(IH) \(\imodels{\Ialt}{\contextapp{C}{\phi}} = \iget[const]{\Ialt}(C)\big(\imodel{\Ialt}{\phi}\big)\)
since \(\iget[state]{\I}=\iget[state]{\Ialt}\) on \(\freevars{\contextapp{C}{\phi}}=\allvars\cup\D{\allvars}\), so \(\iget[state]{\I}=\iget[state]{\Ialt}\),
and \(\iget[const]{\I}=\iget[const]{\Ialt}\) on \(\intsigns{\contextapp{C}{\phi}}=\{C\}\cup\intsigns{\phi}\), so \(\iget[const]{\I}(C)=\iget[const]{\Ialt}(C)\) and, by induction hypothesis, \(\imodel{\I}{\phi}=\imodel{\Ialt}{\phi}\).

\item \(\imodels{\I}{\lnot\phi}\) iff \(\inonmodels{\I}{\phi}\)
iff(IH) \(\inonmodels{\Ialt}{\phi}\) iff \(\imodels{\Ialt}{\lnot\phi}\)
by induction hypothesis as $\freevars{\lnot\phi}=\freevars{\phi}$.

\item \(\imodels{\I}{\phi\land\psi}\) iff \(\imodels{\I}{\phi}\cap\imodel{\I}{\psi}\)
iff(IH) \(\imodels{\Ialt}{\phi}\cap\imodel{\Ialt}{\psi}\) iff \(\imodels{\Ialt}{\phi\land\psi}\)
by induction hypothesis using $\freevars{\phi\land\psi}=\freevars{\phi}\cup\freevars{\psi}$.

\item
{\def\Im{\imodif[state]{\I}{x}{r}}%
\def\Imalt{\dLint[const=I,state=\tilde{\nu}_x^r]}%
\(\imodels{\I}{\lexists{x}{\phi}}\) iff \(\iget[state]{\Im} \in \imodel{\I}{\phi} ~\text{for some}~r\in\reals\)
iff \(\iget[state]{\Imalt} \in \imodel{\I}{\phi} ~\text{for some}~r\in\reals\) iff(H) \(\imodels{\Ialt}{\lexists{x}{\phi}}\)
for the same $r$
by induction hypothesis using that \(\iget[state]{\Im}=\iget[state]{\Imalt}\) on $\freevars{\phi}\subseteq\{x\}\cup\freevars{\lexists{x}{\phi}}$.
}

\item The case \(\lforall{x}{\phi}\) follows from the equivalence \(\lforall{x}{\phi} \mequiv \lnot\lexists{x}{\lnot\phi}\) using \(\freevars{\lnot\lexists{x}{\lnot\phi}} = \freevars{\lforall{x}{\phi}}\).

\item \(\imodels{\I}{\ddiamond{\alpha}{\phi}}\) iff there is a $\iget[state]{\It}$ such that \(\iaccessible[\alpha]{\I}{\It}\) and \(\imodels{\It}{\phi}\).
Since \(\iget[state]{\I}=\iget[state]{\Ialt}\) on $\freevars{\ddiamond{\alpha}{\phi}}\supseteq\freevars{\alpha}$
and \(\iaccessible[\alpha]{\I}{\It}\),
\rref{lem:coincidence-HP} implies with \(\iget[const]{\I}=\iget[const]{\Ialt}\) on $\intsigns{\alpha}$ that there is an $\iget[state]{\Italt}$ such that
  \(\iaccessible[\alpha]{\Ialt}{\Italt}\)
  and \(\iget[state]{\It}=\iget[state]{\Italt}\) on $\freevars{\ddiamond{\alpha}{\phi}}\cup\mustboundvars{\alpha}
  = \freevars{\alpha}\cup(\freevars{\phi}\setminus\mustboundvars{\alpha})\cup\mustboundvars{\alpha}
  = \freevars{\alpha}\cup\freevars{\phi}\cup\mustboundvars{\alpha}
  \supseteq \freevars{\phi}$.

\begin{center}
\includegraphics{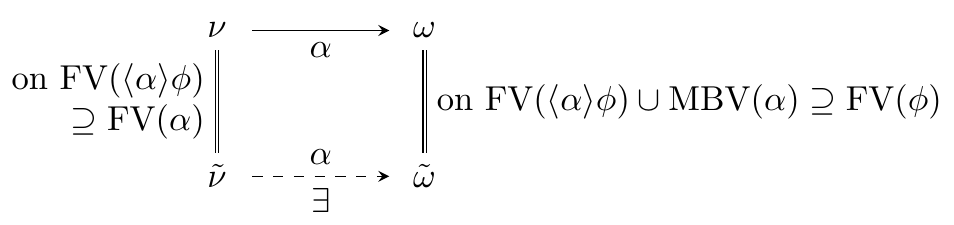}
\end{center}

Since, \(\iget[state]{\It}=\iget[state]{\Italt}\) on $\freevars{\phi}$ and \(\iget[const]{\I}=\iget[const]{\Ialt}\) on $\intsigns{\phi}$,
the induction hypothesis implies that
\(\imodels{\Italt}{\phi}\) since \(\imodels{\It}{\phi}\).
Since \(\iaccessible[\alpha]{\Ialt}{\Italt}\), this implies
\m{\imodels{\Ialt}{\ddiamond{\alpha}{\phi}}}.

\item \(\imodels{\I}{\dbox{\alpha}{\phi}} = \imodel{\I}{\lnot\ddiamond{\alpha}{\lnot\phi}}\)
iff \(\inonmodels{\I}{\ddiamond{\alpha}{\lnot\phi}}\)
iff \(\inonmodels{\Ialt}{\ddiamond{\alpha}{\lnot\phi}}\) iff \(\imodels{\Ialt}{\dbox{\alpha}{\phi}}\)
by induction hypothesis using $\freevars{\ddiamond{\alpha}{\lnot\phi}}=\freevars{\dbox{\alpha}{\phi}}$.
\qedhere
\end{enumerate}
\end{proofatend}

In a sense, the runs of an \HP $\alpha$ also only depend on the values of its free variables, because its behavior cannot depend on the values of variables that it never reads.
That is, if \(\iget[state]{\I}=\iget[state]{\Ialt}\) on $\freevars{\alpha}$
  and \(\iaccessible[\alpha]{\I}{\It}\),
  then there is an $\iget[state]{\Italt}$ such that
  \(\iaccessible[\alpha]{\Ialt}{\Italt}\)
  and $\iget[state]{\It}$ and $\iget[state]{\Italt}$ agree in some sense.
There is a subtlety, though.
The resulting states $\iget[state]{\It}$ and $\iget[state]{\Italt}$ will only continue to agree on $\freevars{\alpha}$ and the variables that are bound on the particular path that $\alpha$ took for the transition \(\iaccessible[\alpha]{\I}{\It}\).
On variables $z$ that are neither free (so the initial states $\iget[state]{\I}$ and $\iget[state]{\Ialt}$ have not been assumed to coincide) nor bound on the particular path that $\alpha$ took, $\iget[state]{\It}$ and $\iget[state]{\Italt}$ may continue to disagree, because $z$ has not been written to.
\iflongversion
\begin{example}
Let \(\iaccessible[\alpha]{\I}{\It}\).
It is not enough to assume \(\iget[state]{\I}=\iget[state]{\Ialt}\) only on $\freevars{\alpha}$ in order to guarantee \(\iget[state]{\It}=\iget[state]{\Italt}\) on $\vars{\alpha}$ for some $\iget[state]{\Italt}$ such that
  \(\iaccessible[\alpha]{\Ialt}{\Italt}\), because
\[
\alpha \mdefequiv \pchoice{\pupdate{\pumod{x}{1}}}{\pupdate{\pumod{y}{2}}}
\]
will force the final states to agree only on either $x$ or on $y$, whichever one was assigned to during the respective run of $\alpha$, not on both \(\boundvars{\alpha}=\{x,y\}\),
even though any initial states \(\iget[state]{\I},\iget[state]{\Ialt}\) agree on $\freevars{\alpha}=\emptyset$.
Note that this can only happen because \(\mustboundvars{\alpha}=\emptyset\neq\boundvars{\alpha}=\{x,y\}\).
\end{example}
\fi
Yet, $\iget[state]{\It}$ and $\iget[state]{\Italt}$ agree on the variables that are bound on \emph{all} paths of $\alpha$, rather than somewhere in $\alpha$.
That is on the must-bound variables of $\alpha$.
If initial states agree on (at least) all free variables $\freevars{\alpha}$ that \HP $\alpha$ may read, then the final states agree on those as well as on all variables that $\alpha$ must write, i.e.\ on $\mustboundvars{\alpha}$.

\begin{lemma}[Coincidence lemma] \label{lem:coincidence-HP}
  If \(\iget[state]{\I}=\iget[state]{\Ialt}\) on $V\supseteq\freevars{\alpha}$
  and \(\iget[const]{\I}=\iget[const]{\Ialt}\) on $\intsigns{\alpha}$
  and \(\iaccessible[\alpha]{\I}{\It}\),
  then there is an $\iget[state]{\Italt}$ such that
  \(\iaccessible[\alpha]{\Ialt}{\Italt}\)
  and \(\iget[state]{\It}=\iget[state]{\Italt}\) on $V\cup\mustboundvars{\alpha}$.
\iflongversion
\begin{center}
  \vspace{-\baselineskip}
\begin{tikzpicture}
  \matrix (m) [matrix of math nodes,row sep=3em,column sep=4em,minimum width=2em]
  {
     \iget[state]{\I} & \iget[state]{\It} \\
     \iget[state]{\Ialt} & \iget[state]{\Italt} \\};
  \path
    (m-1-1) edge [similar state] node [left,align=right] {on $V\supseteq\freevars{\alpha}$} (m-2-1)
            edge [transition] node [below] {$\alpha$} (m-1-2)
    (m-2-1) edge [transition,transition exists] node [above] {$\alpha$}
            node [below] {$\exists$} (m-2-2)
    (m-1-2) edge [similar state] node [right,align=right] {on $V\cup\mustboundvars{\alpha}$} (m-2-2);
  \path
    (m-1-1) edge[similar state,bend left=30] node[above] {on $\scomplement{\boundvars{\alpha}}$} (m-1-2)
    (m-2-1) edge[similar state,bend right=30] node[below] {on $\scomplement{\boundvars{\alpha}}$} (m-2-2);
\end{tikzpicture}
\end{center}
\fi
\end{lemma}
\begin{proofatend}
The proof is by induction on the structural complexity of $\alpha$, where $\prepeat{\alpha}$ is considered to be structurally more complex than \HPs of any length but with less nested repetitions, which induces a well-founded order on \HPs.
For atomic programs $\alpha$, for which \(\boundvars{\alpha}=\mustboundvars{\alpha}\), it is enough to conclude agreement on $\vars{\alpha}\mdefeq \freevars{\alpha}\cup\boundvars{\alpha}=\freevars{\alpha}\cup\mustboundvars{\alpha}$, because any variable in $V\setminus\vars{\alpha}$ is in \(\scomplement{\boundvars{\alpha}}\), which remains unchanged by $\alpha$ according to \rref{lem:bound}.

\begin{compactitem}

\item Since \(\freevars{a} = \allvars\cup\D{\allvars}\) so $\iget[state]{\I}=\iget[state]{\Ialt}$, the statement is vacuously true for program constant $a$.

\item \m{\iaccessible[\pupdate{\pumod{x}{\theta}}]{\I}{\It} = \{(\iget[state]{\I},\iget[state]{\It}) \with \iget[state]{\It}=\iget[state]{\I}~\text{except that}~\ivaluation{\It}{x}=\ivaluation{\I}{\theta}\}}
then there is a transition \(\iaccessible[\pupdate{\pumod{x}{\theta}}]{\Ialt}{\Italt}\)
and \(\iget[state]{\Italt}(x)=\ivaluation{\Italt}{x}=\ivaluation{\Ialt}{\theta}
=\ivaluation{\I}{\theta}=\ivaluation{\It}{x} = \iget[state]{\I}(x)\)
by \rref{lem:coincidence}, since \(\iget[state]{\I}=\iget[state]{\Ialt}\) on $\freevars{\pupdate{\pumod{x}{\theta}}}=\freevars{\theta}$ and \(\iget[const]{\I}=\iget[const]{\Ialt}\) on $\intsigns{\theta}$.
So, \(\iget[state]{\It}=\iget[state]{\Italt}\) on $\boundvars{\pupdate{\pumod{x}{\theta}}}=\{x\}$.
Also, \(\iget[state]{\I}=\iget[state]{\It}\) on $\scomplement{\boundvars{\pupdate{\pumod{x}{\theta}}}}$
and \(\iget[state]{\Ialt}=\iget[state]{\Italt}\) on $\scomplement{\boundvars{\pupdate{\pumod{x}{\theta}}}}$ by \rref{lem:bound}.
Since \(\iget[state]{\I}=\iget[state]{\Ialt}\) on $\freevars{\pupdate{\pumod{x}{\theta}}}$, these imply
\(\iget[state]{\It}=\iget[state]{\Italt}\) on $\freevars{\pupdate{\pumod{x}{\theta}}}\setminus\boundvars{\pupdate{\pumod{x}{\theta}}}$.
Since \(\iget[state]{\It}=\iget[state]{\Italt}\) on $\boundvars{\pupdate{\pumod{x}{\theta}}}$ had been shown already, this implies
\(\iget[state]{\It}=\iget[state]{\Italt}\) on $\vars{\pupdate{\pumod{x}{\theta}}}$.

\item
{\def\Im{\imodif[state]{\I}{x'}{r}}%
\def\Ialtm{\vdLint[const=J,state=\modif{\tilde{\nu}}{x'}{r}]}%
\m{\iaccessible[\Dupdate{\Dumod{\D{x}}{\theta}}]{\I}{\It} = \{(\iget[state]{\I},\iget[state]{\Im}) \with r=\ivaluation{\I}{\theta}\}}.
As \(\ivaluation{\I}{\theta}=\ivaluation{\Ialt}{\theta}\) by \rref{lem:coincidence-term} since \(\freevars{\theta}\subseteq\freevars{\Dupdate{\Dumod{\D{x}}{\theta}}}\),
this implies
\(\iaccessible[\Dupdate{\Dumod{\D{x}}{\theta}}]{\Ialt}{\Ialtm} = \{(\iget[state]{\Ialt},\iget[state]{\Ialtm}) \with r=\ivaluation{\Ialt}{\theta}\}\).
By construction \(\iget[state]{\It}=\iget[state]{\Ialtm}\) on $\boundvars{\Dupdate{\Dumod{\D{x}}{\theta}}}=\{\D{x}\}$
and they continue to agree on $\freevars{\Dupdate{\Dumod{\D{x}}{\theta}}}\setminus\boundvars{{\Dupdate{\Dumod{\D{x}}{\theta}}}}$.
}

\item \m{\iaccessible[\ptest{\ivr}]{\I}{\It} = \{(\iget[state]{\I},\iget[state]{\I}) \with \imodels{\I}{\ivr} ~\text{i.e.}~ \iget[state]{\I}\in\imodel{\I}{\ivr}\}}
then \(\iget[state]{\It}=\iget[state]{\I}\) by \rref{def:HP-transition}.
Since, \(\imodels{\I}{\ivr}\) and \(\iget[state]{\I}=\iget[state]{\Ialt}\) on $\freevars{\ptest{\ivr}}$ and \(\iget[const]{\I}=\iget[const]{\Ialt}\) on $\intsigns{\ivr}$,
\rref{lem:coincidence} implies that \(\imodels{\Ialt}{\ivr}\), so \(\iaccessible[\ptest{\ivr}]{\Ialt}{\Ialt}\).
So \(\iget[state]{\I}=\iget[state]{\Ialt}\) on $\vars{\ptest{\ivr}}$ since \(\boundvars{\ptest{\ivr}}=\emptyset\).

\item
  \m{\iaccessible[\pevolvein{\D{x}=\genDE{x}}{\ivr}]{\I}{\It}} 
  implies that there is an $\iget[state]{\Italt}$ reached from $\iget[state]{\Ialt}$ by following the differential equation for the same amount it took to reach $\iget[state]{\It}$ from $\iget[state]{\I}$.
  The solution will be the same, because \(\iget[const]{\I}=\iget[const]{\Ialt}\) on $\intsigns{\pevolvein{\D{x}=\genDE{x}}{\ivr}}$ and \(\iget[state]{\I}=\iget[state]{\Ialt}\) on $\freevars{\pevolvein{\D{x}=\genDE{x}}{\ivr}}$, which, using \rref{lem:coincidence}, contains all the variables whose values the differential equation solution depends on.
  Thus, both solutions agree on all variables that evolve during the continuous evolution, i.e.\ $\boundvars{\pevolvein{\D{x}=\genDE{x}}{\ivr}}$.
  Thus,
  \m{\iaccessible[\pevolvein{\D{x}=\genDE{x}}{\ivr}]{\Ialt}{\Italt}}
  and \(\iget[state]{\It}=\iget[state]{\Italt}\) on $\vars{\pevolvein{\D{x}=\genDE{x}}{\ivr}}$.

\item \m{\iaccessible[\pchoice{\alpha}{\beta}]{\I}{\It} = \iaccess[\alpha]{\I} \cup \iaccess[\beta]{\I}}
implies \(\iaccessible[\alpha]{\I}{\It}\) or \(\iaccessible[\beta]{\I}{\It}\),
which since \(V\supseteq\freevars{\pchoice{\alpha}{\beta}}\supseteq\freevars{\alpha}\)
and \(V\supseteq\freevars{\pchoice{\alpha}{\beta}}\supseteq\freevars{\beta}\) implies, by induction hypothesis,
that there is an $\iget[state]{\Italt}$ such that
  \(\iaccessible[\alpha]{\Ialt}{\Italt}\)
  and \(\iget[state]{\It}=\iget[state]{\Italt}\) on $V\cup\mustboundvars{\alpha}$
or that there is an $\iget[state]{\Italt}$ such that
  \(\iaccessible[\beta]{\Ialt}{\Italt}\)
  and \(\iget[state]{\It}=\iget[state]{\Italt}\) on $V\cup\mustboundvars{\beta}$, respectively.
  In either case, there is a $\iget[state]{\Italt}$ such that
  \(\iaccessible[\pchoice{\alpha}{\beta}]{\Ialt}{\Italt}\)
  and \(\iget[state]{\It}=\iget[state]{\Italt}\) on $V\cup\mustboundvars{\pchoice{\alpha}{\beta}}$,
  because \(\iaccess[\alpha]{\Ialt}\subseteq\iaccess[\pchoice{\alpha}{\beta}]{\Ialt}\)
  and \(\iaccess[\beta]{\Ialt}\subseteq\iaccess[\pchoice{\alpha}{\beta}]{\Ialt}\) 
  and \(\mustboundvars{\pchoice{\alpha}{\beta}}=\mustboundvars{\alpha}\cap\mustboundvars{\beta}\).

\begin{center}
\includegraphics{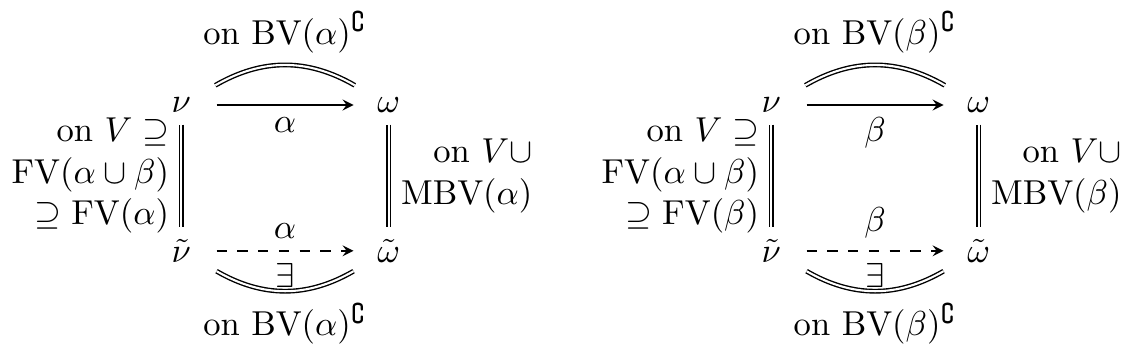}
\end{center}

\item
\newcommand{\Iz}{\iconcat[state=\mu]{\I}}%
\newcommand{\Izalt}{\vdLint[const=J,state=\tilde{\mu}]}%
\m{\iaccessible[\alpha;\beta]{\I}{\It} = \iaccess[\alpha]{\I} \compose \iaccess[\beta]{\I}},
i.e.\ there is a $\iget[state]{\Iz}$ such that \(\iaccessible[\alpha]{\I}{\Iz}\) and \(\iaccessible[\beta]{\Iz}{\It}\).
Since \(V\supseteq\freevars{\alpha;\beta}\supseteq\freevars{\alpha}\), by induction hypothesis, there is a $\iget[state]{\Izalt}$ such that
  \(\iaccessible[\alpha]{\Ialt}{\Izalt}\)
  and \(\iget[state]{\Iz}=\iget[state]{\Izalt}\) on $V\cup\mustboundvars{\alpha}$.
  Since \(V\supseteq\freevars{\alpha;\beta}\),
  so
  \(V\cup\mustboundvars{\alpha} \supseteq \freevars{\alpha;\beta} \cup \mustboundvars{\alpha}
  = \freevars{\alpha}\cup(\freevars{\beta}\setminus\mustboundvars{\alpha}) \cup \mustboundvars{\alpha}
  = \freevars{\alpha}\cup\freevars{\beta}\cup\mustboundvars{\alpha}
  \supseteq \freevars{\beta}\)
  by \rref{def:freevars-HP},
  and since \(\iaccessible[\beta]{\Iz}{\It}\), the induction hypothesis implies that
  there is an $\iget[state]{\Italt}$ such that
  \(\iaccessible[\beta]{\Izalt}{\Italt}\)
  and \(\iget[state]{\It}=\iget[state]{\Italt}\) on $(V\cup\mustboundvars{\alpha})\cup\mustboundvars{\beta} = V\cup\mustboundvars{\alpha;\beta}$.

  \begin{center}
  \includegraphics{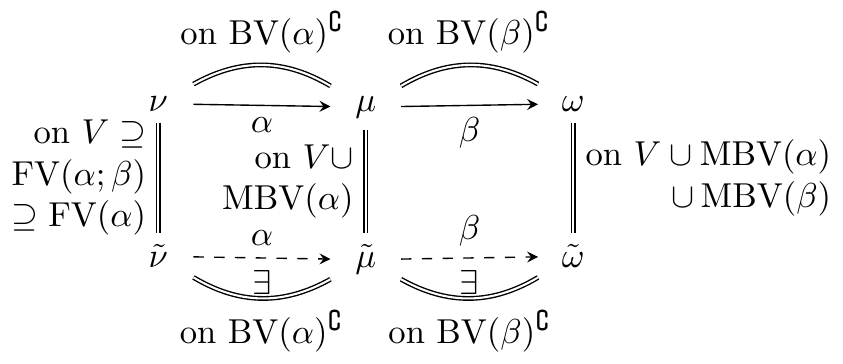}
\end{center}

\item
\renewcommand{\Iz}[1][]{\dLint[state=\nu_{#1}]}%
\m{\iaccessible[\prepeat{\alpha}]{\I}{\It} = \displaystyle\cupfold_{n\in\naturals}\iaccess[{\prepeat[n]{\alpha}}]{\I}}
iff there is an $n\in\naturals$ such that \(\iaccessible[\alpha^n]{\I}{\It}\).
The case $n=0$ follows from the assumption \(\iget[state]{\I}=\iget[state]{\Ialt}\) on $V\supseteq\freevars{\alpha}$, since \(\iget[state]{\It}=\iget[state]{\I}\) holds in that case and $\mustboundvars{\prepeat{\alpha}}=\emptyset$.
The case $n>0$ proceeds as follows.
Since \(\freevars{\prepeat[n]{\alpha}}=\freevars{\prepeat{\alpha}}=\freevars{\alpha}\), the induction hypothesis applied to the structurally simpler \HP $\prepeat[n]{\alpha}$ implies
that there is an $\iget[state]{\Italt}$ such that
  \(\iaccessible[\alpha^n]{\Ialt}{\Italt}\)
  and \(\iget[state]{\It}=\iget[state]{\Italt}\) on $V\cup\mustboundvars{\prepeat[n]{\alpha}} \supseteq V = V\cup\mustboundvars{\prepeat{\alpha}}$,
  since $\mustboundvars{\prepeat{\alpha}}=\emptyset$.
  Since \(\iaccess[{\prepeat[n]{\alpha}}]{\Ialt}\subseteq\iaccess[\prepeat{\alpha}]{\Ialt}\), this concludes the proof.
\qedhere
\end{compactitem}
\end{proofatend}

\iflongversion
When assuming $\iget[state]{\I}$ and $\iget[state]{\Ialt}$ to agree on all $\vars{\alpha}$, whether free or bound, $\iget[state]{\It}$ and $\iget[state]{\Italt}$ will continue to agree on $\vars{\alpha}$:
\begin{corollary}[Coincidence lemma] \label{cor:coincidence-HP}
  If \(\iget[state]{\I}=\iget[state]{\Ialt}\) on $\vars{\alpha}$
  and \(\iget[const]{\I}=\iget[const]{\Ialt}\) on $\intsigns{\alpha}$
  and \(\iaccessible[\alpha]{\I}{\It}\),
  then there is an $\iget[state]{\Italt}$ such that
  \(\iaccessible[\alpha]{\Ialt}{\Italt}\)
  and \(\iget[state]{\It}=\iget[state]{\Italt}\) on $\vars{\alpha}$.
  The same continues to hold if \(\iget[state]{\I}=\iget[state]{\Ialt}\) only on $\vars{\alpha}\setminus\mustboundvars{\alpha}$.
\end{corollary}
\begin{proofatend}
By \rref{lem:coincidence-HP} with \(V=\vars{\alpha}\supseteq\freevars{\alpha}\) or \(V=\vars{\alpha}\setminus\mustboundvars{\alpha}\), respectively.
\qedhere
\end{proofatend}

\begin{remark}
  Using hybrid computation sequences, the agreement in \rref{lem:coincidence-HP} continues to hold for
  \(\iget[state]{\It}=\iget[state]{\Italt}\) on $V\cup W$, where $W$ is the set of must-bound variables on the hybrid computation sequence that $\alpha$ actually took for the transition \(\iaccessible[\alpha]{\I}{\It}\), which could be larger than $\mustboundvars{\alpha}$.
\end{remark}
\fi

\section{Uniform Substitutions} \label{sec:usubst}

The uniform substitution rule \irref{US0} from first-order logic \cite[\S35,40]{Church_1956} substitutes \emph{all} occurrences of predicate $p(\usarg)$ by a formula $\mapply{\psi}{\usarg}$, i.e.\ it replaces all occurrences of $p(\theta)$, for any (vectorial) term $\theta$, by the corresponding $\mapply{\psi}{\theta}$ simultaneously:
\[
      \cinferenceRule[US0|US$_1$]{uniform substitution}
      {\linferenceRule[formula]
        {\preusubst[\phi]{p}}
        {\usubst[\phi]{p}{\psi}}
      }{}%
      \qquad\qquad
      \cinferenceRule[US|US]{uniform substitution}
      {\linferenceRule[formula]
        {\phi}
        {\applyusubst{\sigma}{\phi}}
      }{}%
\]
Rule \irref{US0} \cite{DBLP:journals/corr/Platzer14:dGL} requires all relevant substitutions of $\mapply{\psi}{\theta}$ for $p(\theta)$ to be admissible and requires that no $p(\theta)$ occurs in the scope of a quantifier or modality binding a variable of $\mapply{\psi}{\theta}$ other than the occurrences in $\theta$; see \cite[\S35,40]{Church_1956}.

This section considers a constructive definition of this proof rule that is more general: \irref{US}.
The \dL calculus uses uniform substitutions that affect terms, formulas, and programs.
A \dfn{uniform substitution} $\sigma$ is a mapping
from expressions of the
form \(f(\usarg)\) to terms $\applysubst{\sigma}{f(\usarg)}$,
from \(p(\usarg)\) to formulas $\applysubst{\sigma}{p(\usarg)}$,
from \(\contextapp{C}{\uscarg}\) to formulas $\applysubst{\sigma}{\contextapp{C}{\uscarg}}$,
and from program constants \(a\) to \HPs $\applysubst{\sigma}{a}$.
Vectorial extensions are accordingly for uniform substitutions of other arities $k\geq0$.
Here $\usarg$ is a reserved function symbol of arity zero and $\uscarg$ a reserved \predicational symbol of arity zero.
Figure~\ref{fig:usubst} defines the result $\applyusubst{\sigma}{\phi}$ of applying to a \dL formula~$\phi$ the \dfn{uniform substitution} $\sigma$ that uniformly replaces all occurrences of function~$f$ by a (instantiated) term and all occurrences of predicate~$p$ or \predicational~$C$ by a (instantiated) formula 
as well as of program constant $a$ by a program.
The notation $\applysubst{\sigma}{f(\usarg)}$ denotes the replacement for $f(\usarg)$ according to $\sigma$, i.e.\ the value $\applysubst{\sigma}{f(\usarg)}$ of function $\sigma$ at $f(\usarg)$.
By contrast, $\applyusubst{\sigma}{\phi}$ denotes the result of applying $\sigma$ to $\phi$ according to \rref{fig:usubst} (likewise for $\applyusubst{\sigma}{\theta}$ and $\applyusubst{\sigma}{\alpha}$).
The notation $f\in\replacees{\sigma}$ signifies that $\sigma$ replaces $f$, i.e.\ \(\applysubst{\sigma}{f(\usarg)} \neq f(\usarg)\).
Finally, $\sigma$ is a total function when augmented with \(\applysubst{\sigma}{g(\usarg)}=g(\usarg)\) for all $g\not\in\replacees{\sigma}$.
Accordingly for predicate symbols, \predicational{}s, and program constants.

\begin{definition}[Admissible uniform substitution] \label{def:usubst-admissible}
  \index{admissible|see{substitution, admissible}}
  The uniform substitution~$\sigma$ is $U$-\emph{\admissible} for $\phi$ (or $\theta$ or $\alpha$, respectively) with respect to the set $U\subseteq\allvars\cup\D{\allvars}$ iff
  \(\freevars{\restrict{\sigma}{\intsigns{\phi}}}\cap U=\emptyset\),
  where \({\restrict{\sigma}{\intsigns{\phi}}}\) is the restriction of $\sigma$ that only replaces symbols that occur in $\phi$
  and
  \(\freevars{\sigma}=\cupfold_{f\in\replacees{\sigma}} \freevars{\applysubst{\sigma}{f(\usarg)}} \cup \cupfold_{p\in\replacees{\sigma}} \freevars{\applysubst{\sigma}{p(\usarg)}}\)
  are the \emph{free variables} that $\sigma$ introduces. 
  The uniform substitution~$\sigma$ is \emph{\admissible} for $\phi$ (or $\theta$ or $\alpha$, respectively) iff all admissibility conditions during its application according to \rref{fig:usubst} hold, 
  which check that the bound variables $U$ of each operator are not free in the substitution on its arguments, i.e.\ $\sigma$ is $U$-admissible.
  Otherwise the substitution clashes so its result $\applyusubst{\sigma}{\phi}$ ($\applyusubst{\sigma}{\theta}$ or $\applyusubst{\sigma}{\alpha}$) is not defined.
\end{definition}

\irref{US} is only applicable if $\sigma$ is admissible for $\phi$.
In all subsequent results, all applications of uniform substitutions are required to be defined (no clash).

\begin{figure}[tbhp]
  \begin{displaymath}
    \begin{array}{@{}rcll@{}}
    \applyusubst{\sigma}{x} &=& x & \text{for variable $x\in\allvars$}\\
    \applyusubst{\sigma}{\D{x}} &=& \D{x} & \text{for differential symbol $\D{x}\in\D{\allvars}$}\\
    \applyusubst{\sigma}{f(\theta)} &=& (\applyusubst{\sigma}{f})(\applyusubst{\sigma}{\theta})
  \mdefeq \applyusubst{\{\usarg\mapsto\applyusubst{\sigma}{\theta}\}}{\applysubst{\sigma}{f(\usarg)}} &
  \text{for function symbol}~f\in\replacees{\sigma}
    \\
  \applyusubst{\sigma}{g(\theta)} &=& g(\applyusubst{\sigma}{\theta}) &\text{for function symbol}~g\not\in\replacees{\sigma}
  \\
  \applyusubst{\sigma}{\theta+\eta} &=& \applyusubst{\sigma}{\theta} + \applyusubst{\sigma}{\eta}
  \\
  \applyusubst{\sigma}{\theta\cdot\eta} &=& \applyusubst{\sigma}{\theta} \cdot \applyusubst{\sigma}{\eta}
  \\
  \applyusubst{\sigma}{\der{\theta}} &=& \der{\applyusubst{\sigma}{\theta}} &\text{if $\sigma$ $\allvars\cup\D{\allvars}$-admissible for $\theta$}
  \\
  \hline
  \applyusubst{\sigma}{\theta\geq\eta} &\mequiv& \applyusubst{\sigma}{\theta} \geq \applyusubst{\sigma}{\eta}\\
    \applyusubst{\sigma}{p(\theta)} &\mequiv& (\applyusubst{\sigma}{p})(\applyusubst{\sigma}{\theta})
  \mdefequiv \applyusubst{\{\usarg\mapsto\applyusubst{\sigma}{\theta}\}}{\applysubst{\sigma}{p(\usarg)}} &
  \text{for predicate symbol}~p\in\replacees{\sigma}\\
  \applyusubst{\sigma}{q(\theta)} &\mequiv& q(\applyusubst{\sigma}{\theta}) &\text{for predicate symbol}~q\not\in\replacees{\sigma}\\
  \applyusubst{\sigma}{\contextapp{C}{\phi}} &\mequiv& \contextapp{\applyusubst{\sigma}{C}}{\applyusubst{\sigma}{\phi}}
  \mdefequiv \applyusubst{\{\uscarg\mapsto\applyusubst{\sigma}{\phi}\}}{\applysubst{\sigma}{\contextapp{C}{\uscarg}}} &
  \text{if $\sigma$ $\allvars\cup\D{\allvars}$-admissible for $\phi$, $C\in\replacees{\sigma}$}
  \\
  \applyusubst{\sigma}{\contextapp{C}{\phi}} &\mequiv& \contextapp{C}{\applyusubst{\sigma}{\phi}} &
  \text{if $\sigma$ $\allvars\cup\D{\allvars}$-admissible for $\phi$, $C\not\in\replacees{\sigma}$}
  \\
    \applyusubst{\sigma}{\lnot\phi} &\mequiv& \lnot\applyusubst{\sigma}{\phi}\\
    \applyusubst{\sigma}{\phi\land\psi} &\mequiv& \applyusubst{\sigma}{\phi} \land \applyusubst{\sigma}{\psi}\\
    \applyusubst{\sigma}{\lforall{x}{\phi}} &=& \lforall{x}{\applyusubst{\sigma}{\phi}} & \text{if $\sigma$ $\{x\}$-admissible for $\phi$}\\
    \applyusubst{\sigma}{\lexists{x}{\phi}} &=& \lexists{x}{\applyusubst{\sigma}{\phi}} & \text{if $\sigma$ $\{x\}$-admissible for $\phi$}\\
    \applyusubst{\sigma}{\dbox{\alpha}{\phi}} &=& \dbox{\applyusubst{\sigma}{\alpha}}{\applyusubst{\sigma}{\phi}} & \text{if $\sigma$ $\boundvars{\applyusubst{\sigma}{\alpha}}$-admissible for $\phi$}\\
    \applyusubst{\sigma}{\ddiamond{\alpha}{\phi}} &=& \ddiamond{\applyusubst{\sigma}{\alpha}}{\applyusubst{\sigma}{\phi}} & \text{if $\sigma$ $\boundvars{\applyusubst{\sigma}{\alpha}}$-admissible for $\phi$}
    \\
  \hline
    \applyusubst{\sigma}{a} &\mequiv& \applysubst{\sigma}{a} &\text{for program constant $a\in\replacees{\sigma}$}\\
    \applyusubst{\sigma}{b} &\mequiv& b &\text{for program constant $b\not\in\replacees{\sigma}$}\\
    \applyusubst{\sigma}{\pupdate{\umod{x}{\theta}}} &\mequiv& \pupdate{\umod{x}{\applyusubst{\sigma}{\theta}}}\\
    \applyusubst{\sigma}{\Dupdate{\Dumod{\D{x}}{\theta}}} &\mequiv& \Dupdate{\Dumod{\D{x}}{\applyusubst{\sigma}{\theta}}}\\
    \applyusubst{\sigma}{\pevolvein{\D{x}=\genDE{x}}{\ivr}} &\mequiv&
    \hevolvein{\D{x}=\applyusubst{\sigma}{\genDE{x}}}{\applyusubst{\sigma}{\ivr}} & \text{if $\sigma$ $\{x,\D{x}\}$-admissible for $\genDE{x},\ivr$}\\
    \applyusubst{\sigma}{\ptest{\ivr}} &\mequiv& \ptest{\applyusubst{\sigma}{\ivr}}\\
    \applyusubst{\sigma}{\pchoice{\alpha}{\beta}} &\mequiv& \pchoice{\applyusubst{\sigma}{\alpha}} {\applyusubst{\sigma}{\beta}}\\
    \applyusubst{\sigma}{\alpha;\beta} &\mequiv& \applyusubst{\sigma}{\alpha}; \applyusubst{\sigma}{\beta} &\text{if $\sigma$ $\boundvars{\applyusubst{\sigma}{\alpha}}$-admissible for $\beta$}\\
    \applyusubst{\sigma}{\prepeat{\alpha}} &\mequiv& \prepeat{(\applyusubst{\sigma}{\alpha})} &\text{if $\sigma$ $\boundvars{\applyusubst{\sigma}{\alpha}}$-admissible for $\alpha$}
    \end{array}%
  \end{displaymath}%
  \vspace{-\baselineskip}
  \caption{Recursive application of uniform substitution~$\sigma$}%
  \index{substitution!uniform|textbf}%
  \label{fig:usubst}
\end{figure}
\subsection{Correctness of Uniform Substitutions}

Let
\m{\iget[const]{\imodif[const]{\I}{p}{R}}} denote the interpretation that agrees with interpretation~$\iget[const]{\I}$ except for the interpretation of predicate symbol~$p$, which is changed to~\m{R \subseteq \reals}.
Accordingly for predicate symbols of other arities, for function symbols $f$, and \predicational{s} $C$.

\begin{corollary}[Substitution adjoints] \label{cor:adjointUsubst}
\def\Ialta{\iadjointSubst{\sigma}{\Ialt}}%
The \emph{adjoint interpretation} $\iget[const]{\Ia}$ to substitution $\sigma$ for $\iportray{\I}$ is the interpretation that agrees with $\iget[const]{\I}$ except that for each function symbol $f\in\replacees{\sigma}$, predicate symbol $p\in\replacees{\sigma}$, \predicational $C\in\replacees{\sigma}$, and program constant $a\in\replacees{\sigma}$:
\begin{align*}
  \iget[const]{\Ia}(f) &: \reals\to\reals;\, d\mapsto\ivaluation{\imodif[const]{\I}{\,\usarg}{d}}{\applysubst{\sigma}{f}(\usarg)}\\
  \iget[const]{\Ia}(p) &= \{d\in\reals \with \imodels{\imodif[const]{\I}{\,\usarg}{d}}{\applysubst{\sigma}{p}(\usarg)}\}
  \\
  \iget[const]{\Ia}(C) &: \powerset{\reals}\to\powerset{\reals};\, R\mapsto\imodel{\imodif[const]{\I}{\,\uscarg}{R}}{\applysubst{\sigma}{\contextapp{C}{\uscarg}}}
  \\
  \iget[const]{\Ia}(a) &= \iaccess[\applysubst{\sigma}{a}]{\I}
\end{align*}
If \(\iget[state]{\I}=\iget[state]{\It}\) on \(\freevars{\sigma}\),
then \(\iget[const]{\Ia}=\iget[const]{\Ita}\).
If $\sigma$ is $U$-admissible for $\phi$ (or $\theta$ or $\alpha$, respectively) and \(\iget[state]{\I}=\iget[state]{\It}\) on $\scomplement{U}$, then
\begin{align*}
  \ivalues{\Ia}{\theta} &= \ivalues{\Ita}{\theta}
  ~\text{i.e.}~
  \ivaluation{\iconcat[state=\mu]{\Ia}}{\theta} = \ivaluation{\iconcat[state=\mu]{\Ita}}{\theta} ~\text{for all}~\mu
  \\
  \imodel{\Ia}{\phi} &= \imodel{\Ita}{\phi}\\
  \iaccess[\alpha]{\Ia} &= \iaccess[\alpha]{\Ita}
\end{align*}
\end{corollary}
\begin{proofatend}
For well-definedness of $\iget[const]{\Ia}$, note that $\iget[const]{\Ia}(f)$ is a smooth function since \({\applysubst{\sigma}{f}(\usarg)}\) has smooth values.
First, \(\iget[const]{\Ia}(a) = \iaccess[\applysubst{\sigma}{a}]{\I} = \iget[const]{\Ita}(a)\) holds because the adjoint to $\sigma$ for $\iportray{\I}$ in the case of programs is independent of $\iget[state]{\Ia}$ (the program has access to its respective initial state at runtime).
Likewise \(\iget[const]{\Ia}(C) = \iget[const]{\Ita}(C)\) for \predicational{s}.
By \rref{lem:coincidence-term},
\(\ivaluation{\imodif[const]{\I}{\,\usarg}{d}}{\applysubst{\sigma}{f}(\usarg)}
= \ivaluation{\imodif[const]{\It}{\,\usarg}{d}}{\applysubst{\sigma}{f}(\usarg)}\) when \(\iget[state]{\I}=\iget[state]{\It}\) on \(\freevars{\applysubst{\sigma}{f}(\usarg)}\).
Also \(\imodels{\imodif[const]{\I}{\,\usarg}{d}}{\applysubst{\sigma}{p}(\usarg)}\)
iff \(\imodels{\imodif[const]{\It}{\,\usarg}{d}}{\applysubst{\sigma}{p}(\usarg)}\)
by \rref{lem:coincidence} when \(\iget[state]{\I}=\iget[state]{\It}\) on \(\freevars{\applysubst{\sigma}{p}(\usarg)}\).
Thus, \(\iget[const]{\Ia}=\iget[const]{\Ita}\) when $\iget[state]{\I}=\iget[state]{\It}$ on $\freevars{\sigma}$.

If $\sigma$ is $U$-admissible for $\phi$ (or $\theta$ or $\alpha$), then
  \(\freevars{\applysubst{\sigma}{f(\usarg)}}\cap U=\emptyset\)
  so
  \(\freevars{\applysubst{\sigma}{f(\usarg)}}\subseteq\scomplement{U}\)
  for every function symbol $f\in\intsigns{\phi}$ (or $\theta$ or $\alpha$) and likewise for predicate symbols $p\in\intsigns{\phi}$.
  Since \(\iget[state]{\I}=\iget[state]{\It}\) on $\scomplement{U}$,
  so \(\iget[const]{\Ita}=\iget[const]{\Ia}\) on the function and predicate symbols in $\intsigns{\phi}$ (or $\theta$ or $\alpha$).
  Finally \(\iget[const]{\Ita}=\iget[const]{\Ia}\) implies that
  \(\imodels{\Ita}{\phi}\) iff \(\imodels{\Ia}{\phi}\) by \rref{lem:coincidence}
  and that \(\ivalues{\Ia}{\theta} = \ivalues{\Ita}{\theta}\) by \rref{lem:coincidence-term}
  and that \(\iaccess[\alpha]{\Ita} = \iaccess[\alpha]{\Ia}\) by \rref{lem:coincidence-HP}.
\qedhere
\end{proofatend}

Substituting equals for equals is sound by the compositional semantics of \dL.
The more general uniform substitutions are still sound, because interpretations of uniform substitutes correspond to interpretations of their adjoints.
The semantic modification of adjoint interpretations has the same effect as the syntactic uniform substitution, proved by simultaneous induction.
\iflongversion
Recall that all substitutions in the following are assumed to be defined (not clash), otherwise the lemmas make no claim.
\fi

\begin{lemma}[Uniform substitution lemma] \label{lem:usubst-term}
The uniform substitution $\sigma$ and its adjoint interpretation $\iportray{\Ia}$ to $\sigma$ for $\iportray{\I}$ have the same \emph{term} semantics:
\[\ivaluation{\I}{\applyusubst{\sigma}{\theta}} = \ivaluation{\Ia}{\theta}\]
\end{lemma}
\begin{proofatend}
\def\Im{\vdLint[const=\usebox{\ImnI},state=\nu\oplus(x\mapsto\usebox{\Imnx})]}%

The proof is by structural induction on $\theta$.
\begin{compactitem}
\item
  \(\ivaluation{\I}{\applyusubst{\sigma}{x}} 
  = \ivaluation{\I}{x} = \iget[state]{\I}(x) \)
  = \(\ivaluation{\Ia}{x}\)
  \(\text{since}~ x\not\in\replacees{\sigma}\)
  for variable $x\in\allvars$

\item
  \(\ivaluation{\I}{\applyusubst{\sigma}{\D{x}}} 
  = \ivaluation{\I}{\D{x}} = \iget[state]{\I}(\D{x}) \)
  = \(\ivaluation{\Ia}{\D{x}}\)
  \(\text{as}~ \D{x}\not\in\replacees{\sigma}\)
  for differential symbol $\D{x}\in\D{\allvars}$

\item
  \(\ivaluation{\I}{\applyusubst{\sigma}{f(\theta)}}
  = \ivaluation{\I}{(\applyusubst{\sigma}{f})\big(\applyusubst{\sigma}{\theta}\big)}
  = \ivaluation{\I}{\applyusubst{\{\usarg\mapsto\applyusubst{\sigma}{\theta}\}}{\applysubst{\sigma}{f(\usarg)}}}\)
  \(\stackrel{\text{IH}}{=} \ivaluation{\Iminner}{\applysubst{\sigma}{f(\usarg)}}\)
    \(= (\iget[const]{\Ia}(f))(d)\)
  \(= (\iget[const]{\Ia}(f))(\ivaluation{\Ia}{\theta})
  = \ivaluation{\Ia}{f(\theta)}\)
  with
  \(d\mdefeq\ivaluation{\I}{\applyusubst{\sigma}{\theta}} 
  \stackrel{\text{IH}}{=} \ivaluation{\Ia}{\theta}\)
  by using the induction hypothesis twice,
  once for \(\applyusubst{\sigma}{\theta}\) on the smaller $\theta$ 
  and once for
  \(\applyusubst{\{\usarg\mapsto\applyusubst{\sigma}{\theta}\}}{\applysubst{\sigma}{f(\usarg)}}\)
  on the possibly bigger term
  \({\applysubst{\sigma}{f(\usarg)}}\)
  but the structurally simpler uniform substitution
  \(\applyusubst{\{\usarg\mapsto\applyusubst{\sigma}{\theta}\}}{\dots}\)
  that is a substitution on the symbol $\usarg$ of arity zero, not a substitution of functions with arguments.
  For well-foundedness of the induction note that the $\usarg$ substitution only happens for function symbols $f$ with at least one argument $\theta$
  (\(\text{for}~f\in\replacees{\sigma}\)).
  
\item
  \(\ivaluation{\I}{\applyusubst{\sigma}{g(\theta)}}
  = \ivaluation{\I}{g(\applyusubst{\sigma}{\theta})}\)
  = \(\iget[const]{\I}(g)\big(\ivaluation{\I}{\applyusubst{\sigma}{\theta}}\big)\)
  \(\stackrel{\text{IH}}{=} \iget[const]{\I}(g)\big(\ivaluation{\Ia}{\theta}\big)
  = \iget[const]{\Ia}(g)\big(\ivaluation{\Ia}{\theta}\big)
  = \ivaluation{\Ia}{g(\theta)}\)
  by induction hypothesis and since \(\iget[const]{\I}(g)=\iget[const]{\Ia}(g)\) as the interpretation of $g$ does not change in $\iget[const]{\Ia}$
  \(\text{for}~g\not\in\replacees{\sigma}\).

\item
  \(\ivaluation{\I}{\applyusubst{\sigma}{\theta+\eta}}
  = \ivaluation{\I}{\applyusubst{\sigma}{\theta} + \applyusubst{\sigma}{\eta}}
  = \ivaluation{\I}{\applyusubst{\sigma}{\theta}} + \ivaluation{\I}{\applyusubst{\sigma}{\eta}}\)
  \(\stackrel{\text{IH}}{=} \ivaluation{\Ia}{\theta} + \ivaluation{\Ia}{\eta}
  = \ivaluation{\Ia}{\theta+\eta}\)
  by induction hypothesis.

\item
  \(\ivaluation{\I}{\applyusubst{\sigma}{\theta\cdot\eta}}
  = \ivaluation{\I}{\applyusubst{\sigma}{\theta} \cdot \applyusubst{\sigma}{\eta}}
  = \ivaluation{\I}{\applyusubst{\sigma}{\theta}} \cdot \ivaluation{\I}{\applyusubst{\sigma}{\eta}}\)
  \(\stackrel{\text{IH}}{=} \ivaluation{\Ia}{\theta} \cdot \ivaluation{\Ia}{\eta}
  = \ivaluation{\Ia}{\theta\cdot\eta}\)
  by induction hypothesis.

\item
\def\Im{\imodif[state]{\I}{x}{X}}%
\def\Ima{\iadjointSubst{\sigma}{\Im}}%
\def\Iam{\imodif[state]{\Ia}{x}{X}}%
\(
\ivaluation{\I}{\applyusubst{\sigma}{\der{\theta}}}
= \ivaluation{\I}{\der{\applyusubst{\sigma}{\theta}}}
= \sum_x \iget[state]{\I}(\D{x}) \itimes \Dp[X]{\ivaluation{\Im}{\applyusubst{\sigma}{\theta}}}
\stackrel{\text{IH}}{=} \sum_x \iget[state]{\I}(\D{x}) \itimes \Dp[X]{\ivaluation{\Ima}{\theta}}
= \sum_x \iget[state]{\I}(\D{x}) \itimes \Dp[X]{\ivaluation{\Iam}{\theta}}
= \ivaluation{\Ia}{\der{\theta}}
\)
by induction hypothesis,
provided $\sigma$ is $\allvars\cup\D{\allvars}$-admissible for $\theta$, i.e. does not introduce any variables or differential symbols, 
so that \rref{cor:adjointUsubst} implies \(\iget[const]{\Ia}=\iget[const]{\Ita}\) for all $\iget[state]{\I},\iget[state]{\It}$ (that agree on $\scomplement{(\allvars\cup\D{\allvars})}=\emptyset$, which imposes no condition on $\iget[state]{\I},\iget[state]{\It}$).
\qedhere
\end{compactitem}
\end{proofatend}

\begin{lemma}[Uniform substitution lemma] \label{lem:usubst}
The uniform substitution $\sigma$ and its adjoint interpretation $\iportray{\Ia}$ to $\sigma$ for $\iportray{\I}$ have the same \emph{formula} semantics:
\[\imodels{\I}{\applyusubst{\sigma}{\phi}} ~\text{iff}~ \imodels{\Ia}{\phi}\]
\end{lemma}
\begin{proofatend}
The proof is by structural induction on $\phi$.
\begin{compactitem}
\item
  \(\imodels{\I}{\applyusubst{\sigma}{\theta\geq\eta}}\)
  iff \(\imodels{\I}{\applyusubst{\sigma}{\theta} \geq \applyusubst{\sigma}{\eta}}\)
  iff \(\ivaluation{\I}{\applyusubst{\sigma}{\theta}} \geq \ivaluation{\I}{\applyusubst{\sigma}{\eta}}\),
  by \rref{lem:usubst-term},
  iff \(\ivaluation{\Ia}{\theta} \geq \ivaluation{\Ia}{\eta}\)
  iff \(\ivaluation{\Ia}{\theta\geq\eta}\)

\item
  \(\imodels{\I}{\applyusubst{\sigma}{p(\theta)}}\)
  iff \(\imodels{\I}{(\applyusubst{\sigma}{p})\big(\applyusubst{\sigma}{\theta}\big)}\)
  iff \(\imodels{\I}{\applyusubst{\{\usarg\mapsto\applyusubst{\sigma}{\theta}\}}{\applysubst{\sigma}{p(\usarg)}}}\)
  iff, by IH, \(\imodels{\Iminner}{\applysubst{\sigma}{p(\usarg)}}\)
  iff \(d \in \iget[const]{\Ia}(p)\)
  iff \((\ivaluation{\Ia}{\theta}) \in \iget[const]{\Ia}(p)\)
  iff \(\imodels{\Ia}{p(\theta)}\)
  with \(d\mdefeq\ivaluation{\I}{\applyusubst{\sigma}{\theta}} = \ivaluation{\Ia}{\theta}\)
  by using \rref{lem:usubst-term} for \(\applyusubst{\sigma}{\theta}\)
  and by using the induction hypothesis
  for \(\applyusubst{\{\usarg\mapsto\applyusubst{\sigma}{\theta}\}}{\applysubst{\sigma}{p(\usarg)}}\)
  on the possibly bigger formula \({\applysubst{\sigma}{p(\usarg)}}\) but the structurally simpler uniform substitution \(\applyusubst{\{\usarg\mapsto\applyusubst{\sigma}{\theta}\}}{\dots}\) that is a mere substitution on symbol $\usarg$ of arity zero, not a substitution of predicates
  (\(\text{for}~p\in\replacees{\sigma}\)).

\item
  \(\imodels{\I}{\applyusubst{\sigma}{q(\theta)}}\)
  iff \(\imodels{\I}{q(\applyusubst{\sigma}{\theta})}\)
  iff \(\big(\ivaluation{\I}{\applyusubst{\sigma}{\theta}}\big)\in \iget[const]{\I}(q)\)
  so, by \rref{lem:usubst-term}, iff \(\big(\ivaluation{\Ia}{\theta}\big) \in \iget[const]{\I}(q)\)
  iff \(\big(\ivaluation{\Ia}{\theta}\big) \in \iget[const]{\Ia}(q)\)
  iff \(\imodels{\Ia}{q(\theta)}\)
  since \(\iget[const]{\I}(q)=\iget[const]{\Ia}(q)\) as the interpretation of $q$ does not change in $\iget[const]{\Ia}$
  (\(\text{for}~q\not\in\replacees{\sigma}\))

\item
\def\ImM{\imodif[const]{\I}{\uscarg}{R}}%
\let\ImN\ImM%
\def\IaM{\imodif[const]{\Ia}{\uscarg}{R}}%
  For the case \({\applyusubst{\sigma}{\contextapp{C}{\phi}}}\), first show 
  \(\imodel{\I}{\applyusubst{\sigma}{\phi}} = \imodel{\Ia}{\phi}\).
  By induction hypothesis for the smaller $\phi$:
  \(\imodels{\It}{\applyusubst{\sigma}{\phi}}\)
  iff
  \(\imodels{\Ita}{\phi}\),
  where 
  \(\imodel{\Ita}{\phi}=\imodel{\Ia}{\phi}\)
  by \rref{cor:adjointUsubst}
  for all $\iget[state]{\Ia},\iget[state]{\Ita}$
  (that agree on $\scomplement{(\allvars\cup\D{\allvars})}=\emptyset$, which imposes no condition on $\iget[state]{\I},\iget[state]{\It}$)
  since $\sigma$ is $\allvars\cup\D{\allvars}$-admissible for $\phi$.
  The proof then proceeds:

  \(\imodels{\I}{\applyusubst{\sigma}{\contextapp{C}{\phi}}}\)
  \(=\imodel{\I}{\contextapp{\applyusubst{\sigma}{C}}{\applyusubst{\sigma}{\phi}}}\)
  \(= \imodel{\I}{\applyusubst{\{\uscarg\mapsto\applyusubst{\sigma}{\phi}\}}{\applysubst{\sigma}{\contextapp{C}{\uscarg}}}}\),
  so, by induction hypothesis for the structurally simpler uniform substitution ${\{\uscarg\mapsto\applyusubst{\sigma}{\phi}\}}$ that is a mere substitution on symbol $\uscarg$ of arity zero, iff
  \(\imodels{\ImM}{\applysubst{\sigma}{\contextapp{C}{\uscarg}}}\)
  since the adjoint to \(\{\uscarg\mapsto\applyusubst{\sigma}{\phi}\}\) is $\iget[const]{\ImM}$ with \(R\mdefeq\imodel{\I}{\applyusubst{\sigma}{\phi}}\).
  
  Also
  \(\imodels{\Ia}{\contextapp{C}{\phi}}\)
  \(= \iget[const]{\Ia}(C)\big(\imodel{\Ia}{\phi}\big)\)
  \(= \imodel{\ImN}{\applysubst{\sigma}{\contextapp{C}{\uscarg}}}\)
  for \(R=\imodel{\Ia}{\phi}=\imodel{\I}{\applyusubst{\sigma}{\phi}}\) by induction hypothesis.
  Both sides are, thus, equivalent.
  
\item
  The case \({\applyusubst{\sigma}{\contextapp{C}{\phi}}}\) for $C\not\in\replacees{\sigma}$ again first shows
  \(\imodel{\I}{\applyusubst{\sigma}{\phi}} = \imodel{\Ia}{\phi}\)
  for all $\iget[state]{\I}$ using that $\sigma$ is $\allvars\cup\D{\allvars}$-admissible for $\phi$.
  Then
  \(\imodels{\I}{\applyusubst{\sigma}{\contextapp{C}{\phi}}}\)
  \(= \imodel{\I}{\contextapp{C}{\applyusubst{\sigma}{\phi}}}\)
  \(= \iget[const]{\I}(C)\big(\imodel{\I}{\applyusubst{\sigma}{\phi}}\big)\)
  \(= \iget[const]{\I}(C)\big(\imodel{\Ia}{\phi}\big)\)
  \(= \iget[const]{\Ia}(C)\big(\imodel{\Ia}{\phi}\big)\)
  \(= \imodel{\Ia}{\contextapp{C}{\phi}}\)
  iff \(\imodels{\Ia}{\contextapp{C}{\phi}}\)

\item
  \(\imodels{\I}{\applyusubst{\sigma}{\lnot\phi}}\)
  iff \(\imodels{\I}{\lnot\applyusubst{\sigma}{\phi}}\)
  iff \(\inonmodels{\I}{\applyusubst{\sigma}{\phi}}\),
  by induction hypothesis,
  iff \(\inonmodels{\Ia}{\phi}\)
  iff \(\imodels{\Ia}{\lnot\phi}\)

\item
  \(\imodels{\I}{\applyusubst{\sigma}{\phi\land\psi}}\)
  iff \(\imodels{\I}{\applyusubst{\sigma}{\phi} \land \applyusubst{\sigma}{\psi}}\)
  iff \(\imodels{\I}{\applyusubst{\sigma}{\phi}}\) and \(\imodels{\I}{\applyusubst{\sigma}{\psi}}\),
  by induction hypothesis,
  iff \(\imodels{\Ia}{\phi}\) and \(\imodels{\Ia}{\psi}\)
  iff \(\imodels{\Ia}{\phi\land\psi}\)

\item
\def\Imd{\imodif[state]{\I}{x}{d}}%
\def\Iamd{\imodif[state]{\Ia}{x}{d}}%
\def\Imda{\iadjointSubst{\sigma}{\Imd}}%
  \(\imodels{\I}{\applyusubst{\sigma}{\lexists{x}{\phi}}}\)
  iff \(\imodels{\I}{\lexists{x}{\applyusubst{\sigma}{\phi}}}\)
  (provided $\sigma$ is $\{x\}$-admissible for $\phi$)
  iff \(\imodels{\Imd}{\applyusubst{\sigma}{\phi}}\) for some $d$,
  so, by induction hypothesis,
  iff \(\imodels{\Imda}{\phi}\) for some $d$,
  which is equivalent to
  \(\imodels{\Iamd}{\phi}\) by \rref{cor:adjointUsubst} as $\sigma$ is $\{x\}$-admissible for $\phi$ and $\iget[state]{\I}=\iget[state]{\Imd}$ on $\scomplement{\{x\}}$.
  Thus, this is equivalent to
  \(\imodels{\Ia}{\lexists{x}{\phi}}\).
  
\item The case
  \(\imodels{\I}{\applyusubst{\sigma}{\lforall{x}{\phi}}}\)
  follows by duality \(\lforall{x}{\phi} \mequiv \lnot\lexists{x}{\lnot\phi}\), which is respected in the definition of uniform substitutions.

\item
  \newcommand{\Iat}{\iconcat[state=\omega]{\Ia}}%
  \(\imodels{\I}{\applyusubst{\sigma}{\ddiamond{\alpha}{\phi}}}\)
  iff \(\imodels{\I}{\ddiamond{\applyusubst{\sigma}{\alpha}}{\applyusubst{\sigma}{\phi}}}\)
  (provided $\sigma$ is $\boundvars{\applyusubst{\sigma}{\alpha}}$-admissible for $\phi$)
  iff there is a $\iget[state]{\It}$ such that
  \(\iaccessible[\applyusubst{\sigma}{\alpha}]{\I}{\It}\) and \(\imodels{\It}{\applyusubst{\sigma}{\phi}}\),
  which, by \rref{lem:usubst-HP} and induction hypothesis, respectively, is equivalent to:
  there is a $\iget[state]{\Ita}$ such that
  \(\iaccessible[\alpha]{\Ia}{\Ita}\) and \(\imodels{\Ita}{\phi}\),
  which is equivalent to
  \(\imodels{\Ia}{\ddiamond{\alpha}{\phi}}\),
  because \(\imodels{\Ita}{\phi}\) is equivalent to \(\imodels{\Iat}{\phi}\) by \rref{cor:adjointUsubst}
  as $\sigma$ is $\boundvars{\applyusubst{\sigma}{\alpha}}$-admissible for $\phi$ and \(\iget[state]{\Ia}=\iget[state]{\Iat}\) on $\scomplement{\boundvars{\applyusubst{\sigma}{\alpha}}}$ by \rref{lem:bound} since
  \(\iaccessible[\applyusubst{\sigma}{\alpha}]{\I}{\It}\).

\item The case
  \(\imodels{\I}{\applyusubst{\sigma}{\dbox{\alpha}{\phi}}}\)
  follows by duality \(\dbox{\alpha}{\phi} \mequiv \lnot\ddiamond{\alpha}{\lnot\phi}\), which is respected in the definition of uniform substitutions.
\qedhere
\end{compactitem}
\end{proofatend}

\begin{lemma}[Uniform substitution lemma] \label{lem:usubst-HP}
The uniform substitution $\sigma$ and its adjoint interpretation $\iportray{\Ia}$ to $\sigma$ for $\iportray{\I}$ have the same \emph{program} semantics:
\[
\iaccessible[{\applyusubst{\sigma}{\alpha}}]{\I}{\It}
~\text{iff}~
\iaccessible[\alpha]{\Ia}{\Ita}
\]
\end{lemma}
\begin{proofatend}
The proof is by structural induction on $\alpha$.
\begin{compactitem}
\item
  \(\iaccessible[\applyusubst{\sigma}{a}]{\I}{\It} = \iaccess[\applysubst{\sigma}{a}]{\I} = \iget[const]{\Ia}(a) = \iaccess[a]{\Ia}\)
  for program constant $a\in\replacees{\sigma}$
  (the proof is accordingly for $a\not\in\replacees{\sigma}$).

\item 
  \(\iaccessible[\applyusubst{\sigma}{\pumod{x}{\theta}}]{\I}{\It}
  = \iaccess[\pumod{x}{\applyusubst{\sigma}{\theta}}]{\I}\)
  iff \(\iget[state]{\It} = \modif{\iget[state]{\I}}{x}{\ivaluation{\I}{\applyusubst{\sigma}{\theta}}}\)
  = \(\modif{\iget[state]{\I}}{x}{\ivaluation{\Ia}{\theta}}\)
  by \rref{lem:usubst}, which is, thus, equivalent to
  \(\iaccessible[\pumod{x}{\theta}]{\Ia}{\Ita}\).

\item
  \(\iaccessible[\applyusubst{\sigma}{\Dumod{\D{x}}{\theta}}]{\I}{\It}
  = \iaccess[\Dumod{\D{x}}{\applyusubst{\sigma}{\theta}}]{\I}\)
  iff \(\iget[state]{\It} = \modif{\iget[state]{\I}}{\D{x}}{\ivaluation{\I}{\applyusubst{\sigma}{\theta}}}\)
  = \(\modif{\iget[state]{\I}}{\D{x}}{\ivaluation{\Ia}{\theta}}\)
  by \rref{lem:usubst}, which is, thus, equivalent to
  \(\iaccessible[\Dumod{\D{x}}{\theta}]{\Ia}{\Ita}\).

\item
  \(\iaccessible[\applyusubst{\sigma}{\ptest{\ivr}}]{\I}{\It}
  = \iaccess[\ptest{\applyusubst{\sigma}{\ivr}}]{\I}\)
  iff \(\iget[state]{\It}=\iget[state]{\I}\)
  and \(\imodels{\I}{\applyusubst{\sigma}{\ivr}}\),
  iff, by \rref{lem:usubst},
  \(\iget[state]{\Ita}=\iget[state]{\Ia}\)     and
  \(\imodels{\Ia}{\ivr}\),
  which is equivalent to
  \(\iaccessible[\ptest{\ivr}]{\Ia}{\Ita}\).

\item
\newcommand{\Izeta}{\iconcat[state=\varphi(t)]{\I}}
\def\Izetaa{\iadjointSubst{\sigma}{\Izeta}}%
\newcommand{\Iazeta}{\iconcat[state=\varphi(t)]{\Ia}}
  \(\iaccessible[\applyusubst{\sigma}{\pevolvein{\D{x}=\genDE{x}}{\ivr}}]{\I}{\It}
  = \iaccess[\pevolvein{\D{x}=\applyusubst{\sigma}{\genDE{x}}}
  {\applyusubst{\sigma}{\ivr}}]{\I}\)
  (provided $\sigma$ $\{x,\D{x}\}$-admissible for $\genDE{x},\ivr$)
  iff \(\mexists{\varphi:[0,T]\to\linterpretations{\Sigma}{V}}\)
  with \(\varphi(0)=\iget[state]{\I}, \varphi(T)=\iget[state]{\It}\) and for all $t\geq0$:
  \(\D{\varphi}(t) = \ivaluation{\Izeta}{\applyusubst{\sigma}{\genDE{x}}}
  = \ivaluation{\Izetaa}{\genDE{x}}\) by \rref{lem:usubst-term}
  as well as
  \(\imodels{\Izeta}{\applyusubst{\sigma}{\ivr}}\),
  which, by \rref{lem:usubst}, is equivalent to
  \(\imodels{\Izetaa}{\ivr}\).
  
  Also
  \(\iaccessible[\pevolvein{\D{x}=\genDE{x}}{\ivr}]{\Ia}{\Ita}\)
  iff \(\mexists{\varphi:[0,T]\to\linterpretations{\Sigma}{V}}\)
  with \(\varphi(0)=\iget[state]{\I}, \varphi(T)=\iget[state]{\It}\) and for all $t\geq0$:
  \(\D{\varphi}(t) = \ivaluation{\Iazeta}{\genDE{x}}\)
  and
  \(\imodels{\Iazeta}{\ivr}\).
  Finally,
  \(\ivalues{\Iazeta}{\genDE{x}}=\ivalues{\Izetaa}{\genDE{x}}\) and
  \(\imodel{\Izetaa}{\ivr}=\imodel{\Iazeta}{\ivr}\)
  by \rref{cor:adjointUsubst}
  since $\sigma$ is $\{x,\D{x}\}$-admissible for $\genDE{x},\ivr$ and \(\iget[state]{\I}=\iget[state]{\Iazeta}\) on $\scomplement{\boundvars{\pevolvein{\D{x}=\genDE{x}}{\ivr}}}=\scomplement{\{x,\D{x}\}}$ by \rref{lem:bound}.
  
\item  
  \(\iaccessible[\applyusubst{\sigma}{\pchoice{\alpha}{\beta}}]{\I}{\It}
  = \iaccess[\pchoice{\applyusubst{\sigma}{\alpha}}{\applyusubst{\sigma}{\beta}}]{\I}\)
  = \(\iaccess[\applyusubst{\sigma}{\alpha}]{\I} \cup \iaccess[\applyusubst{\sigma}{\beta}]{\I}\),
  which, by induction hypothesis, is equivalent to
  \(\iaccessible[\alpha]{\Ia}{\Ita}\) or \(\iaccessible[\beta]{\Ia}{\Ita}\),
  which is equivalent to
  \(\iaccessible[\alpha]{\Ia}{\Ita} \cup \iaccess[\beta]{\Ia} = \iaccess[\pchoice{\alpha}{\beta}]{\Ia}\).

\item
{\newcommand{\Iz}{\iconcat[state=\mu]{\I}}%
\newcommand{\Iza}{\iadjointSubst{\sigma}{\Iz}}%
\newcommand{\Iaz}{\iconcat[state=\mu]{\Ia}}%
  \(\iaccessible[\applyusubst{\sigma}{\alpha;\beta}]{\I}{\It}
  = \iaccess[\applyusubst{\sigma}{\alpha}; \applyusubst{\sigma}{\beta}]{\I}\)
  = \(\iaccess[\applyusubst{\sigma}{\alpha}]{\I} \compose \iaccess[\applyusubst{\sigma}{\beta}]{\I}\)
  (provided $\sigma$ is $\boundvars{\applyusubst{\sigma}{\alpha}}$-admissible for $\beta$)
  iff there is a $\iget[state]{\Iz}$ such that
  \(\iaccessible[\applyusubst{\sigma}{\alpha}]{\I}{\Iz}\) and \(\iaccessible[\applyusubst{\sigma}{\beta}]{\Iz}{\It}\),
  which, by induction hypothesis, is equivalent to
  \(\iaccessible[\alpha]{\Ia}{\Iza}\) and \(\iaccessible[\beta]{\Iza}{\Ita}\).
  Yet, \(\iaccess[\beta]{\Iza}=\iaccess[\beta]{\Ia}\)
  by \rref{cor:adjointUsubst}, because $\sigma$ is $\boundvars{\applyusubst{\sigma}{\alpha}}$-admissible for $\beta$ and \(\iget[state]{\I}=\iget[state]{\It}\) on $\scomplement{\boundvars{\applyusubst{\sigma}{\alpha}}}$ by \rref{lem:bound} since \(\iaccessible[\applyusubst{\sigma}{\alpha}]{\I}{\Iz}\).
  Finally,
  \(\iaccessible[\alpha]{\Ia}{\Iaz}\) and \(\iaccessible[\beta]{\Iaz}{\Ita}\) for some $\iget[state]{\Iaz}$
  is equivalent to \(\iaccessible[\alpha;\beta]{\Ia}{\Ita}\).

}

\item
\newcommand{\Iz}[1][]{\iconcat[state=\nu_{#1}]{\I}}%
\newcommand{\Iza}[1][]{\iadjointSubst{\sigma}{\Iz[#1]}}%
\newcommand{\Iaz}[1][]{\iconcat[state=\nu_{#1}]{\Ia}}%
  \(\iaccessible[\applyusubst{\sigma}{\prepeat{\alpha}}]{\I}{\It}
  = \iaccess[\prepeat{(\applyusubst{\sigma}{\alpha})}]{\I}
  = \closureTransitive{\big(\iaccess[\applyusubst{\sigma}{\alpha}]{\I}\big)}
  = \cupfold_{n\in\naturals} (\iaccess[\applyusubst{\sigma}{\alpha}]{\I})^n
  \)
  (provided $\sigma$ is $\boundvars{\applyusubst{\sigma}{\alpha}}$-admissible for $\alpha$)
  iff there are $n\in\naturals$ and \(\iget[state]{\Iz[0]}=\iget[state]{\Ia},\iget[state]{\Iz[1]},\dots,\iget[state]{\Iz[n]}=\iget[state]{\Ita}\) such that
  \(\iaccessible[\applyusubst{\sigma}{\alpha}]{\Iz[i]}{\Iz[i+1]}\) for all $i<n$.
  By $n$ uses of the induction hypothesis, this is equivalent to
  \(\iaccessible[\alpha]{\Iza[i]}{\Iza[i+1]}\) for all $i<n$,
  which is equivalent to
  \(\iaccessible[\alpha]{\Iaz[i]}{\Iza[i+1]}\) 
  by \rref{cor:adjointUsubst}
  since $\sigma$ is $\boundvars{\applyusubst{\sigma}{\alpha}}$-admissible for $\alpha$
  and \(\iget[state]{\Iza[i+1]}=\iget[state]{\Iza[i]}\) on $\scomplement{\boundvars{\applyusubst{\sigma}{\alpha}}}$ by \rref{lem:bound} as \(\iaccessible[\applyusubst{\sigma}{\alpha}]{\Iz[i]}{\Iz[i+1]}\) for all $i<n$.
  Thus, this is equivalent to
  \(\iaccessible[\prepeat{\alpha}]{\Ia}{\Ita}
  = \closureTransitive{\big(\iaccess[\alpha]{\Ia}\big)}\).
\qedhere
\end{compactitem}
\end{proofatend}

\subsection{Soundness}

The uniform substitution lemmas are the key insights for the soundness of \irref{US}.
\irref{US} is only applicable if the uniform substitution is defined (does not clash).
\begin{theorem}[Soundness of uniform substitution] \label{thm:usubst-sound}
  \irref{US} is sound and so is its special case \irref{US0}.
  That is, if their premise is valid, then so is their conclusion.
\end{theorem}
\begin{proofatend}
\def\Ia{\iadjointSubst{\sigma}{\I}}%
Let the premise $\phi$ of \irref{US} be valid, i.e.\ \m{\imodels{\I}{\phi}} for all interpretations and states $\iportray{\I}$.
To show that the conclusion is valid, consider any interpretation and state $\iportray{\I}$ and show \(\imodels{\I}{\applyusubst{\sigma}{\phi}}\).
By \rref{lem:usubst}, \(\imodels{\I}{\applyusubst{\sigma}{\phi}}\) iff \(\imodels{\Ia}{\phi}\).
The latter holds, because \(\imodels{\I}{\phi}\) for all $\iportray{\I}$, including for $\iportray{\Ia}$, by premise.
The rule \irref{US0} is the special case of \irref{US} where $\sigma$ only substitutes predicate symbol $p$.
\qedhere
\end{proofatend}

\section{Differential Dynamic Logic Axioms} \label{sec:dL-axioms}

Proof rules and axioms for a Hilbert-type axiomatization of \dL from prior work \cite{DBLP:conf/lics/Platzer12b} are shown in \rref{fig:dL}, except that, thanks to rule \irref{US}, axioms and rules now comprise the finite list of \dL formulas in \rref{fig:dL} as opposed to an infinite collection of axioms from a finite list of axiom schemata along with schema variables, side conditions, and implicit instantiation rules.
\begin{figure}[tbhp]
  \renewcommand*{\irrulename}[1]{\text{#1}}%
    \iflongversion\else\linferenceRulevskipamount=0.4em\fi
  \begin{calculuscollections}{\columnwidth}
    \begin{calculus}
      \cinferenceRule[diamond|$\didia{\cdot}$]{diamond axiom}
      {\linferenceRule[equiv]
        {\lnot\dbox{a}{\lnot p(\usall)}}
        {\ddiamond{a}{p(\usall)}}
      }
      {}
      \cinferenceRule[assignb|$\dibox{:=}$]{assignment / substitution axiom}
      {\linferenceRule[equiv]
        {p(f)}
        {\dbox{\pupdate{\umod{x}{f}}}{p(x)}}
      }
      {}%
      \cinferenceRule[testb|$\dibox{?}$]{test}
      {\linferenceRule[equiv]
        {(q \limply p)}
        {\dbox{\ptest{q}}{p}}
      }{}
      \cinferenceRule[choiceb|$\dibox{\cup}$]{axiom of nondeterministic choice}
      {\linferenceRule[equiv]
        {\dbox{a}{p(\usall)} \land \dbox{b}{p(\usall)}}
        {\dbox{\pchoice{a}{b}}{p(\usall)}}
      }{}
      \cinferenceRule[composeb|$\dibox{{;}}$]{composition} %
      {\linferenceRule[equiv]
        {\dbox{a}{\dbox{b}{p(\usall)}}}
        {\dbox{a;b}{p(\usall)}}
      }{}
      \cinferenceRule[iterateb|$\dibox{{}^*}$]{iteration/repeat unwind} %
      {\linferenceRule[equiv]
        {p(\usall) \land \dbox{a}{\dbox{\prepeat{a}}{p(\usall)}}}
        {\dbox{\prepeat{a}}{p(\usall)}}
      }{}
      \cinferenceRule[K|K]{K axiom / modal modus ponens} %
      {\linferenceRule[impl]
        {\dbox{a}{(p(\usall)\limply q(\usall))}}
        {(\dbox{a}{p(\usall)}\limply\dbox{a}{q(\usall)})}
      }{}
      \cinferenceRule[I|I]{loop induction}
      {\linferenceRule[impl]
        {\dbox{\prepeat{a}}{(p(\usall)\limply\dbox{a}{p(\usall)})}}
        {(p(\usall)\limply\dbox{\prepeat{a}}{p(\usall)})}
      }{}
      \cinferenceRule[V|V]{vacuous $\dbox{}{}$}
      {\linferenceRule[impl]
        {p}
        {\dbox{a}{p}}
      }{}%
    \end{calculus}
    \qquad
    \begin{calculus}
      \cinferenceRule[G|G]{$\dbox{}{}$ generalisation} %
      {\linferenceRule[formula]
        {p(\usall)}
        {\dbox{a}{p(\usall)}}
      }{}
      \cinferenceRule[gena|$\forall{}$]{$\forall{}$ generalisation}
      {\linferenceRule[formula]
        {p(x)}
        {\lforall{x}{p(x)}}
      }{}%
      \cinferenceRule[MP|MP]{modus ponens}
      {\linferenceRule[formula]
        {p\limply q \quad p}
        {q}
      }{}%
      \cinferenceRule[CT|CT]{congterm congruence on terms}
      {\linferenceRule[formula]
        {f(\usall) = g(\usall)}
        {c(f(\usall)) = c(g(\usall))}
      }{}%
      \cinferenceRule[CQ|CQ]{congequal congruence of equations on formulas (convert term congruence to formula congruence: term congruence on formulas)}
      {\linferenceRule[formula]
        {f(\usall) = g(\usall)}
        {p(f(\usall)) \lbisubjunct p(g(\usall))}
      }{}%
      \cinferenceRule[CE|CE]{congequiv congruence of equivalences on formulas}
      {\linferenceRule[formula]
        {p(\usall) \lbisubjunct q(\usall)}
        {\contextapp{C}{p(\usall)} \lbisubjunct \contextapp{C}{q(\usall)}}
      }{}%
      \iflongversion
      \cinferenceRule[USagain|US]{uniform substitution}
      {\linferenceRule[formula]
        {\phi}
        {\applyusubst{\sigma}{\phi}}
      }{}%
      \fi
    \end{calculus}%
  \end{calculuscollections}
  \caption{Differential dynamic logic axioms and proof rules}
  \label{fig:dL}
\end{figure}
Soundness of the axioms in \rref{fig:dL} follows from the soundness of corresponding axiom schemata \cite{DBLP:conf/lics/Platzer12b}, but would be easier to prove standalone, because it is a finite list of formulas without the need to prove soundness for all their instantiations.
The rules in \rref{fig:dL} are \emph{axiomatic rules}, i.e.\ pairs of concrete formulas instantiated by \irref{US}.
Further, $\usall$ is the vector of all relevant variables, which is finite-dimensional, or, in practice, considered as a built-in vectorial term.
Proofs in the uniform substitution \dL calculus use \irref{US} (and bound renaming such as \(\lforall{x}{p(x)}\lbisubjunct\lforall{y}{p(y)}\)) to instantiate the axioms from \rref{fig:dL} to the required form.
\irref{CT+CQ+CE} are congruence rules, which are included for efficiency to use axioms in any context even if not needed for completeness.

\paragraph{Real Quantifiers.}

Besides (decidable) real arithmetic (whose use is denoted \irref{qear}), complete axioms for first-order logic can be adopted to express
universal instantiation \irref{allinst}\iflongversion (if $p$ is true of all $x$ it is also true of constant symbol $f$)\fi,
distributivity \irref{alldist},
and vacuous quantification \irref{vacuousall}\iflongversion (predicate $p$ of arity zero does not depend on $x$)\fi.

  \begin{calculuscollections}{\columnwidth}
    \iflongversion\else\linferenceRulevskipamount=0.2em\fi
    \begin{calculus}
      \cinferenceRule[allinst|$\forall$i]{universal instantiation}
       {(\lforall{x}{p(x)}) \limply p(f)}
       {}
       \cinferenceRule[alldist|$\forall{\limply}$]{$\forall$ distributes over $\limply$}
       {\lforall{x}{(p(x)\limply q(x))} \limply (\lforall{x}{p(x)} \limply \lforall{x}{q(x)})}
       {}
       \cinferenceRule[vacuousall|V$_\forall$]{vacuous universal quantifier}
       {p \limply \lforall{x}{p}}
       {}%
    \end{calculus}
  \end{calculuscollections}
\vspace{-\baselineskip}

\paragraph{The Significance of Clashes.}

\iflongversion
This section illustrates how soundness-critical it is for \irref{US} to produce substitution clashes by showing unsound na\"ive proof attempts that \irref{US} prevents successfully.
\fi
\irref{US} clashes for substitutions that introduce a free variable into a bound context.
Even an occurrence of $p(x)$ in a context where $x$ is bound does not allow mentioning $x$ in the replacement except in the $\usarg$ places:
\[
\linfer[clash]
{\dbox{\pupdate{\umod{x}{f}}}{p(x)} \lbisubjunct p(f)}
{\dbox{\pupdate{\umod{x}{x+1}}}{x\neq x} \lbisubjunct x+1\neq x}
\qquad
\sigma=\usubstlist{\usubstmod{f}{x+1},\usubstmod{p(\usarg)}{(\usarg\neq x)}}
\]
\irref{US} can directly handle even nontrivial binding structures, though, e.g. from \irref{assignb}
with the substitution \(\sigma=\usubstlist{\usubstmod{f}{x^2},\usubstmod{p(\usarg)}{\dbox{\prepeat{(z:=\usarg+z)};z:=\usarg+yz}{y\geq\usarg}}}\):
\[
\linfer[US]
{\dbox{\pupdate{\umod{x}{f}}}{p(x)} \lbisubjunct p(f)}
{\dbox{x:=x^2}{\dbox{\prepeat{(z:=x{+}z)};z:=x{+}yz}{y{\geq}x}} \lbisubjunct
\dbox{\prepeat{(z:=x^2{+}z)};z:=x^2{+}yz}{y{\geq}x^2}
}
\]

\iflongversion
Similarly from \irref{assignb}
with \(\usubstlist{\usubstmod{f}{x^2},\usubstmod{p(\usarg)}{\dbox{(\pchoice{y:=y+1}{\prepeat{z:=\usarg+z}});z:=\usarg+yz}{y>\usarg}}}\):
{\footnotesize
\[
\hspace{-1cm}
\linfer[US]
{\dbox{\pupdate{\umod{x}{f}}}{p(x)} \lbisubjunct p(f)}
{\dbox{x:=x^2}{\dbox{(\pchoice{y:=y{+}1}{\prepeat{z:=x{+}z}});z:=x{+}yz}{y{>}x}} \lbisubjunct
\dbox{(\pchoice{y:=y{+}1}{\prepeat{z:=x^2{+}z}});z:=x^2{+}yz}{y{>}x^2}
}
\]
}%
It is soundness-critical that \irref{US} clashes when trying to instantiate $p$ in \irref{vacuousall} with a formula that mentions the bound variable $x$:
\[
\linfer[clash]
{p \limply \lforall{x}{p}}
{x\geq0 \limply \lforall{x}{x\geq0}}
\qquad
\usubstlist{\usubstmod{p}{x\geq0}}
\]
It is soundness-critical that \irref{US} clashes when substituting $p$ in vacuous program axiom \irref{V} with a formula with a free occurrence of a variable bound by $a$:
\[
\linfer[clash]
{p \limply \dbox{a}{p}}
{x\geq0 \limply \dbox{x:=x-1}{x\geq0}}
\qquad
\usubstlist{\usubstmod{a}{x:=x-1},\usubstmod{p}{x\geq0}}
\]
G\"odel's generalization rule \irref{G} uses $p(\usall)$ instead of $p$ from \irref{V}, so allows the proof:
\[
\linfer[US]
{(-x)^2\geq0}
{\dbox{x:=x-1}{(-x)^2\geq0}}
\]

\noindent
Let \(\usall=(x,y)\), 
\(\usubstlist{\usubstmod{a}{x:=x+1},\usubstmod{b}{x:=0;y:=0},\usubstmod{p(\usall)}{x\geq y}}\), \irref{US} derives:%
\vspace{-\baselineskip}
\begin{sequentdeduction}
  \linfer[US]
  {\linfer[choiceb]
    {\lclose}
    {\dbox{\pchoice{a}{b}}{p(\usall)} \lbisubjunct \dbox{a}{p(\usall)} \land \dbox{b}{p(\usall)}}
  }
  {\dbox{\pchoice{x:=x+1}{(x:=0;y:=0)}}{x\geq y} \lbisubjunct \dbox{x:=x+1}{x\geq0} \land \dbox{x:=0;y:=0}{x\geq y}}
\end{sequentdeduction}

\noindent
With \(\usall=(x,y)\) and
\(\usubstlist{\usubstmod{a}{\pchoice{x:=x+1}{y:=0}},\usubstmod{b}{y:=y+1},\usubstmod{p(\usall)}{x\geq y}}\) \irref{US} derives:
\vspace{-\baselineskip}
\begin{sequentdeduction}
\hspace{-0.5cm}
  \linfer[US]
  {\linfer[composeb]
    {\lclose}
    {\dbox{a;b}{p(\usall)} \lbisubjunct \dbox{a}{\dbox{b}{p(\usall)}}}
  }
  {\dbox{(\pchoice{x:=x+1}{y:=0});y:=y+1}{x\geq y} \lbisubjunct \dbox{\pchoice{x:=x+1}{y:=0}}{\dbox{y:=y+1}{x\geq y}}}
\end{sequentdeduction}

Not all axioms fit to the uniform substitution framework.
The Barcan axiom was used in a completeness proof for the Hilbert-type calculus for differential dynamic logic \cite{DBLP:conf/lics/Platzer12b} (but not in the completeness proof for its sequent calculus \cite{DBLP:journals/jar/Platzer08}):
\[
      \cinferenceRule[B|B]{Barcan$\dbox{}{}\forall{}$} %
      {\linferenceRule[impl]
        {\lforall{x}{\dbox{\alpha}{p(x)}}}
        {\dbox{\alpha}{\lforall{x}{p(x)}}}
      }{\m{x\not\in\alpha}}
\]
\irref{B} is unsound without the restriction \(x\not\in\alpha\), though, so that the following would be an unsound axiom:
\begin{equation}
{\lforall{x}{\dbox{a}{p(x)}}\limply{\dbox{a}{\lforall{x}{p(x)}}}}
\label{eq:unsound-B-attempt}
\end{equation}
because $x\not\in a$ cannot be enforced for program constants, since their effect might very well depend on the value of $x$ or since they might write to $x$.
In \rref{eq:unsound-B-attempt}, $x$ cannot be written by $a$ without violating soundness:
\[
\linfer[unsound]
  {\lforall{x}{\dbox{a}{p(x)}}\limply{\dbox{a}{\lforall{x}{p(x)}}}}
  {\lforall{x}{\dbox{x:=0}{x\geq0}}\limply{\dbox{x:=0}{\lforall{x}{x\geq0}}}}
\qquad
\usubstlist{\usubstmod{a}{x:=0},\usubstmod{p(\usarg)}{\usarg\geq0}}
\]
nor can $x$ be read by $a$ in \rref{eq:unsound-B-attempt} without violating soundness:
\[
\linfer[unsound]
  {\lforall{x}{\dbox{a}{p(x)}}\limply{\dbox{a}{\lforall{x}{p(x)}}}}
  {\lforall{x}{\dbox{\ptest{y=x^2}}{y=x^2}}\limply{\dbox{\ptest{y=x^2}}{\lforall{x}{y=x^2}}}}
\qquad
\usubstlist{\usubstmod{a}{\ptest{y=x^2}},\usubstmod{p(\usarg)}{y=\usarg^2}}
\]

Thus, the completeness proof for differential dynamic logic from prior work \cite{DBLP:conf/lics/Platzer12b} does not directly carry over.
A more general completeness result for differential game logic \cite{DBLP:journals/corr/Platzer14:dGL} implies, however, that \irref{B} is unnecessary for completeness.
\fi

\section{Differential Equations and Differential Axioms} \label{sec:differential}

\rref{sec:dL-axioms} leverages the first-order features of \dL and \irref{US} to obtain a finite list of axioms without side-conditions.
They lack axioms for differential equations, though.
Classical calculi for \dL have axioms for replacing differential equations with a quantifier for time $t\geq0$ and an assignment for their solutions $\solf(t)$ \cite{DBLP:journals/jar/Platzer08,DBLP:conf/lics/Platzer12b}.
Besides being limited to simple differential equations, such axioms have the inherent side-condition ``if $\solf(t)$ is a solution of the differential equation \(\pevolve{\D{x}=\genDE{x}}\) with symbolic initial value $x$''.
Such a side-condition is more difficult than occurrence and read/write conditions, but equally soundness-critical.
This section leverages \irref{US} and the new differential forms in \dL to obtain a logically internalized version of differential invariants and related proof rules for differential equations \cite{DBLP:journals/logcom/Platzer10,DBLP:journals/lmcs/Platzer12} as axioms (without schema variables and free of side-conditions).
These axioms can prove properties of more general ``unsolvable'' differential equations. They can also prove all properties of differential equations that can be proved with solutions \cite{DBLP:journals/lmcs/Platzer12} while guaranteeing correctness of the solution as part of the proof.

\subsection{Differentials: Invariants, Cuts, Effects, and Ghosts} \label{sec:diffind}

Figure~\ref{fig:dL-ODE} shows differential equation axioms for differential weakening (\irref{DW}), differential cuts (\irref{DC}), differential effect (\irref{DE}), differential invariants (\irref{DI}) \cite{DBLP:journals/logcom/Platzer10}, differential ghosts (\irref{DG}) \cite{DBLP:journals/lmcs/Platzer12}, solutions (\irref{DS}), differential substitutions (\irref{Dassignb}), and differential axioms (\irref{Dplus+Dtimes+Dcompose}).
Axioms identifying \(\der{x}=\D{x}\) for variables $x\in\allvars$ and \(\der{f}=0\) for functions $f$ and number literals of arity 0 are used implicitly.
Some axioms use reverse implications \((\phi\lylpmi\psi)\mequiv(\psi\limply\phi)\) for emphasis.%
\begin{figure}[tbhp]
  \begin{calculuscollections}{\columnwidth}
  \renewcommand*{\irrulename}[1]{\text{#1}}%
    \iflongversion\else\linferenceRulevskipamount=0.4em\fi
    \begin{calculus}
      \cinferenceRule[DW|DW]{differential evolution domain} %
      {\dbox{\pevolvein{\D{x}=f(x)}{q(x)}}{q(x)}}
      {}
      \cinferenceRule[DC|DC]{differential cut} %
      {\linferenceRule[lpmi]
        {\big(\dbox{\pevolvein{\D{x}=f(x)}{q(x)}}{p(x)} \lbisubjunct \dbox{\pevolvein{\D{x}=f(x)}{q(x)\land r(x)}}{p(x)}\big)}
        {\dbox{\pevolvein{\D{x}=f(x)}{q(x)}}{r(x)}}
      }
      {}%
      \cinferenceRule[DE|DE]{differential effect} %
      {\linferenceRule[viuqe]
        {\dbox{\pevolvein{\D{x}=f(x)}{q(x)}}{p(x,\D{x})}}
        {\dbox{\pevolvein{\D{x}=f(x)}{q(x)}}{\dbox{\Dupdate{\Dumod{\D{x}}{f(x)}}}{p(x,\D{x})}}}
      }
      {}%
      \cinferenceRule[DI|DI]{differential induction} %
      {\linferenceRule[lpmi]
        {\dbox{\pevolvein{\D{x}=f(x)}{q(x)}}{p(x)}}
        {\big(q(x)\limply p(x)\land\dbox{\pevolvein{\D{x}=f(x)}{q(x)}}{\der{p(x)}}}\big)
      }
      {}%
      \cinferenceRule[DG|DG]{differential ghost variables} %
      {\linferenceRule[viuqe]
        {\dbox{\pevolvein{\D{x}=f(x)}{q(x)}}{p(x)}}
        {\lexists{y}{\dbox{\pevolvein{\D{x}=f(x)\syssep\D{y}=a(x)y+b(x)}{q(x)}}{p(x)}}}
      }
      {}%
      \cinferenceRule[DS|DS]{(constant) differential equation solution} %
      {\linferenceRule[viuqe]
        {\dbox{\pevolvein{\D{x}=f}{q(x)}}{p(x)}}
        {\lforall{t{\geq}0}{\big((\lforall{0{\leq}s{\leq}t}{q(x+f\itimes s)}) \limply \dbox{\pupdate{\pumod{x}{x+f\itimes t}}}{p(x)}\big)}}
      }
      {}%
      \cinferenceRule[Dassignb|$\dibox{':=}$]{differential assignment}
      {\linferenceRule[equiv]
        {p(f)}
        {\dbox{\Dupdate{\Dumod{\D{x}}{f}}}{p(\D{x})}}
      }
      {}%
      \cinferenceRule[Dplus|$+'$]{derive sum}
      {\linferenceRule[eq]
        {\der{f(\usall)}+\der{g(\usall)}}
        {\der{f(\usall)+g(\usall)}}
      }
      {}
      \cinferenceRule[Dtimes|$\cdot'$]{derive product}
      {\linferenceRule[eq]
        {\der{f(\usall)}\cdot g(\usall)+f(\usall)\cdot\der{g(\usall)}}
        {\der{f(\usall)\cdot g(\usall)}}
      }
      {}
      \cinferenceRule[Dcompose|$\compose'$]{derive composition}
      {
        \dbox{\pupdate{\pumod{y}{g(x)}}}{\dbox{\Dupdate{\Dumod{\D{y}}{1}}}
        {\big( \der{f(g(x))} = \der{f(y)}\stimes\der{g(x)}\big)}}
      }
      {}%
   \end{calculus}%
\end{calculuscollections}%
  \caption{Differential equation axioms and differential axioms}
  \label{fig:dL-ODE}
\end{figure}

\emph{Differential weakening} axiom \irref{DW} internalizes that differential equations can never leave their evolution domain $q(x)$.
\irref{DW} implies\footnote{%
\(\dbox{\pevolvein{\D{x}=f(x)}{q(x)}}{(q(x)\limply p(x))} \limply \dbox{\pevolvein{\D{x}=f(x)}{q(x)}}{p(x)}\) derives by \irref{K} from \irref{DW}.
The converse
\(\dbox{\pevolvein{\D{x}=f(x)}{q(x)}}{p(x)} \limply \dbox{\pevolvein{\D{x}=f(x)}{q(x)}}{(q(x)\limply p(x))}\)
derives by \irref{K} since \irref{G} derives \(\dbox{\pevolvein{\D{x}=f(x)}{q(x)}}{\big(p(x)\limply(q(x)\limply p(x))\big)}\).
}
\({\dbox{\pevolvein{\D{x}=f(x)}{q(x)}}{p(x)}} \lbisubjunct
        {\dbox{\pevolvein{\D{x}=f(x)}{q(x)}}{(q(x)\limply p(x))}} \irlabel{diffweaken|DW}\)
also called \irref{diffweaken},
whose (right) assumption is best proved by \irref{G}\iflongversion yielding premise \(q(x)\limply p(x)\)\fi.
The \emph{differential cut} axiom \irref{DC} is a cut for differential equations.
It internalizes that differential equations staying in $r(x)$ stay in $p(x)$ iff $p(x)$ always holds after the differential equation that is restricted to the smaller evolution domain \(\pevolvein{}{q(x)\land r(x)}\).
\irref{DC} is a differential variant of modal modus ponens \irref{K}.

\emph{Differential effect} axiom \irref{DE} internalizes that the effect on differential symbols along a differential equation is a differential assignment assigning the right-hand side $f(x)$ to the left-hand side $\D{x}$.
Axiom \irref{DI} internalizes \emph{differential invariants},
i.e.\ that a differential equation stays in $p(x)$ if it starts in $p(x)$ and if its differential $\der{p(x)}$ always holds after the differential equation \(\pevolvein{\D{x}=f(x)}{q(x)}\).
The differential equation also vacuously stays in $p(x)$ if it starts outside $q(x)$, since it is stuck then.
The (right) assumption of \irref{DI} is best proved by \irref{DE} to select the appropriate vector field \(\D{x}=f(x)\) for the differential $\der{p(x)}$
and a subsequent \irref{diffweaken+G} to make the evolution domain constraint $q(x)$ available as an assumption.
For simplicity, this paper focuses on atomic postconditions for which \(\der{\theta\geq\eta} \mequiv \der{\theta>\eta} \mequiv \der{\theta}\geq\der{\eta}\)
and \(\der{\theta=\eta} \mequiv \der{\theta\neq\eta} \mequiv \der{\theta}=\der{\eta}\), etc.
Axiom \irref{DG} internalizes \emph{differential ghosts},
i.e. that additional differential equations can be added if their solution exists long enough.
Axiom \irref{DS} solves differential equations with the help of \irref{DG+DC}.
Vectorial generalizations to systems of differential equations are possible for the axioms in \rref{fig:dL-ODE}.

The following proof proves a property of a differential equation using differential invariants without having to solve that differential equation.
One use of \irref{US} is shown explicitly, other uses of \irref{US} are similar for \irref{DI+DE+Dassignb} instances.
\vspace{-\baselineskip}
{\footnotesize
\renewcommand{\linferPremissSeparation}{~}%
\let\orgcdot\cdot%
\def\cdot{{\orgcdot}}%
\begin{sequentdeduction}[array]
\linfer[DI]
{\linfer[DE]
 {\linfer[CE]%
  {\linfer[G]
    {\linfer[Dassignb]
      {\linfer[qear]
        {\lclose}
        {\lsequent{}{x^3\cdot x + x\cdot x^3\geq0}}
      }%
      {\lsequent{}{\dbox{\Dupdate{\Dumod{\D{x}}{x^3}}}{\D{x}\cdot x+x\cdot\D{x}\geq0}}}
    }%
    {\lsequent{}{\dbox{\pevolve{\D{x}=x^3}}{\dbox{\Dupdate{\Dumod{\D{x}}{x^3}}}{\D{x}\cdot x+x\cdot\D{x}\geq0}}}}
    !
    \linfer%
    {\linfer[CQ] %
      {\linfer%
        {\linfer[US]
          {\linfer[Dtimes]
            {\lclose}
            {\lseqalign{\der{f(\usall)\cdot g(\usall)}} {= \der{f(\usall)}\cdot g(\usall) + f(\usall)\cdot\der{g(\usall)}}}
          }
        {\lseqalign{\der{x\cdot x}} {= \der{x}\cdot x + x\cdot\der{x}}}
      }
      {\lseqalign{\der{x\cdot x}} {= \D{x}\cdot x + x\cdot\D{x}}}
      }
      {\lseqalign{\der{x\cdot x}\geq0}{\lbisubjunct\D{x}\cdot x+x\cdot\D{x}\geq0}}
    }
    {\lseqalign{\der{x\cdot x\geq1}}{\lbisubjunct\D{x}\cdot x+x\cdot\D{x}\geq0}}
  }%
  {\lsequent{}{\dbox{\pevolve{\D{x}=x^3}}{\dbox{\Dupdate{\Dumod{\D{x}}{x^3}}}{\der{x\cdot x\geq1}}}}}
 }%
  {\lsequent{}{\dbox{\pevolve{\D{x}=x^3}}{\der{x\cdot x\geq1}}}}
}%
{\lsequent{x\cdot x\geq1} {\dbox{\pevolve{\D{x}=x^3}}{x\cdot x\geq1}}}
\end{sequentdeduction}
}%
Previous calculi \cite{DBLP:journals/logcom/Platzer10,DBLP:journals/lmcs/Platzer12} collapse this proof into a single proof step with complicated built-in operator implementations that silently perform the same reasoning in an opaque way.
The approach presented here combines separate axioms to achieve the same effect in a modular way, because they have individual responsibilities internalizing separate logical reasoning principles in \emph{differential-form} \dL.
\iflongversion Tactics combining the axioms as indicated make the axiomatic way equally convenient.
Clever cuts or \irref{MP} enable proofs in which the main argument remains as fast \cite{DBLP:journals/logcom/Platzer10,DBLP:journals/lmcs/Platzer12} while the additional premises subsequently check soundness.
Both \irref{CQ} and also \irref{CE} simplify the proof substantially but are not necessary:
\vspace{-\baselineskip}
{\footnotesize
\begin{sequentdeduction}%
\hspace*{-0.7cm}
\linfer[MP]
  {\linfer
    {\lclose}
    {\sdots\limply(\der{x\cdot x}\geq0 \lbisubjunct \D{x}\cdot x + x\cdot\D{x}\geq0)}
  &\linfer%
    {\linfer[US]
      {\linfer[Dtimes]
        {\lclose}
        {\der{f(\usall)\cdot g(\usall)} = \der{f(\usall)}\cdot g(\usall) + f(\usall)\cdot\der{g(\usall)}}
      }
      {\der{x\cdot x} = \der{x}\cdot x + x\cdot\der{x}}
    }
    {\der{x\cdot x} = \D{x}\cdot x + x\cdot\D{x}}
  }
  {\linfer[G]
    {\der{x\cdot x}\geq0 \lbisubjunct \D{x}\cdot x + x\cdot\D{x}\geq0}
    {\linfer[K]
      {\dbox{\Dupdate{\Dumod{\D{x}}{x^3}}}{(\der{x\cdot x}\geq0 \lbisubjunct \D{x}\cdot x + x\cdot\D{x}\geq0)}}
      {\dbox{\Dupdate{\Dumod{\D{x}}{x^3}}}{\der{x\cdot x}\geq0} \lbisubjunct \dbox{\Dupdate{\Dumod{\D{x}}{x^3}}}{\D{x}\cdot x + x\cdot\D{x}\geq0}}
    }
}
\end{sequentdeduction}
\vspace{-\baselineskip}
\begin{sequentdeduction}[array]
\linfer[DI]
{\linfer[DE]
 {\linfer[G]
  {\linfer[MP]
    {\text{use proof above}
    !\linfer[Dassignb]
      {\linfer[qear]
        {\lclose}
        {\lsequent{}{x^3\cdot x + x\cdot x^3\geq0}}
      }%
      {\lsequent{}{\dbox{\Dupdate{\Dumod{\D{x}}{x^3}}}{\D{x}\cdot x + x\cdot\D{x}\geq0}}}
    }%
    {\lsequent{}{\dbox{\Dupdate{\Dumod{\D{x}}{x^3}}}{\der{x\cdot x}\geq0}}}
  }%
  {\lsequent{}{\dbox{\pevolve{\D{x}=x^3}}{\dbox{\Dupdate{\Dumod{\D{x}}{x^3}}}{\der{x\cdot x}\geq0}}}}
 }%
  {\lsequent{}{\dbox{\pevolve{\D{x}=x^3}}{\der{x\cdot x}\geq0}}}
}%
{\lsequent{x\cdot x\geq1} {\dbox{\pevolve{\D{x}=x^3}}{x\cdot x\geq1}}}
\end{sequentdeduction}
}%
The proof uses (implicit) cuts with equivalences predicting the outcome of the right branch, which is simple but inconvenient.
\newcommand{\mydiffcond}[1][x,\D{x}]{j(#1)}%
A constructive direct proof uses a free function symbol $\mydiffcond$, instead, which is ultimately instantiated by \irref{US} as in \rref{thm:dL-sound}.

The same technique is helpful for invariant search, in which case a free predicate symbol $p(\usall)$ is used and instantiated by \irref{US} lazily when the proof closes.
{\footnotesize
\begin{sequentdeduction}[array]
\linfer[DI]
{\linfer[DE]
 {\linfer[CE]%
  {\linfer[G]
    {\linfer[Dassignb]
      {\linfer%
        {\linfer[qear]
          {\lclose}
          {\lsequent{}{x^3\cdot x + x\cdot x^3\geq0}}
        }
        {\lsequent{}\mydiffcond[x,x^3]\geq0}
      }%
      {\lsequent{}{\dbox{\Dupdate{\Dumod{\D{x}}{x^3}}}{\mydiffcond\geq0}}}
    }%
    {\lsequent{}{\dbox{\pevolve{\D{x}=x^3}}{\dbox{\Dupdate{\Dumod{\D{x}}{x^3}}}{\mydiffcond\geq0}}}}
    !
    \linfer%
    {\linfer[CQ] %
      {\linfer%
        {\linfer%
        {\linfer[US]
          {\linfer[Dtimes]
            {\lclose}
            {\lseqalign{\der{f(\usall)\cdot g(\usall)}} {= \der{f(\usall)}\cdot g(\usall) + f(\usall)\cdot\der{g(\usall)}}}
          }
        {\lseqalign{\der{x\cdot x}} {= \der{x}\cdot x + x\cdot\der{x}}}
        }
      {\lseqalign{\der{x\cdot x}} {= \D{x} \cdot x + x\cdot\D{x}}}
      }
      {\lseqalign{\der{x\cdot x}} {= \mydiffcond}}
      }
      {\lseqalign{\der{x\cdot x}\geq0}{\lbisubjunct\mydiffcond\geq0}}
    }
    {\lseqalign{\der{x\cdot x\geq1}}{\lbisubjunct\mydiffcond\geq0}}
  }%
  {\lsequent{}{\dbox{\pevolve{\D{x}=x^3}}{\dbox{\Dupdate{\Dumod{\D{x}}{x^3}}}{\der{x\cdot x\geq1}}}}}
 }%
  {\lsequent{}{\dbox{\pevolve{\D{x}=x^3}}{\der{x\cdot x\geq1}}}}
}%
{\lsequent{x\cdot x\geq1} {\dbox{\pevolve{\D{x}=x^3}}{x\cdot x\geq1}}}
\end{sequentdeduction}
}%
\fi

\iflongversion
Proofs based entirely on equivalences for solving differential equations involve \irref{DG} for introducing a time variable, \irref{DC} to cut the solutions in, \irref{DW} to export the solution to the postcondition, inverse \irref{DC} to remove the evolution domain constraints again, inverse \irref{DG} to remove the original differential equations, and finally \irref{DS} to solve the differential equation for time:
\def\prem{\phi}%
{\footnotesize
\begin{sequentdeduction}[array]
\linfer[DG]
 {\linfer%
   {\linfer[DC]
     {\linfer[DC]
       {\linfer[diffweaken]
         {\linfer[G+K]%
           {\linfer[DC]
             {\linfer[DC]
               {\linfer[DG]
                 {\linfer[DG]
                   {\linfer[DS]
                     {\linfer[assignb]
                       {\linfer[qear]
                         {\lclose}
                         {\lsequent{\prem} {\lforall{s{\geq}0}{(x_0+\frac{a}{2}s^2+v_0s\geq0)}}}
                       }%
                       {\lsequent{\prem} {\lforall{s{\geq}0}{\dbox{\pupdate{\pumod{t}{0+1s}}}{x_0+\frac{a}{2}t^2+v_0t\geq0}}}}
                     }%
                     {\lsequent{\prem} {\dbox{\pevolve{\D{t}=1}}{x_0+\frac{a}{2}t^2+v_0t\geq0}}}
                   }%
                   {\lsequent{\prem} {\dbox{\pevolve{\D{v}=a\syssep\D{t}=1}}{x_0+\frac{a}{2}t^2+v_0t\geq0}}}
                 }%
                 {\lsequent{\prem} {\dbox{\pevolve{\D{x}=v\syssep\D{v}=a\syssep\D{t}=1}}{x_0+\frac{a}{2}t^2+v_0t\geq0}}}
               }%
               {\lsequent{\prem} {\dbox{\pevolvein{\D{x}=v\syssep\D{v}=a\syssep\D{t}=1}{v=v_0+at}}{x_0+\frac{a}{2}t^2+v_0t\geq0}}}
             }%
             {\lsequent{\prem} {\dbox{\pevolvein{\D{x}=v\syssep\D{v}=a\syssep\D{t}=1}{v=v_0+at\land x=x_0+\frac{a}{2}t^2+v_0t}}{x_0+\frac{a}{2}t^2+v_0t\geq0}}} 
           }%
           {\lsequent{\prem} {\dbox{\pevolvein{\D{x}=v\syssep\D{v}=a\syssep\D{t}=1}{v=v_0+at\land x=x_0+\frac{a}{2}t^2+v_0t}}{(x{=}x_0{+}\frac{a}{2}t^2{+}v_0t\limply x{\geq}0)}}}
         }%
         {\lsequent{\prem} {\dbox{\pevolvein{\D{x}=v\syssep\D{v}=a\syssep\D{t}=1}{v=v_0+at\land x=x_0+\frac{a}{2}t^2+v_0t}}{x\geq0}}} 
       }%
       {\lsequent{\prem} {\dbox{\pevolvein{\D{x}=v\syssep\D{v}=a\syssep\D{t}=1}{v=v_0+at}}{x\geq0}}} 
     }%
     {\lsequent{\prem} {\dbox{\pevolve{\D{x}=v\syssep\D{v}=a\syssep\D{t}=1}}{x\geq0}}} 
   }%
   {\lsequent{\prem} {\lexists{t}{\dbox{\pevolve{\D{x}=v\syssep\D{v}=a\syssep\D{t}=1}}{x\geq0}}}}
 }%
 {\lsequent{\prem} {\dbox{\pevolve{\D{x}=v\syssep\D{v}=a}}{x\geq0}}} 
\end{sequentdeduction}
}%
where $\prem$ is \({a\geq0\land v=v_0\geq0 \land x=x_0\geq0}\). %
The existential quantifier for $t$ is instantiated by $0$, leading to $\dbox{\pupdate{\pumod{t}{0}}}{}$ (suppressed in the proof for readability reasons).
The 4 uses of \irref{DC} lead to 2 additional premises proving that \(v=v_0+at\) and then \(x=x_0+\frac{a}{2}t^2+v_0t\) are differential invariants (using \irref{DI+DE+diffweaken}).
Shortcuts using \irref{diffweaken} are possible but the above proof generalize to $\ddiamond{}{}$ because it is an equivalence proof.
The additional premise for \irref{DC} with \(v=v_0+at\) proves as follows:
{\footnotesize
\begin{sequentdeduction}[array]
\linfer[DI]
{\linfer[DE]
 {\linfer[G]
   {\linfer[CE]%
    {\linfer[Dassignb]
      {\linfer[qear]
        {\lclose}
        {\lsequent{}{a=0+a\cdot1}}
      }%
      {\lsequent{} {\dbox{\Dupdate{\Dumod{\D{v}}{a}}}{\dbox{\Dupdate{\Dumod{\D{t}}{1}}}{\D{v}=0+a\D{t}}}}}
    !
    \linfer%
    {\linfer[CQ] %
      {\linfer[Dtimes]%
        {\linfer[US]
          {\linfer[Dplus]
            {\lclose}
            {\lseqalign{\der{f(\usall)+ g(\usall)}} {= \der{f(\usall)} + \der{g(\usall)}}}
          }
        {\lseqalign{\der{v_0+at}} {= \der{v_0}+\der{a t}}}
      }
      {\lseqalign{\der{v_0+at}} {= 0+a(\D{t})}}
      }
      {\lseqalign{\D{v}=\der{v_0+at}}{\lbisubjunct\D{v}=0+a\D{t}}}
    }
    {\lseqalign{\der{v=v_0+at}}{\lbisubjunct\D{v}=0+a\D{t}}}
    }%
    {\lsequent{} {\dbox{\Dupdate{\Dumod{\D{v}}{a}}}{\dbox{\Dupdate{\Dumod{\D{t}}{1}}}{\der{v=v_0+at}}}}}
  }%
  {\lsequent{} {\dbox{\pevolve{\D{x}=v\syssep\D{v}=a\syssep\D{t}=1}}{\dbox{\Dupdate{\Dumod{\D{v}}{a}}}{\dbox{\Dupdate{\Dumod{\D{t}}{1}}}{\der{v=v_0+at}}}}}}
 }%
  {\lsequent{} {\dbox{\pevolve{\D{x}=v\syssep\D{v}=a\syssep\D{t}=1}}{\der{v=v_0+at}}}}
}%
{\lsequent{\prem} {\dbox{\pevolve{\D{x}=v\syssep\D{v}=a\syssep\D{t}=1}}{v=v_0+at}}}
\end{sequentdeduction}
}
The additional premise for \irref{DC} with \(x=x_0+\frac{a}{2}t^2+v_0t\) proves as follows:

{\tiny%
\begin{sequentdeduction}[array]
\linfer[DI]
{\linfer[DE]
 {\linfer[diffweaken]
 {\linfer[G]
   {\linfer[CE]%
    {\linfer[Dassignb]
      {\linfer[qear]
        {\lclose}
        {\lsequent{}{{v=v_0+at} \limply v=at\cdot1+v_0\cdot1}}
      }%
      {\lsequent{} {{v=v_0+at} \limply \dbox{\Dupdate{\Dumod{\D{x}}{v}}}{\dbox{\Dupdate{\Dumod{\D{t}}{1}}}{\D{x}=at\D{t}+v_0\D{t}}}}}
    !
    \linfer%
    {\linfer[CQ] %
      {\linfer[Dplus+Dtimes]%
        {\linfer%
          {\lclose}
          {\lseqalign{2\frac{a}{2}t\D{t}+v_0\D{t}} {= at\D{t}+v_o\D{t}}}
        }
      {\lseqalign{\der{x_0+\frac{a}{2}t^2+v_0t}} {= at\D{t}+v_o\D{t}}}
      }
      {\lseqalign{\D{x}=\der{x_0+\frac{a}{2}t^2+v_0t}}{\lbisubjunct\D{x}=at\D{t}+v_o\D{t}}}
    }
    {\lseqalign{\der{x=x_0+\frac{a}{2}t^2+v_0t}}{\lbisubjunct\D{x}=at\D{t}+v_o\D{t}}}
    }%
    {\lsequent{} {{v=v_0+at} \limply \dbox{\Dupdate{\Dumod{\D{x}}{v}}}{\dbox{\Dupdate{\Dumod{\D{t}}{1}}}{\der{x=x_0+\frac{a}{2}t^2+v_0t}}}}}
  }%
  {\lsequent{} {\dbox{\pevolvein{\D{x}=v\syssep\D{v}=a\syssep\D{t}=1}{v=v_0+at}}{({v=v_0+at}\limply\dbox{\Dupdate{\Dumod{\D{x}}{v}}}{\dbox{\Dupdate{\Dumod{\D{t}}{1}}}{\der{x=x_0+\frac{a}{2}t^2+v_0t}}})}}}
  }%
  {\lsequent{} {\dbox{\pevolvein{\D{x}=v\syssep\D{v}=a\syssep\D{t}=1}{v=v_0+at}}{\dbox{\Dupdate{\Dumod{\D{x}}{v}}}{\dbox{\Dupdate{\Dumod{\D{t}}{1}}}{\der{x=x_0+\frac{a}{2}t^2+v_0t}}}}}}
 }%
  {\lsequent{} {\dbox{\pevolvein{\D{x}=v\syssep\D{v}=a\syssep\D{t}=1}{v=v_0+at}}{\der{x=x_0+\frac{a}{2}t^2+v_0t}}}}
}%
{\lsequent{\prem} {\dbox{\pevolvein{\D{x}=v\syssep\D{v}=a\syssep\D{t}=1}{v=v_0+at}}{x=x_0+\frac{a}{2}t^2+v_0t}}}
\end{sequentdeduction}
}%
\fi

\subsection{Differential Substitution Lemmas}

The key insight for the soundness of \irref{DI} is that the analytic time-derivative of the value of a term $\eta$ along a differential equation \(\pevolvein{\D{x}=\genDE{x}}{\ivr}\) agrees with the values of its differential $\der{\eta}$ along the vector field of that differential equation.

\begin{lemma}[Differential lemma] \label{lem:differentialLemma}
  If \m{\imodels{\If}{\D{x}=\genDE{x}\land\ivr}}
  holds for some flow \m{\iget[flow]{\If}:[0,r]\to\linterpretations{\Sigma}{V}} 
of any duration $r>0$,
  then for all $0\leq\zeta\leq r$:
  \[
  \ivaluation{\Iff[\zeta]}{\der{\eta}}
  = \D[t]{\ivaluation{\Iff[t]}{\eta}} (\zeta)
  \]
\end{lemma}
\begin{proofatend}
\def\Ifzm{\imodif[state]{\Iff[\zeta]}{x}{X}}%
\newcommand{\Idot}{\vdLint[const=I,state=]}
By chain rule \cite[\S3.10]{Walter:Ana2}:
\[
\D[t]{\ivaluation{\Iff[t]}{\eta}} (\zeta)
=
\D{(\ivaluation{\Idot}{\eta} \compose \iget[flow]{\If})}(\zeta) = (\gradient{\ivaluation{\Idot}{\eta}})\big(\iget[flow]{\If}(\zeta)\big)
\stimes \D{\iget[flow]{\If}}(\zeta)
= \sum_x \Dp[x]{\ivaluation{\Idot}{\eta}}\big(\iget[flow]{\If}(\zeta)\big) \D{\iget[flow]{\If}}(\zeta)(x)
\]
where \((\gradient{\ivaluation{\Idot}{\eta}})\big(\iget[flow]{\If}(\zeta)\big)\), the spatial gradient \(\gradient{\ivaluation{\Idot}{\eta}}\)
at \(\iget[flow]{\If}(\zeta)\),
is the vector of
\(\Dp[x]{\ivaluation{\Idot}{\eta}}\big(\iget[flow]{\If}(\zeta)\big)
= \Dp[X]{\ivaluation{\Ifzm}{\eta}}\).
Chain rule and \rref{def:dL-valuationTerm} and\rref{def:HP-transition}, thus, imply:
\[
\ivaluation{\Iff[\zeta]}{\der{\eta}}
= \sum_x \iget[state]{\Iff[\zeta]}(\D{x}) \itimes \Dp[X]{\ivaluation{\Ifzm}{\eta}}
= \sum_x \Dp[X]{\ivaluation{\Ifzm}{\eta}} \itimes \D[t]{\iget[state]{\Iff[t]}(x)}(\zeta)
= \D[t]{\ivaluation{\Iff[t]}{\eta}} (\zeta)
\qedhere
\]
\end{proofatend}

The key insight for the soundness of differential effects \irref{DE} is that differential assignments mimicking the differential equation are vacuous along that differential equation.
The differential substitution resulting from a subsequent use of \irref{Dassignb} is crucial to relay the values of the time-derivatives of the state variables $x$ along a differential equation by way of their corresponding differential symbol $\D{x}$.
In combination, this makes it possible to soundly substitute the right-hand side of a differential equation for its left-hand side in a proof.

\begin{lemma}[Differential assignment] \label{lem:differentialAssignLemma}
  If \m{\imodels{\If}{\D{x}=\genDE{x}\land\ivr}}
  for some flow \m{\iget[flow]{\If}:[0,r]\to\linterpretations{\Sigma}{V}} 
of any duration $r\geq0$,
  then
  \[
  \imodels{\If}{\phi \lbisubjunct \dbox{\Dupdate{\Dumod{\D{x}}{\genDE{x}}}}{\phi}}
  \]
\end{lemma}
\begin{proofatend}
\m{\imodels{\If}{\D{x}=\genDE{x}\land\ivr}} implies
\(\imodels{\Iff[\zeta]}{\D{x}=\genDE{x}\land\ivr}\),
i.e. \(\iget[state]{\Iff[\zeta]}(\D{x}) = \ivaluation{\Iff[\zeta]}{\genDE{x}}\) and \(\imodels{\Iff[\zeta]}{\ivr}\)
for all $0\leq \zeta\leq r$.
Consequently \(\iaccessible[\Dupdate{\Dumod{\D{x}}{\genDE{x}}}]{\Iff[\zeta]}{\Iff[\zeta]}\)
does not change the state, so that $\phi$ and \(\dbox{\Dupdate{\Dumod{\D{x}}{\genDE{x}}}}{\phi}\) are equivalent along $\iget[flow]{\If}$.
\qedhere
\end{proofatend}

The final insights for differential invariant reasoning for differential equations are syntactic ways of computing differentials, which can be internalized as axioms (\irref{Dplus+Dtimes+Dcompose}), since differentials are syntactically represented in differential-form \dL. 
\begin{lemma}[Derivations] \label{lem:derivationLemma}
  The following equations of differentials are valid:
  \begin{align}%
    \der{f} & = 0
      &&\text{for arity 0 functions/numbers}~f
    \label{eq:Dconstant}\\
    \der{x} & =  \D{x}
      &&\text{for variables}~x\in\allvars\label{eq:Dpolynomial}\\
    \der{\theta+\eta} & = \der{\theta} + \der{\eta}
    \label{eq:Dadditive}\\
    \der{\theta\cdot \eta} & = \der{\theta}\cdot \eta + \theta\cdot\der{\eta}
    \label{eq:DLeibniz}
    \\
    \dbox{\pupdate{\pumod{y}{\theta}}}{\dbox{\Dupdate{\Dumod{\D{y}}{1}}}
    {}&{\big( \der{f(\theta)} = \der{f(y)}\stimes\der{\theta}\big)}}
    &&\text{for $y,\D{y}\not\in\theta$}
    \label{eq:Dcompose}
  \end{align}%
\end{lemma}
\begin{proofatend}
\def\Im{\imodif[state]{\I}{x}{X}}%
\edef\Imyypyval{\lenvelope\theta\renvelope^I \nu}%
\def\Imyyp{\vdLint[const=I,state=\modif{\nu}{y}{\Imyypyval}\modif{}{\D{y}}{1}]{\I}}%
\newcommand{\Idot}{\vdLint[const=I,state=]}%
The proof shows each equation separately.
The first parts consider any constant function (i.e. arity 0) or number literal $f$ for \rref{eq:Dconstant} and align the differential \(\der{x}\) of a term that happens to be a variable $x\in\allvars$ with its corresponding differential symbol $\D{x}\in\D{\allvars}$ for \rref{eq:Dpolynomial}.
The other cases exploit linearity for \rref{eq:Dadditive} and Leibniz properties of partial derivatives for \rref{eq:DLeibniz}.
Case \rref{eq:Dcompose} exploits the chain rule and assignments and differential assignments for the fresh $y,\D{y}$ to mimic partial derivatives.
Equation \rref{eq:Dcompose} generalizes to functions $f$ of arity $n>1$, in which case $\stimes$ is the (definable) Euclidean scalar product.
\iflongversion
\def\aptag#1{\tag{#1}}%
\else%
\def\aptag#1{\notag&\hspace*{-8pt}(#1)}%
\fi
\begin{align}
\ivaluation{\I}{\der{f}}
&= \sum_x \iget[state]{\I}(\D{x}) \itimes \Dp[X]{\ivaluation{\Im}{f}}
= \sum_x \iget[state]{\I}(\D{x}) \itimes \Dp[X]{\iget[const]{\Im}(f)}
= 0
\aptag{\ref{eq:Dconstant}}
\\
\ivaluation{\I}{\der{x}}
&=
\def\Imy{\imodif[state]{\I}{y}{X}}%
\sum_y \iget[state]{\I}(\D{y}) \itimes \Dp[X]{\ivaluation{\Imy}{x}}
= \iget[state]{\I}(\D{x}) \itimes \Dp[X]{\ivaluation{\Im}{x}}
= \iget[state]{\I}(\D{x}) \itimes \Dp[X]{X}
= \iget[state]{\I}(\D{x})
= \ivaluation{\I}{\D{x}}
\aptag{\ref{eq:Dpolynomial}}
\\
\ivaluation{\I}{\der{\theta+\eta}}
&= \sum_x \iget[state]{\I}(\D{x}) \itimes \Dp[X]{\ivaluation{\Im}{\theta+\eta}}
= \sum_x \iget[state]{\I}(\D{x}) \itimes \Dp[X]{(\ivaluation{\Im}{\theta}+\ivaluation{\Im}{\eta})}
\notag
\\&= \sum_x \iget[state]{\I}(\D{x}) \itimes \Big(\Dp[X]{\ivaluation{\Im}{\theta}} + \Dp[X]{\ivaluation{\Im}{\eta}}\Big)
\notag
\\&= \sum_x \iget[state]{\I}(\D{x}) \itimes \Dp[X]{\ivaluation{\Im}{\theta}} + \sum_x \iget[state]{\I}(\D{x}) \itimes \Dp[X]{\ivaluation{\Im}{\eta}}
\notag
\\&= \ivaluation{\I}{\der{\theta}} + \ivaluation{\I}{\der{\eta}}
= \ivaluation{\I}{\der{\theta} + \der{\eta}}
\aptag{\ref{eq:Dadditive}}
\\
\ivaluation{\I}{\der{\theta\cdot\eta}}
&= \sum_x \iget[state]{\I}(\D{x}) \itimes \Dp[X]{\ivaluation{\Im}{\theta\cdot\eta}}
= \sum_x \iget[state]{\I}(\D{x}) \itimes \Dp[X]{(\ivaluation{\Im}{\theta}\cdot\ivaluation{\Im}{\eta})}
\notag
\\
&= \sum_x \iget[state]{\I}(\D{x}) \itimes \Big(\ivaluation{\I}{\eta} \itimes \Dp[X]{\ivaluation{\Im}{\theta}}
+ \ivaluation{\I}{\theta} \itimes \Dp[X]{\ivaluation{\Im}{\eta}}\Big)
\notag
\\
&=
\ivaluation{\I}{\eta} \sum_x \iget[state]{\I}(\D{x}) \itimes \Dp[X]{\ivaluation{\Im}{\theta}}
+ \ivaluation{\I}{\theta} \sum_x \iget[state]{\I}(\D{x}) \itimes \Dp[X]{\ivaluation{\Im}{\eta}}
\notag
\\
&= 
\ivaluation{\I}{\der{\theta}}\cdot \ivaluation{\I}{\eta} + \ivaluation{\I}{\theta}\cdot\ivaluation{\I}{\der{\eta}}
= \ivaluation{\I}{\der{\theta}\cdot \eta + \theta\cdot\der{\eta}}
\aptag{\ref{eq:DLeibniz}}
\intertext{Proving that
\(\imodels{\I}{\dbox{\pupdate{\pumod{y}{\theta}}}{\dbox{\Dupdate{\Dumod{\D{y}}{1}}}{\big(\der{f(\theta)} = \der{f(y)}\stimes\der{\theta}\big)}}}\)
requires showing that
\newline
\(\imodels{\Imyyp}{\der{f(\theta)} = \der{f(y)}\stimes\der{\theta}}\),
i.e.\
\(\ivaluation{\Imyyp}{\der{f(\theta)}} = \ivaluation{\Imyyp}{\der{f(y)}\stimes\der{\theta}}\).
This is equivalent to
\(\ivaluation{\I}{\der{f(\theta)}} = \ivaluation{\Imyyp}{\der{f(y)}}\stimes\ivaluation{\I}{\der{\theta}}\)
by \rref{lem:coincidence-term} since
\(\iget[state]{\I}=\iget[state]{\Imyyp}\) on $\scomplement{\{y,\D{y}\}}$ and
\(y,\D{y}\not\in\freevars{\theta}\) by assumption,
so \(y,\D{y}\not\in\freevars{\der{f(\theta)}}\) and \(y,\D{y}\not\in\freevars{\der{\theta}}\).
The latter equation proves using the chain rule and a fresh variable $z$ when denoting \(\ivaluation{\Idot}{f} \mdefeq \iget[const]{\Idot}(f)\):}
    \ivaluation{\I}{\der{f(\theta)}} & =
    \sum_x \iget[state]{\I}(\D{x}) \Dp[x]{\ivaluation{\Idot}{f(\theta)}}(\iget[state]{\I})
    = \sum_x \iget[state]{\I}(\D{x}) \Dp[x]{(\ivaluation{\Idot}{f}\compose\ivaluation{\Idot}{\theta})}(\iget[state]{\I})
  \notag
    \\&
     \stackrel{\text{chain}}{=} \sum_x \iget[state]{\I}(\D{x}) \Dp[y]{\ivaluation{\Idot}{f}}\big(\ivaluation{\I}{\theta}\big) \stimes \Dp[x]{\ivaluation{\Idot}{\theta}}(\iget[state]{\I})
\notag
     \\&
     = \Dp[y]{\ivaluation{\Idot}{f}}\big(\ivaluation{\I}{\theta}\big) \stimes \sum_x \iget[state]{\I}(\D{x}) \Dp[x]{\ivaluation{\Idot}{\theta}}(\iget[state]{\I})
     = \Dp[y]{\ivaluation{\Idot}{f}}\big(\ivaluation{\I}{\theta}\big) \stimes \ivaluation{\I}{\der{\theta}}
\notag
     \\&= \Dp[y]{\iget[const]{\Idot}(f)}\big(\ivaluation{\I}{\theta}\big) \stimes \ivaluation{\I}{\der{\theta}}
\notag
    \\&=
    \Dp[z]{\iget[const]{\Idot}(f)}(\Imyypyval) \itimes 1 \stimes \ivaluation{\I}{\der{\theta}}
\notag
    \\&=
    \Dp[z]{\iget[const]{\Idot}(f)}\big(\ivaluation{\Imyyp}{y}\big) \itimes \Dp[y]{\ivaluation{\Idot}{y}}(\iget[state]{\Imyyp}) \stimes \ivaluation{\I}{\der{\theta}}
\notag
    \\&\stackrel{\text{chain}}{=}
    \Dp[y]{(\iget[const]{\Idot}(f)\compose\ivaluation{\Idot}{y})}(\iget[state]{\Imyyp}) \stimes \ivaluation{\I}{\der{\theta}}
\notag
    \\&=
    \left(\Dp[y]{\ivaluation{\Idot}{f(y)}}(\iget[state]{\Imyyp})\right) \stimes \ivaluation{\I}{\der{\theta}}
\notag
    \\&=
    \left(\iget[state]{\Imyyp}(\D{y}) \Dp[y]{\ivaluation{\Idot}{f(y)}}(\iget[state]{\Imyyp})\right) \stimes \ivaluation{\I}{\der{\theta}}
\notag
    \\&=
    \left(\sum_{x\in\{y\}} \iget[state]{\Imyyp}(\D{x}) \Dp[x]{\ivaluation{\Idot}{f(y)}}(\iget[state]{\Imyyp})\right) \stimes \ivaluation{\I}{\der{\theta}}
\notag
    \\&=
    \ivaluation{\Imyyp}{\der{f(y)}} \stimes \ivaluation{\I}{\der{\theta}}
    \aptag{\ref{eq:Dcompose}}
    \qedhere
\end{align}
\qedhere
\end{proofatend}

\subsection{Soundness}

\begin{theorem}[Soundness] \label{thm:dL-sound}
  The \dL axioms and proof rules in \rref{fig:dL}, \ref{fig:dL-ODE} are sound, i.e.\ the axioms are valid formulas and the conclusion of a rule is valid if its premises are.
  All \irref{US} instances of the proof rules (with $\freevars{\sigma}=\emptyset$) are sound.
\end{theorem}
\begin{proofatend}
The axioms (and most proof rules) in \rref{fig:dL} are special instances of corresponding axiom schemata and proof rules for differential dynamic logic \cite{DBLP:conf/lics/Platzer12b} and, thus, sound.
All proof rules except \irref{US} are even \emph{locally sound}, i.e. for all $\iget[const]{\I}$: if all their premises $\phi_j$ are valid in $\iget[const]{\I}$ (\m{\iget[const]{\I}\models{\phi_j}}) then their conclusion $\psi$ is, too (\m{\iget[const]{\I}\models{\psi}}).
Local soundness implies soundness.
In addition, local soundness implies that \irref{US} can be used to soundly instantiate proof rules just like it soundly instantiates axioms (\rref{thm:usubst-sound}).
If
\begin{equation}
\linfer
{\phi_1 \quad \dots \quad \phi_n}
{\psi}
\label{eq:proofrule}
\end{equation}
is a locally sound proof rule then its substitution instance is locally sound:
\begin{equation}
\linfer
{\applyusubst{\sigma}{\phi_1} \quad \dots \quad \applyusubst{\sigma}{\phi_n}}
{\applyusubst{\sigma}{\psi}}
\label{eq:usubstituted-proofrule}
\end{equation}
where $\sigma$ is any uniform substitution (for which the above results are defined, i.e.\ no clash) with $\freevars{\sigma}=\emptyset$.
To show this, consider any $\iget[const]{\I}$ in which all premises of \rref{eq:usubstituted-proofrule} are valid, i.e.\
\(\iget[const]{\I}\models{\applyusubst{\sigma}{\phi_j}}\) for all $j$.
That is, \(\imodels{\I}{\applyusubst{\sigma}{\phi_j}}\) for all $\iget[state]{\I}$ and all $j$.
By \rref{lem:usubst}, \(\imodels{\I}{\applyusubst{\sigma}{\phi_j}}\) is equivalent to
\(\imodels{\Ia}{\phi_j}\),
which, thus, also holds for all $\iget[state]{\I}$ and all $j$.
By \rref{cor:adjointUsubst}, \(\imodel{\Ia}{\phi_j}=\imodel{\Ita}{\phi_j}\) for any $\iget[state]{\Ita}$, since $\freevars{\sigma}=\emptyset$.
Consequently, all premises of \rref{eq:proofrule} are valid in $\iget[const]{\Ita}$, i.e. \(\iget[const]{\Ita}\models{\phi_j}\) for all $j$.
Thus, \(\iget[const]{\Ita}\models{\psi}\) by local soundness of \rref{eq:proofrule}.
That is, \(\imodels{\Ia}{\psi}=\imodel{\Ita}{\psi}\) by \rref{cor:adjointUsubst} for all $\iget[state]{\Ia}$.
By \rref{lem:usubst}, \(\imodels{\Ia}{\psi}\) is equivalent to \(\imodels{\I}{\applyusubst{\sigma}{\psi}}\),
which continues to hold for all $\iget[state]{\I}$.
Thus, \(\iget[const]{\I}\models{\applyusubst{\sigma}{\psi}}\), i.e.\ the conclusion of \rref{eq:usubstituted-proofrule} is valid in $\iget[const]{\I}$, hence \rref{eq:usubstituted-proofrule} locally sound.
Consequently, all \irref{US} instances of the locally sound proof rules of \dL with $\freevars{\sigma}=\emptyset$ are locally sound.
Note that \irref{gena+MP} can be augmented soundly to use $p(\usall)$ instead of $p(x)$ or $p$, respectively, such that the $\freevars{\sigma}=\emptyset$ requirement will be met during \irref{US} instances of all rules.

\begin{compactitem}
\item[\irref{DW}]
Soundness of \irref{DW} uses that differential equations can never leave their evolution domain by \rref{def:HP-transition}.
To show \(\imodels{\I}{\dbox{\pevolvein{\D{x}=f(x)}{q(x)}}{q(x)}}\), consider any $\iget[flow]{\If}$ of any duration $r\geq0$ solving \(\imodels{\If}{\D{x}=f(x)\land q(x)}\).
Then \(\imodels{\If}{q(x)}\) hence \(\imodels{\Iff[r]}{q(x)}\).

\item[\irref{DC}]
Soundness of \irref{DC} is a stronger version of  soundness for the differential cut rule \cite{DBLP:journals/logcom/Platzer10}.
\irref{DC} is a differential version of the modal modus ponens \irref{K}.
The core is that if $r(x)$ always holds after the differential equation
and $p(x)$ always holds after the differential equation \(\pevolvein{\D{x}=f(x)}{q(x)\land r(x)}\) that is restricted to $r(x)$,
then $p(x)$ always holds after the differential equation \(\pevolvein{\D{x}=f(x)}{q(x)}\) without that additional restriction.
Let \(\imodels{\I}{\dbox{\pevolvein{\D{x}=f(x)}{q(x)}}{r(x)}}\).
Since all restrictions of solutions are solutions, this is equivalent to
\(\imodels{\If}{r(x)}\) for all $\iget[flow]{\If}$ of any duration solving \(\imodels{\If}{\D{x}=f(x)\land q(x)}\) and starting in \(\iget[state]{\Iff[0]}=\iget[state]{\I}\) on $\scomplement{\{\D{x}\}}$.
Consequently, for all $\iget[flow]{\If}$ starting in \(\iget[state]{\Iff[0]}=\iget[state]{\I}\) on $\scomplement{\{\D{x}\}}$:
\(\imodels{\If}{\D{x}=f(x)\land q(x)}\) is equivalent to
\(\imodels{\If}{\D{x}=f(x)\land q(x) \land r(x)}\).
Hence, \(\imodels{\I}{\dbox{\pevolvein{\D{x}=f(x)}{q(x)\land r(x)}}{p(x)}}\)
is equivalent to \(\imodels{\I}{\dbox{\pevolvein{\D{x}=f(x)}{q(x)}}{p(x)}}\).

\item[\irref{DE}]
Soundness of \irref{DE} is genuine to differential-form \dL leveraging \rref{lem:differentialAssignLemma}.
Consider any state $\iget[state]{\I}$.
Then
\(\imodels{\I}{\dbox{\pevolvein{\D{x}=f(x)}{q(x)}}{p(x,\D{x})}}\)
iff
\(\imodels{\Iff[r]}{p(x,\D{x})}\)
for all solutions $\iget[flow]{\If}:[0,r]\to\linterpretations{\Sigma}{V}$ of
\(\imodels{\If}{\D{x}=f(x)\land q(x)}\) of any duration $r$ starting in \(\iget[state]{\Iff[0]}=\iget[state]{\I}\) on $\scomplement{\{\D{x}\}}$.
That is equivalent to: for all $\iget[flow]{\If}$,
if
\(\imodels{\If}{\D{x}=f(x)\land q(x)}\)
then
\(\imodels{\If}{p(x,\D{x})}\).
By \rref{lem:differentialAssignLemma}, \(\imodels{\If}{p(x,\D{x})}\) iff
\(\imodels{\If}{\dbox{\Dupdate{\Dumod{\D{x}}{f(x)}}}{p(x,\D{x})}}\),
so, that is equivalent to
\(\imodels{\Iff[r]}{\dbox{\Dupdate{\Dumod{\D{x}}{f(x)}}}{p(x,\D{x})}}\)
for all solutions $\iget[flow]{\If}:[0,r]\to\linterpretations{\Sigma}{V}$ of
\(\imodels{\If}{\D{x}=f(x)\land q(x)}\) of any duration $r$ starting in \(\iget[state]{\Iff[0]}=\iget[state]{\I}\) on $\scomplement{\{\D{x}\}}$,
which is, consequently, equivalent to
\(\imodels{\I}{\dbox{\pevolvein{\D{x}=f(x)}{q(x)}}{\dbox{\Dupdate{\Dumod{\D{x}}{f(x)}}}{p(x,\D{x})}}}\).

\item[\irref{DI}]
Soundness of \irref{DI} has some relation to the soundness proof for differential invariants \cite{DBLP:journals/logcom/Platzer10}, yet is generalized to leverage differentials.
The proof is only shown for \m{p(x)\mdefequiv g(x)\geq0}, in which case \(\der{p(x)} \mequiv (\der{g(x)}\geq0)\).
Consider a state $\iget[state]{\I}$ in which
\\
\(\imodels{\I}{q(x) \limply (p(x) \land \dbox{\pevolvein{\D{x}=f(x)}{q(x)}}{\der{p(x)}}})\).
If \(\inonmodels{\I}{q(x)}\), there is nothing to show, because there is no solution of \({\pevolvein{\D{x}=f(x)}{q(x)}}\) for any duration, so the consequence holds vacuously.
Otherwise, \(\imodels{\I}{q(x)}\) implies \(\imodels{\I}{p(x) \land \dbox{\pevolvein{\D{x}=f(x)}{q(x)}}{\der{p(x)}}}\).
To show that
\(\imodels{\I}{\dbox{\pevolvein{\D{x}=f(x)}{q(x)}}{p(x)}}\) consider any solution $\iget[flow]{\If}$ of any duration $r\geq0$.
The case $r=0$ follows from \(\imodels{\I}{p(x)}\) by \rref{lem:coincidence} since \(\freevars{p(x)}=\{x\}\) is disjoint from \(\{\D{x}\}\), which is changed by evolutions of any duration.
That leaves the case \(r>0\).

Let \(\imodels{\If}{\pevolvein{\D{x}=f(x)}{q(x)}}\),
which, by
\(\imodels{\I}{\dbox{\pevolvein{\D{x}=f(x)}{q(x)}}{\der{p(x)}}}\),
implies
\(\imodels{\If}{\der{p(x)}}\).
Since \(r>0\), \rref{lem:differentialLemma} implies
\(0\leq\ivaluation{\Iff[\zeta]}{\der{g(x)}}
= \D[t]{\ivaluation{\Iff[t]}{g(x)}} (\zeta)\)
for all $\zeta$.
Together with \(\imodels{\Iff[0]}{p(x)}\) (by \rref{lem:coincidence} and \(\freevars{p(x)}\cap\{\D{x}\}=\emptyset\)), i.e.\
\(\imodels{\Iff[0]}{g(x)\geq0}\),
this implies \(\imodels{\Iff[\zeta]}{g(x)\geq0}\) for all $\zeta$, including $r$,
by the mean-value theorem since \(\ivaluation{\Iff[t]}{g(x)}\) is continuous in $t$ on $[0,r]$ and differentiable on $(0,r)$.

\item[\irref{DG}]
\def\Imd{\imodif[state]{\I}{x}{d}}%
\newcommand{\Ify}{\DALint[const=I,flow=\tilde{\varphi}]}
\newcommand{\Iffy}[1][t]{\vdLint[const=I,state=\tilde{\varphi}(#1)]}
Soundness of \irref{DG} is a constructive variation of the soundness proof for differential auxiliaries \cite{DBLP:journals/lmcs/Platzer12}.
Let \(\imodels{\I}{\lexists{y}{\dbox{\pevolvein{\D{x}=f(x)\syssep\D{y}=a(x)y+b(x)}{q(x)}}{p(x)}}}\),
that is,
\\
\(\imodels{\Imd}{\dbox{\pevolvein{\D{x}=f(x)\syssep\D{y}=a(x)y+b(x)}{q(x)}}{p(x)}}\) for some $d$.
In order to show that
\\
\(\imodels{\I}{\dbox{\pevolvein{\D{x}=f(x)}{q(x)}}{p(x)}}\),
consider any \(\iget[flow]{\If}:[0,r]\to\linterpretations{\Sigma}{V}\) such that
\(\imodels{\If}{\D{x}=f(x)\land q(x)}\) and \(\iget[state]{\Iff[0]}=\iget[state]{\I}\) on $\scomplement{\{\D{x}\}}$.
By modifying the values of $y,\D{y}$ along $\iget[flow]{\If}$, this function can be augmented to a solution 
\(\iget[flow]{\Ify}:[0,r]\to\linterpretations{\Sigma}{V}\) such that
\(\imodels{\Ify}{\D{x}=f(x)\land\D{y}=a(x)y+b(x)\land q(x)}\) and \(\iget[state]{\Iffy[0]}(y)=d\).
The assumption then implies \(\imodels{\Iffy[r]}{p(x)}\), which, by \rref{lem:coincidence}, is equivalent to \(\imodels{\Iff[r]}{p(x)}\) since \(y,\D{y}\not\in\freevars{p(x)}\) and \(\iget[state]{\Iff[r]}=\iget[state]{\Iffy[r]}\) on $\scomplement{\{y,\D{y}\}}$,
which implies \(\imodels{\I}{\dbox{\pevolvein{\D{x}=f(x)}{q(x)}}{p(x)}}\), since $\iget[flow]{\If}$ was arbitrary.
The construction of the modification $\iget[flow]{\Ify}$ of $\iget[flow]{\If}$ on $\{y,\D{y}\}$ proceeds as follows.
By Picard-Lindel\"of \cite[\S10.VII]{Walter:ODE}, there is a solution \(y:[0,r]\to\reals\) of the initial-value problem
\begin{equation}
\begin{aligned}
  y(0) &=d\\
  \D{y}(t) &= F(t,y(t)) \mdefeq y(t) \ivaluation{\Iff[t]}{a(x)} + \ivaluation{\Iff[t]}{b(x)}
\end{aligned}
\label{eq:diffghost-extra-ODE}
\end{equation}
because \(F(t,y)\) is continuous on \([0,r]\times\reals\)
(since \(\ivaluation{\Iff[t]}{a(x)}\) and \(\ivaluation{\Iff[t]}{b(x)}\) are continuous in $t$ as compositions of the, by \rref{def:dL-valuationTerm} smooth, evaluation function and the continuous solution $\iget[state]{\Iff[t]}$ of a differential equation)
and because \(F(t,y)\) satisfies the Lipschitz condition
\[
\norm{F(t,y)-F(t,z)} = \norm{(y-z)\ivaluation{\Iff[t]}{a(x)}} \leq \norm{y-z} \max_{t\in[0,r]} \ivaluation{\Iff[t]}{a(x)}
\]
where the maximum exists, because it is a maximum of a continuous function on the compact set $[0,r]$.
The modification $\iget[flow]{\Ify}$ agrees with $\iget[flow]{\If}$ on $\scomplement{\{y,\D{y}\}}$ and is defined as \(\iget[state]{\Iffy[t]}(y) = y(t)\) and \(\iget[state]{\Iffy[t]}(\D{y}) = F(t,y(t)) = \D{y}(t)\) on $\{y,\D{y}\}$, respectively, for the solution $y(t)$ of \rref{eq:diffghost-extra-ODE}.
By construction, \(\iget[state]{\Iffy[0]}(y)=d\) and
\(\imodels{\Ify}{\D{x}=f(x)\land\D{y}=a(x)y+b(x)\land q(x)}\),
because \(\iget[state]{\Iff[t]}=\iget[state]{\Iffy[t]}\) on $\scomplement{\{y,\D{y}\}}$ so that
\(\pevolvein{\D{x}=f(x)}{q(x)}\) continues to hold along $\iget[flow]{\Ify}$ by \rref{lem:coincidence-term} because \(y,\D{y}\not\in\freevars{\pevolvein{\D{x}=f(x)}{q(x)}}\),
and because \(\D{y}=a(x)y+b(x)\) holds along $\iget[flow]{\Ify}$ by \rref{eq:diffghost-extra-ODE}.

\def\Imyd{\imodif[state]{\I}{y}{d}}%
\newcommand{\Ifrest}{\DALint[const=I,flow=\restrict{\varphi}{\scomplement{\{y,\D{y}\}}}]}%
\newcommand*{\Iffrest}[1][\zeta]{\vdLint[const=I,state=\restrict{\varphi}{\scomplement{\{y,\D{y}\}}}(#1)]}%
Conversely, let \(\imodels{\I}{\dbox{\pevolvein{\D{x}=f(x)}{q(x)}}{p(x)}}\).
This direction shows a stronger version of
\(\imodels{\I}{\lexists{y}{\dbox{\pevolvein{\D{x}=f(x)\syssep\D{y}=a(x)y+b(x)}{q(x)}}{p(x)}}}\)
by showing that
\\
\(\imodels{\Imyd}{\dbox{\pevolvein{\D{x}=f(x)\syssep\D{y}=\eta}{q(x)}}{p(x)}}\)
for all $d\in\reals$ and all terms $\eta$.
Consider any \(\iget[flow]{\If}:[0,r]\to\linterpretations{\Sigma}{V}\) such that
\(\imodels{\If}{\D{x}=f(x)\land\D{y}=\eta\land q(x)}\)
with \(\iget[state]{\Iff[0]}=\iget[state]{\Imyd}\) on $\scomplement{\{\D{x},\D{y}\}}$.
Then the restriction \(\iget[flow]{\Ifrest}\) of $\iget[flow]{\If}$ to $\scomplement{\{y,\D{y}\}}$ with \(\iget[state]{\Iffrest[t]}=\iget[state]{\Imyd}\) on $\{y,\D{y}\}$ for all $t\in[0,r]$ still solves
\(\imodels{\Ifrest}{\D{x}=f(x)\land q(x)}\) by \rref{lem:coincidence-term} since \(\iget[flow]{\Ifrest}=\iget[flow]{\If}\) on $\scomplement{\{y,\D{y}\}}$ and \(y,\D{y}\not\in\freevars{\pevolvein{\D{x}=f(x)}{q(x)}}\).
It also satisfies \(\iget[state]{\Iffrest[0]}=\iget[state]{\Imyd}\) on $\scomplement{\{\D{x}\}}$,
because \(\iget[state]{\Iff[0]}=\iget[state]{\Imyd}\) on $\scomplement{\{\D{x},\D{y}\}}$ yet \(\iget[state]{\Iffrest[t]}(\D{y})=\iget[state]{\Imyd}(\D{y})\).
Thus, by assumption, \(\imodels{\Iffrest[r]}{p(x)}\),
which implies \(\imodels{\Iff[r]}{p(x)}\)
by \rref{lem:coincidence}, because
\(\iget[flow]{\If}=\iget[flow]{\Ifrest}\) on $\scomplement{\{y,\D{y}\}}$ and $y,\D{y}\not\in\freevars{p(x)}$,

\item[\irref{DS}]
\def\Izeta{\imodif[state]{\I}{t}{\zeta}}%
\def\constODE{f}%
Soundness of the solution axiom \irref{DS} follows from existence and uniqueness of global solutions of constant differential equations.
Consider any state $\iget[state]{\I}$.
There is a unique \cite[\S10.VII]{Walter:ODE} global solution $\iget[flow]{\If}:[0,\infty)\to\linterpretations{\Sigma}{V}$ defined as \(\iget[state]{\Iff[\zeta]}(x) \mdefeq \ivaluation{\Izeta}{x+\constODE t}\)
and \(\iget[state]{\Iff[\zeta]}(\D{x}) \mdefeq \D[t]{\iget[state]{\Iff[t]}(x)}(\zeta) = \iget[const]{\Izeta}(\constODE)\)
and \(\iget[state]{\Iff[\zeta]} = \iget[state]{\I}\) on $\scomplement{\{x,\D{x}\}}$.
This solution satisfies
\(\iget[state]{\Iff[0]}=\iget[state]{\I}(x)\) on $\scomplement{\{\D{x}\}}$ 
and
  \m{\imodels{\If}{\D{x}=\constODE}},
  i.e.
  \(\imodels{\Iff[\zeta]}{\D{x}=\constODE}\)
  for all \(0\leq \zeta\leq r\).
All solutions of \(\pevolve{\D{x}=\constODE}\) from initial state $\iget[state]{\I}$ are restrictions of $\iget[flow]{\If}$ to subintervals of $[0,\infty)$.
The (unique) state $\iget[state]{\It}$ that satisfies \(\iaccessible[\pupdate{\pumod{x}{x+\constODE t}}]{\Izeta}{\It}\) agrees with \(\iget[state]{\It}=\iget[state]{\Iff[\zeta]}\) on $\scomplement{\{\D{x}\}}$, so that, by $\D{x}\not\in\freevars{p(x)}$, \rref{lem:coincidence} implies that \(\imodels{\It}{p(x)}\) iff \(\imodels{\Iff[\zeta]}{p(x)}\).

First consider axiom
\({\dbox{\pevolve{\D{x}=\constODE}}{p(x)}} \lbisubjunct {\lforall{t{\geq}0}{\dbox{\pupdate{\pumod{x}{x+\constODE t}}}{p(x)}}}\) for $q(x)\mequiv\ltrue$.
If
\\
\(\imodels{\I}{\dbox{\pevolve{\D{x}=\constODE}}{p(x)}}\),
then \(\imodels{\Iff[\zeta]}{p(x)}\) for all $\zeta\geq0$,
because the restriction of $\iget[flow]{\If}$ to $[0,\zeta)$ solves \(\D{x}=\constODE\) from $\iget[state]{\I}$,
thus \(\imodels{\It}{p(x)}\),
which implies
\(\imodels{\Izeta}{\dbox{\pupdate{\pumod{x}{x+\constODE t}}}{p(x)}}\),
so \(\imodels{\I}{\lforall{t{\geq}0}{\dbox{\pupdate{\pumod{x}{x+\constODE t}}}{p(x)}}}\) as $\zeta\geq0$ was arbitrary.

Conversely, \(\imodels{\I}{\lforall{t{\geq}0}{\dbox{\pupdate{\pumod{x}{x+\constODE t}}}{p(x)}}}\)
implies \(\imodels{\Izeta}{\dbox{\pupdate{\pumod{x}{x+\constODE t}}}{p(x)}}\)
for all $\zeta\geq0$,
i.e. $\imodels{\It}{p(x)}$ when \(\iaccessible[\pupdate{\pumod{x}{x+\constODE t}}]{\Izeta}{\It}\).
\rref{lem:coincidence} again implies \(\imodels{\Iff[\zeta]}{p(x)}\) for all $\zeta\geq0$, so \(\imodels{\I}{\dbox{\pevolve{\D{x}=\constODE}}{p(x)}}\), since all solutions are restrictions of $\iget[flow]{\If}$.

Soundness of \irref{DS} now follows using that all solutions $\iget[flow]{\If}:[0,r]\to\linterpretations{\Sigma}{V}$ of \(\pevolvein{\D{x}=f(x)}{q(x)}\) satisfy \(\imodels{\Iff[\zeta]}{q(x)}\) for all $0\leq\zeta\leq r$, which, using \rref{lem:coincidence} as above, is equivalent to \(\imodels{\I}{\lforall{0{\leq}s{\leq}t}{q(x+\constODE s)}}\) when \(\iget[state]{\I}(t)=r\).

\item[\irref{Dassignb}]
Soundness of \irref{Dassignb} follows from the semantics of differential assignments (\rref{def:HP-transition}) and compositionality.
In detail: \m{\Dupdate{\Dumod{\D{x}}{f}}} changes the value of symbol $\D{x}$ to the value of $f$.
The predicate $p$ has the same value for arguments $\D{x}$ and $f$ that have the same value.

\item[\irref{Dplus+Dtimes+Dcompose}] Soundness of the derivation axioms \irref{Dplus+Dtimes+Dcompose} follows from \rref{lem:derivationLemma}, since they are special instances of \rref{eq:Dadditive} and \rref{eq:DLeibniz} and \rref{eq:Dcompose}, respectively.
For \irref{Dcompose} observe that $y,\D{y}\not\in g(x)$.

\item[\irref{G}]
Let the premise \(p(\usall)\) be valid in some $\iget[const]{\I}$, i.e.\ \m{\iget[const]{\I}\models{p(\usall)}}, i.e.\ \(\imodels{\It}{p(\usall)}\) for all $\iget[state]{\It}$.
Then, the conclusion \(\dbox{a}{p(\usall)}\) is valid in the same $\iget[const]{\I}$,
i.e.\ \(\imodels{\I}{\dbox{a}{p(\usall)}}\) for all $\iget[state]{\I}$,
because \(\imodels{\It}{p(\usall)}\) for all $\iget[state]{\It}$, so also for all $\iget[state]{\It}$ with \(\iaccessible[a]{\I}{\It}\).
Thus, \irref{G} is locally sound.

\item[\irref{gena}]
\def\Imd{\imodif[state]{\I}{x}{d}}%
Let the premise \(p(x)\) be valid in some $\iget[const]{\I}$, i.e.\ \m{\iget[const]{\I}\models{p(x)}}, i.e.\ \(\imodels{\It}{p(x)}\) for all $\iget[state]{\It}$.
Then, the conclusion \(\lforall{x}{p(x)}\) is valid in the same $\iget[const]{\I}$,
i.e.\ \(\imodels{\I}{\lforall{x}{p(x)}}\) for all $\iget[state]{\I}$, i.e.\ \(\imodels{\Imd}{p(x)}\) for all $d\in\reals$,
because \(\imodels{\It}{p(x)}\) for all $\iget[state]{\It}$, so in particular for all $\iget[state]{\It}=\iget[state]{\Imd}$ for any $d\in\reals$.
Thus, \irref{gena} is locally sound.

\item[\irref{CQ}]
Let the premise \(f(\usall)=g(\usall)\) be valid in some $\iget[const]{\I}$, i.e. \m{\iget[const]{\I}\models{f(\usall)=g(\usall)}}, i.e.\ \(\imodels{\I}{f(\usall)=g(\usall)}\) for all $\iget[state]{\I}$,
i.e.\ \(\ivaluation{\I}{f(\usall)}=\ivaluation{\I}{g(\usall)}\) for all $\iget[state]{\I}$.
Consequently, \(\ivaluation{\I}{f(\usall)}\in\iget[const]{\I}(p)\) iff \(\ivaluation{\I}{g(\usall)}\in\iget[const]{\I}(p)\).
So, \(\iget[const]{\I}\models {p(f(\usall)) \lbisubjunct p(g(\usall))}\).
Thus, \irref{CQ} is locally sound.

\item[\irref{CE}]
Let the premise \(p(\usall)\lbisubjunct q(\usall)\) be valid in some $\iget[const]{\I}$, i.e. \m{\iget[const]{\I}\models{p(\usall) \lbisubjunct q(\usall)}}, i.e.\ \(\imodels{\I}{p(\usall) \lbisubjunct q(\usall)}\) for all $\iget[state]{\I}$.
Consequently, \(\imodel{\I}{p(\usall)} = \imodel{\I}{q(\usall)}\).
Thus,
\(\imodel{\I}{\contextapp{C}{p(\usall)}}
= \iget[const]{\I}(C)\big(\imodel{\I}{p(\usall)}\big)
= \iget[const]{\I}(C)\big(\imodel{\I}{q(\usall)}\big)
= \imodel{\I}{\contextapp{C}{q(\usall)}}\).
This implies
\(\iget[const]{\I}\models{\contextapp{C}{p(\usall)} \lbisubjunct \contextapp{C}{q(\usall)}}\),
hence the conclusion is valid in $\iget[const]{\I}$.
Thus, \irref{CE} is locally sound.

\item[\irref{CT}]
Rule \irref{CT} is a (locally sound) derived rule and only included for comparison.
\irref{CT} is derivable from \irref{CQ} using \(p(\usarg) \mdefequiv (c(\usarg)=c(g(\usall)))\) and reflexivity of $=$.

\item[\irref{MP}]
Modus ponens \irref{MP} is locally sound with respect to the interpretation $\iget[const]{\I}$ and the state $\iget[state]{\I}$, which implies local soundness and thus soundness.
If \(\imodels{\I}{p\limply q}\) and \(\imodels{\I}{p}\) then \(\imodels{\I}{q}\).

\item[\irref{US}]
Uniform substitution is sound by \rref{thm:usubst-sound}, just not necessarily locally sound.
\qedhere
\end{compactitem}
\end{proofatend}

\section{Conclusions}

With differential forms for local reasoning about differential equations, uniform substitutions lead to a simple and modular proof calculus for differential dynamic logic that is entirely based on axioms and axiomatic rules, instead of soundness-critical schema variables with side-conditions in axiom schemata.
The \irref{US} calculus is straightforward to implement and enables flexible reasoning with axioms by contextual equivalence.
Efficiency can be regained by tactics that combine multiple axioms and rebalance the proof to obtain short proof search branches.
Contextual equivalence rewriting for implications is possible when adding monotone \predicational{s} $C$ whose substitution instances limit $\uscarg$ to positive polarity.

\paragraph*{Acknowledgment.}
I thank the anonymous reviewers of the conference version \cite{DBLP:conf/cade/Platzer15} for their helpful feedback.

This material is based upon work supported by the National Science Foundation by 
NSF CAREER Award CNS-1054246.
The views and conclusions contained in this document are those of the author and should not be interpreted as representing the official policies, either expressed or implied, of any sponsoring institution or government. Any opinions, findings, and conclusions or recommendations expressed in this publication are those of the author(s) and do not necessarily reflect the views of any sponsoring institution or government.

\appendix

\section{Appendix}
\irlabel{cut|cut}

This appendix briefly discusses generalized uses and forms of the differential ghost axioms and how it generalizes the differential auxiliaries proof rule \cite{DBLP:journals/lmcs/Platzer12}.

\paragraph{Differential Lipschitz Ghosts}
The differential ghost axiom \irref{DG} generalizes to arbitrary Lipschitz-continuous differential equations \m{\pevolve{\D{y}=g(x,y)}}:
\[
      \cinferenceRule[DLG|DG$_\ell$]{differential Lipschitz ghost variables} %
      {\linferenceRule[lpmil]
        {\big({\dbox{\pevolvein{\D{x}=f(x)}{q(x)}}{p(x)}} \lbisubjunct 
        {\lexists{y}{\dbox{\pevolvein{\D{x}=f(x)\syssep\D{y}=g(x,y)}{q(x)}}{p(x)}}}\big)}
        {\lexists{\ell}{\lforall{x,y,z}{\abs{g(x,y)-g(x,z)} \leq \ell\abs{y-z}}}}
      }
      {}%
\]
The soundness argument for \irref{DLG} is an extension of the soundness proof for \irref{DG}.
The direction ``$\lylpmi$'' of \irref{DG} is sound for all differential equations.
The proof for the direction ``$\limply$'' extends the proof for \irref{DG} with an adaptation of the function $F$ from \rref{eq:diffghost-extra-ODE} to the differential equation \m{\pevolve{\D{y}=g(x,y)}}:
\begin{equation}
\begin{aligned}
  y(0) &=d\\
  \D{y}(t) &= F(t,y(t)) \mdefeq \ivaluation{\imodif[state]{\Iff[t]}{y}{y(t)}}{g(x,y)}
\end{aligned}
\label{eq:Lipschitz-diffghost-extra-ODE}
\end{equation}
This function $F(t,\delta)$ is still continuous on $[0,r]\times\reals$ since it is a composition of the continuous evaluation (of the, by assumption, continuous term $g(x,y)$) with the (continuous) composition of the continuous function $\iget[state]{\Iff[t]}$ of $t$ with the continuous modification of the value of variable $y$ to $\delta$.
By assumption $F(t,y)$ is Lipschitz in $y$, since there is an $\ell\in\reals$ such that for all $t,a,b\in\reals$:
\begin{multline*}
\abs{F(t,a)-F(t,b)} = \abs{ \ivaluation{\imodif[state]{\Iff[t]}{y}{a}}{g(x,y)} - \ivaluation{\imodif[state]{\Iff[t]}{y}{b}}{g(x,y)} }
= \abs{\ivaluation{\imodif[state]{\imodif[state]{\Iff[t]}{y}{a}}{z}{b}}{g(x,y) - g(x,z)}}
\\
= \ivaluation{\imodif[state]{\imodif[state]{\Iff[t]}{y}{a}}{z}{b}}{\underbrace{\abs{g(x,y) - g(x,z)}}_{\leq\ell\ivaluation{\imodif[state]{\imodif[state]{\Iff[t]}{y}{a}}{z}{b}}{\abs{y-z}}}}
\leq \ell \abs{a-b}
\end{multline*}
This establishes the only two properties of $F$ that the soundness proof of \irref{DG} was based on.
The existence of a solution \(y:[0,r]\to\reals\) of \rref{eq:Lipschitz-diffghost-extra-ODE} is, thus, established again by Picard-Lindel\"of as needed for the soundness proof.

\paragraph{Differential Auxiliaries Rule}
The differential auxiliaries proof rule \cite{DBLP:journals/lmcs/Platzer12} is derivable from \irref{DG} and monotonicity \irref{M}.

\begin{calculuscollections}{10cm}
\begin{calculus}
\cinferenceRule[diffaux|$DA$]{differential auxiliary variables}
{\linferenceRule[sequent]
  {\lsequent[s]{}{p(x)\lbisubjunct\lexists{y}{r(x,y)}}
  &\lsequent{r(x,y)} {\dbox{\hevolvein{\D{x}=f(x)\syssep\D{y}=g(x,y)}{q(x)}}{r(x,y)}}}
  {\lsequent{p(x)} {\dbox{\hevolvein{\D{x}=f(x)}{q(x)}}{p(x)}}}
}{}%
\end{calculus}
\end{calculuscollections}

\noindent
where $y$ is new and \m{\hevolve{\D{y}=g(x,y),y(0)=y_0}} has a solution $y:[0,\infty)\to\reals^n$ for each $y_0$.

The derivation proceeds as follows (the middle premise uses \irref{vacuousexists} with $y\not\in p(x)$):

{%
\renewcommand{\linferPremissSeparation}{~}%
\begin{sequentdeduction}[array]
\linfer[cut]
  {\lsequent{} {p(x)\lbisubjunct\lexists{y}{r(x,y)}}
  !\linfer[existsinst]%
   {\linfer[DG]
    {\linfer[existsinst]%
     {\linfer[M]
        {\linfer[vacuousexists]%
         {\lsequent{\lexists{y}{r(x,y)}} {p(x)}}
         {\lsequent{r(x,y)} {p(x)}}
        !\lsequent{r(x,y)} {\dbox{\pevolvein{\D{x}=f(x)\syssep\D{y}=g(x,y)}{q(x)}}{r(x,y)}}
        }
        {\lsequent{r(x,y)} {\dbox{\pevolvein{\D{x}=f(x)\syssep\D{y}=g(x,y)}{q(x)}}{p(x)}}}
      }
      {\lsequent{r(x,y)} {\lexists{y}{\dbox{\pevolvein{\D{x}=f(x)\syssep\D{y}=g(x,y)}{q(x)}}{p(x)}}}}
      }
    {\lsequent{r(x,y)} {\dbox{\pevolvein{\D{x}=f(x)}{q(x)}}{p(x)}}}
  }
  {\lsequent{\lexists{y}{r(x,y)}} {\dbox{\pevolvein{\D{x}=f(x)}{q(x)}}{p(x)}}}
  }
  {\lsequent{p(x)} {\dbox{\pevolvein{\D{x}=f(x)}{q(x)}}{p(x)}}}
\end{sequentdeduction}
}%
Using the following duals of \irref{allinst} and  \irref{vacuousall} as well as monotonicity rule \irref{M} \cite{DBLP:journals/tocl/Platzer15} that derives from \irref{G+K}:

\begin{calculuscollections}{\columnwidth}
  \begin{calculus}
      \cinferenceRule[existsinst|$\exists$i]{existential instantiation}
       {p(f) \limply (\lexists{x}{p(x)})}
       {}
       \cinferenceRule[vacuousexists|V$_\exists$]{vacuous existential quantifier}
       {\lexists{x}{p} \limply p}
       {}%
      \cinferenceRule[M|M]{$\dbox{}{}$ monotonic / $\dbox{}{}$-generalization} %
      {\linferenceRule[formula]
        {\phi\limply\psi}
        {\dbox{\alpha}{\phi}\limply\dbox{\alpha}{\psi}}
      }{}
  \end{calculus}
\end{calculuscollections}

\bibliographystyle{plainurl}
\bibliography{platzer,bibliography}

\ifkeepproof
\else
\appendix
\section{Proofs}
\printproofs
\fi
\end{document}